\documentclass[10pt,onecolumn, doublespace]{article}
\usepackage{subfigure}
\usepackage[latin1]{inputenc}
\usepackage{multirow}
\usepackage{graphicx}
\date{}
\begin{document}
\title{Technical Report v2\\Characterization of P2P IPTV Traffic:\\Scaling Analysis}
\author{Thomas Silverston, Olivier Fourmaux and Kav\'e Salamatian\\
Universit\'e Pierre et Marie Curie - Paris VI\\
Laboratoire LIP6/CNRS, UMR 7606\\
104 avenue du pr\'esident Kennedy
75016 Paris, France\\
Email: \{thomas.silverston, olivier.fourmaux, kave.salamatian\}@lip6.fr}
\maketitle
\begin{abstract}
P2P IPTV applications arise on the Internet and will be massively used in the future. 
It is expected that P2P IPTV will contribute to increase the overall Internet traffic. 
In this context, it is important to measure the impact of P2P IPTV on the networks and to characterize this traffic.
During the 2006 FIFA World Cup, we performed an extensive measurement campaign. 
We measured network traffic generated by broadcasting soccer games by the most popular P2P IPTV applications, namely PPLive, PPStream, SOPCast and TVAnts. 
From the collected data, we characterized the P2P IPTV traffic structure at different time scales by using wavelet based transform method.
To the best of our knowledge, this is the first work, which presents a complete multiscale analysis of the P2P IPTV traffic.\\
Our results show that the scaling properties of the TCP traffic present periodic behavior whereas the UDP traffic is stationary and lead to long-range depedency characteristics.
For all the applications, the download traffic has different characteristics than the upload traffic.
The signaling traffic has a significant impact on the download traffic but it has negligible impact on the upload.
Both sides of the traffic and its granularity has to be taken into account to design accurate P2P IPTV traffic models.
\end{abstract}
\section{Introduction}
P2P live streaming applications like P2P IPTV are emerging on the Internet and will be massively used in the future.
The P2P traffic counts already for a large part of the Internet traffic and this is mainly due to P2P file-sharing applications as BitTorrent~\cite{bittorrent} or eDonkey~\cite{edonkey}.
Video streaming services like Youtube~\cite{youtube} appeared only a few months ago but contribute already to an important part of the Internet traffic.
It is expected that P2P IPTV will largely contribute to increase the overall Internet traffic.
It is therefore important to study P2P IPTV traffic and to characterize its properties.\\

The characterization of P2P IPTV traffic will allow us to understand its impact on the network.
P2P IPTV applications have stringent QoS constraints (e.g. bandwidth, delay, jitter) and their traffic characterization will enable to understand their exact needs in network resources.
The knowledge of the traffic properties enables the development of synthetic traffic generation models that are
key input parameters when modeling or simulating these systems.
Indeed, the modeling or simulating steps are necessary to design judiciously applications. 
From a traffic engineering point of view, well understanding P2P IPTV traffic is essential for Internet service providers to forecast their internal traffic and ensure a good provisioning of their network.
And last but not least, global knowledge of the traffic properties will highlight some drawbacks of the applications and will make it possible to improve the design of new P2P IPTV architectures.
For instance, an important concern of these systems is the scalability. 
The traffic characterization may help estimate the impact of overhead traffic generated by the signaling.\\ 

In this paper, we present a multiscale analysis of the structure of the traffic generated by the most popular P2P IPTV applications, namely PPLive~\cite{pplive}, PPStream~\cite{ppstream}, SOPCast~\cite{sopcast}
 and  TVAnts~\cite{tvants}.\\
During the 2006 FIFA World Cup, we performed an extensive measurement campaign. 
We measured the network traffic generated by broadcasting soccer games by the previously mentioned applications. 
The multiscale behavior of the collected traffic is analyzed using a wavelet transform based tool.
In this paper, we  characterize the network traffic of P2P IPTV systems at different time scales and compare their properties.
To the best of our knowledge, this is the first work that does a comparative multiscale characterization of P2P IPTV traffic.\\

Our multiscale P2P IPTV traffic analysis shows significant differences in the scaling behaviors of TCP and UDP traffic.
the TCP traffic presents periodic behavior while the UDP traffic is stationary and presents long-range dependency characteristics, which will affect the quality of the video reception.
The signaling traffic has an impact on the download traffic but it has negligible impact on the upload traffic.
The upload traffic generated by P2P IPTV systems have different scaling characteristics compare to the download traffic 
and both sides of the traffic has to be taken into account to design judiciously P2P IPTV traffic models. 
Moreover, the traffic granularity has to be considered while using traffic models to simulate these systems.

The rest of the paper is organized as follows. 
Firstly, we present the related work in section~\ref{sec:related}.
In section~\ref{sec:experiments}, we give an overview of the measured applications and describe our measurement experiment setup.
In section~\ref{sec:metho}, we present our methodology to analyze the traffic at different time scales.
We present our P2P IPTV traffic analysis in section~\ref{sec:results} and discuss the results in section~\ref{sec:discussion}.
Finally, we conclude the paper and give perspectives in section~\ref{sec:conclusion}. 
\section{Related Work}
\label{sec:related}
Nowadays, an increasing number of  P2P IPTV measurement studies is conducted to analyze the mechanisms of such systems.\\
Zhang~et.~al~\cite{invited:coolstreaming} present the first measurement results about their protocol Donet~\cite{infocom05:donet}, which were deployed on the Internet and called Coolstreaming.
They provide network statistics, like user's behavior in the whole system and the quality of video reception.\\
Hei~et al.~\cite{pplive-mesure}~\cite{pplive-tom07} made a complete measurement of the popular PPLive application.
They made active measurements by instrumentalizing their own crawler and give many architecture and overlay details like buffer size or number of peers in the networks.
Vu~et al.~\cite{vu-qshine07} made also active measurements of the PPLive system and derive mathematical models for the distributions of channel population size or session length.\\
In our previous work~\cite{silverston-nossdav07}, we passively measured the network traffic generated by several popular applications during a worldwide event. 
We compared the measured applications by inferring their underlying mechanisms and highlight their design differences and similarities. 
Ali~et al.~\cite{commercial-measurement} made passive measurements of PPLive and SOPCast applications and analyze the performance and characteristics of such systems.\\
Still in their previously mentioned works, Ali~et al. provide their own methodology to study the data exchanges of such P2P applications.
Based on their measurement studies, Hei~et al.~\cite{pplive-methodology} developed also a methodology to estimate the overall perceived video quality throughout the network.\\
All these works studied P2P IPTV systems by measuring the traffic and tried to infer their mechanisms, 
but they did not characterize the correlation structure of the generated traffic at different time scales to understand its properties and its impact on the network.
\section{Experiments}
\label{sec:experiments}
\subsection{P2P IPTV applications overview}
For our P2P IPTV traffic measurement experiments,
we chose four applications, namely PPLive, PPStream, SOPCast and TVAnts because they were very popular on the Internet.
Whenever these applications are freely available,  their source codes are not open and their exact implementation details and used protocols are still widely unknown.
We can only rely on reverse engineering to understand their transmission mechanisms.\\
All these applications claim to use swarming protocol like Donet. Similarly to BitTorrent,
video data flows are divided into data chunks and each peer downloads the chunck of data to other peers concurrently. 
The peers know how to download the video data chunks by exchanging randomly with other peers information about the data chunks they have or neighbor peers they know.
With this signaling traffic, each peer discovers iteratively new peers, new available data chunks and is able to download video from several peers. 
In these P2P protocols, there are two kinds of traffic: video traffic where peers exchange data chunks with each other and signaling traffic where peers exchange information to get the data.\\
As we show in~\cite{silverston-nossdav07}, all the applications transport video and signaling traffics differently:
PPStream uses exclusively TCP for all traffics while PPLive adds UDP for some signaling traffic. SOPCast uses almost entirely UDP and TVAnts is more balanced between TCP ($\approx$ 75\%) and UDP for all kinds of traffic.\\
In the next section, we will present the measurement experiments platform we used to collect the P2P IPTV traffic.
\subsection{Measurement experiments platform}
\begin{figure}[!t]
\centering
\includegraphics[height=3.00 cm]{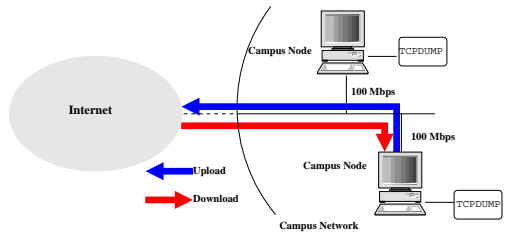}
\caption{Measurement experiments platform. Each node is a common PC directly connected to the Internet via campus network.}
\label{fig:exp}
\end{figure}
\begin{table}[t]
\begin{tiny}
\renewcommand{\arraystretch}{1.7}
\caption{Packet traces summary}
\label{tab:summary}
\begin{center}
\begin{tabular}[b]{|c|c|c|c|c|c|c|c|c|c|}
\hline 
 & \small{PPLive} & \small{PPStream} & \small{SOPCast} & \small{TVAnts} \\ \hline
\small{Duration} (s) & \small{13,321} & \small{12,375} & \small{12,198} & \small{13,358} \\ \hline
\small{Size} (MB) & \small{6,339} & \small{4,121} & \small{5,475} & \small{3,992} \\ \hline \hline
\small{Download}(\%) & \small{14.11} & \small{20.50} & \small{16.13} & \small{24.76} \\ \hline
TCP & 14.09 & 20.50 & 0.23 & 14.71 \\ \hline
UDP & 0.02 & 0 & 15.90 & 10.05 \\ \hline \hline
\small{Upload}(\%) & \small{85.89} & \small{79.50} & \small{83.87} & \small{75.24} \\ \hline
TCP & 85.81 & 79.50 & 3.89 & 61.67 \\ \hline
UDP & 0.08 & 0 & 79.98 & 13.57 \\ \hline
\end{tabular}
\end{center}
\end{tiny}
\end{table}
Our measurement experiments take place during the 2006 FIFA World Cup from 09 June to 09 July.
We collected a huge amount of data, measuring most of the World Cup soccer games with different applications at the same time and under different network environments: campus Ethernet access and residential ADSL access.\\
In this paper, we focus on four packet traces collected on June 30 in our campus network: one for each measured application.
In all our data, we selected these packet traces because they are well representative of all of them.
The traces are made available on our traces sharing service~\cite{content-traces}. 
Two soccer games were scheduled, one in the afternoon (Germany vs. Argentine) measured by PPStream and SOPCast 
and one in the evening (Italy vs. Ukraine) measured by PPLive and TVAnts.
We selected the four traces from different applications to be able to characterize the P2P IPTV traffic without being closely related to the design of the applications.\\

Our measurement experiment set up is described on Fig.~\ref{fig:exp}.
To collect the packets, we used two personal computers with 1.8GHz CPU and common graphic card capabilities. 
The operating system running on the PCs was \texttt{Windows XP}. 
The PCs (\textit{nodes}) were situated in our campus network and were directly connected to the Internet with  100Mbps \texttt{Ethernet} access. 
During a game, each node was running a P2P IPTV application and we used \texttt{tcpdump} on each measuring node to collect all the packets.
For all the measurement experiments, the consumed bandwidth was always relatively low and does not exceed 10Mbps.
The Ethernet cards did not suffer any packet loss and captured all the packets.
For all the experiments, nodes were watching CCTV5, a Chinese TV channel available for all the measured applications.
It was important to watch the same TV channel with all the applications to assure that the behavior of users will be similar in each trace.
For example, during the advertisement, whatever the applications, an user may stop watching the channel and switches the application off and then switch it on a few minutes later.
All the applications used MPEG4 video encoding.\\
 
Our platform has high-speed access and our observations can not be directly 
generalized to residential peers with common access to the Internet (e.g. 20/1 Mbps or 512/128 Kbps). 
However, residential network capacities are quickly increasing 
and will have such high-speed access in only a few years when P2P IPTV would be commonly used. \\

Table~\ref{tab:summary} summarizes the four presented traces.
At first, we can notice in these traces that PPLive, TVAnts and PPStream use massively TCP whereas only SOPCast uses mainly UDP.
The duration of the traces is longer than the duration of a soccer game ($\approx$ 105 minutes). 
We chose to collect the traffic a few minutes before and after the games to capture all the effects that the live interest of a soccer  game could produce on the behavior of users (e.g. flash crowds).
\section{Analysis Methodology}
\label{sec:metho}
\begin{figure*}[!t]
\begin{center}
\subfigure[PPLive]{
\includegraphics[width=5.4 cm]{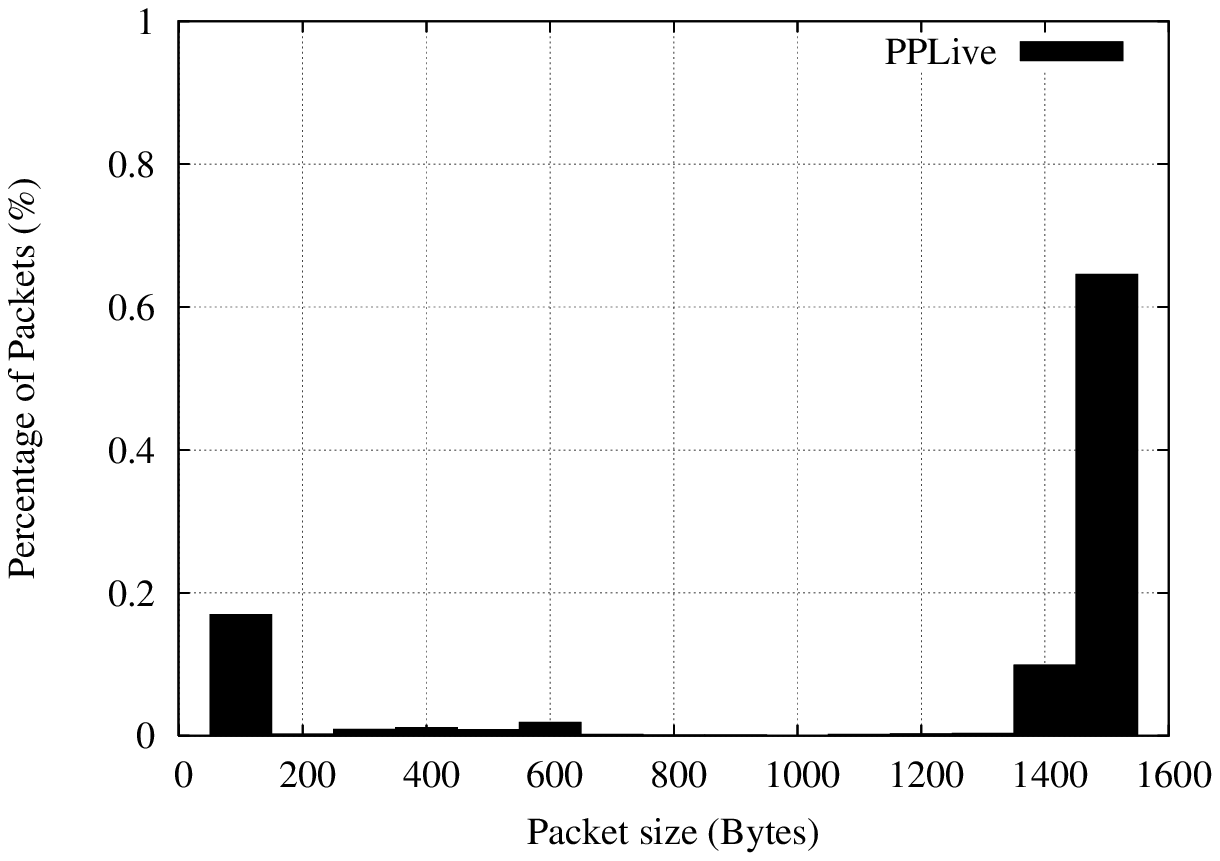}
\label{subfig:histopplive}
}
\subfigure[SOPCast]{
\includegraphics[width=5.4 cm]{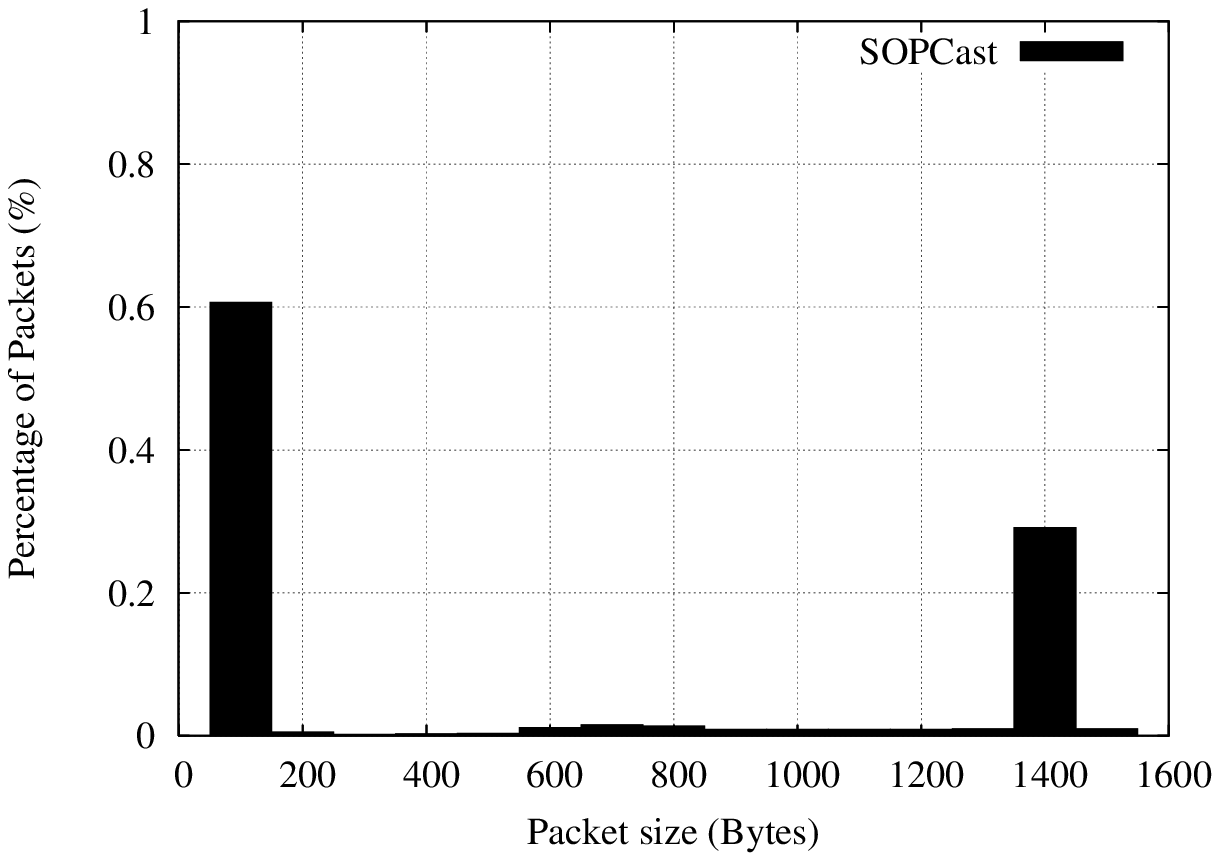}
\label{subfig:histosopcast}
}
\subfigure[PPStream]{
\includegraphics[width=5.4 cm]{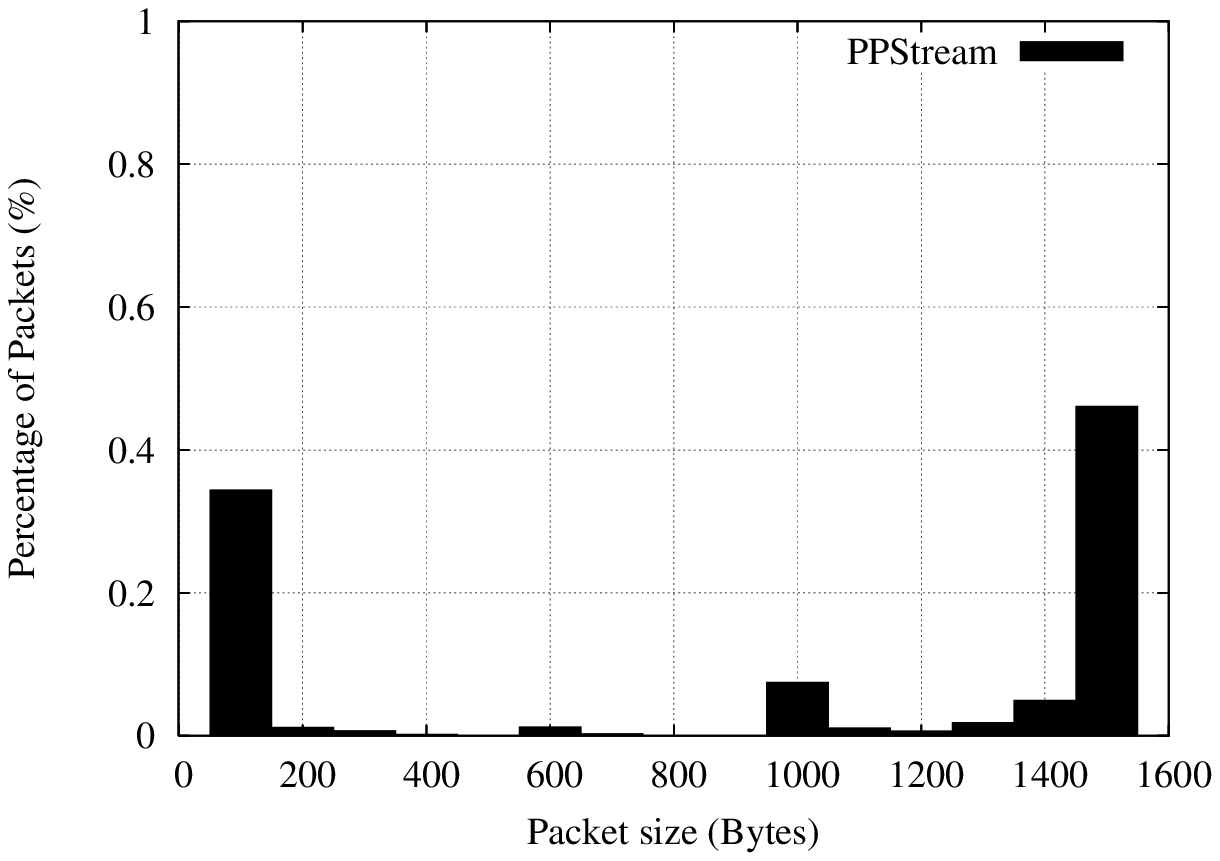}
\label{subfig:histoppstream}
}
\subfigure[TVAnts]{
\includegraphics[width=5.4 cm]{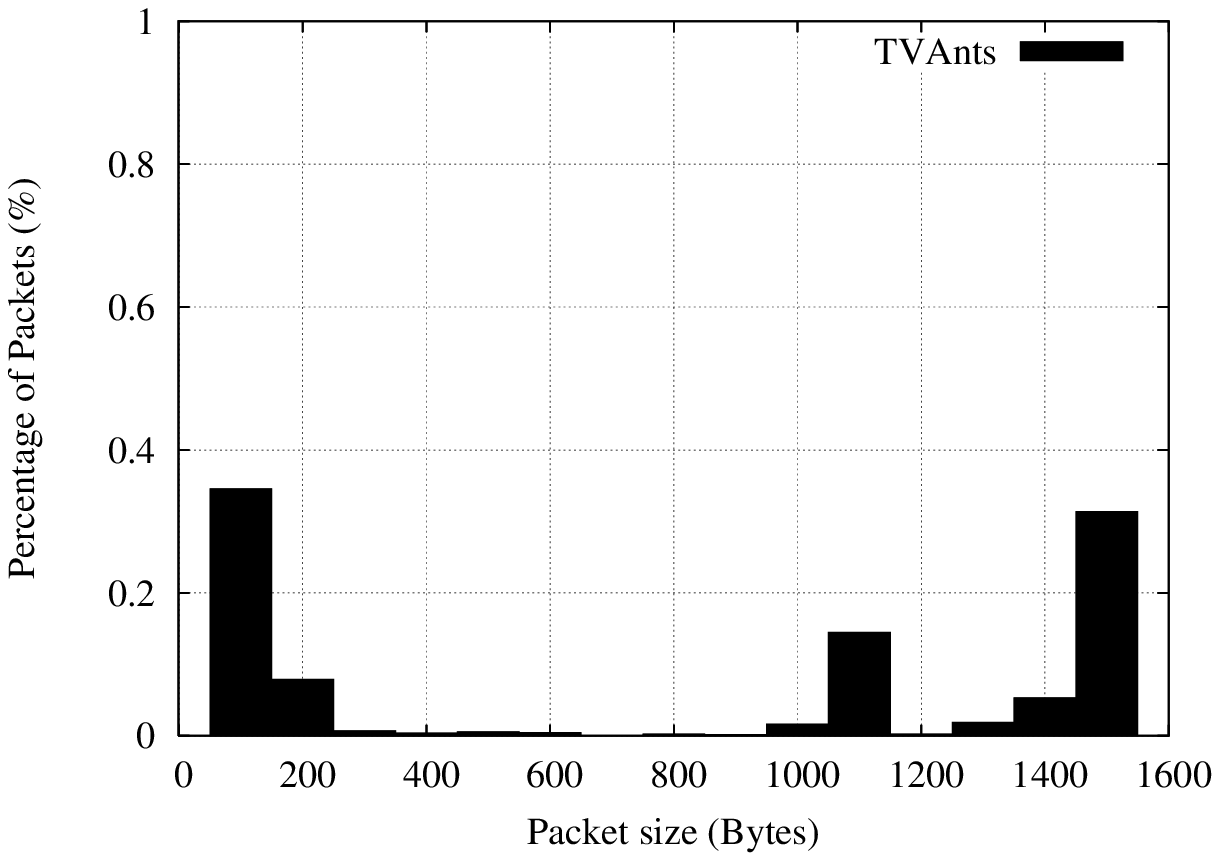}
\label{subfig:histotvants}
}
\caption{Packet distributions for all the applications}
\label{fig:histo}
\end{center}
\end{figure*}
\subsection{Video and signaling traffics}
The packet size distributions for all the applications are presented in Fig.~\ref{fig:histo}.
PPLive has 75\% of its  packets larger than 1300~Bytes and the other packets are very small ($\approx$ 17\% are 100 Bytes length). 
Differently, SOPCast counts only about 30\% of large packets ($>$ 1300 Bytes) and 60\% of its packets are very small ($<$100 Bytes).
PPStream counts more than 40\% of packets bigger than 1400 Bytes, almost 40\% of its packets are 100~Bytes and a small part is 1000 Bytes.
TVAnts packets seems more balanced in three equal parts: one part is 100 Bytes, a second part is 1100 Bytes and the third part is 1400 Bytes.\\
 
All the packets distributions of these applications are different but we can distinguish two sets of packets within these packet distributions:
small-size packets ($<$ 200 Bytes) and large-size packets ($>$ 1000 Bytes). 
As we explained in section~\ref{sec:experiments}, the studied P2P applications generate two kinds of traffic: video and signaling.\\
It is expected that the video traffic is essentially composed of large-size packets. Most of the video packets should belong to the large-size packets set ($>$ 1000 Bytes). 
The video data are also delay sensitive since video packets have to respect stringent playback delay for the users  get a smooth quality video.\\
The signaling packets should belong to the small-size packets set ($<$ 200 Bytes).
The signaling traffic of P2P IPTV systems is not expected to be delay sensitive because it is used to exchange information about peers or data availability 
but not for interactive commands as for Video on-Demand systems like Joost~\cite{joost}.
In video on-demand systems, the users may want to read the video forward or backward instantaneously. 
In the case of P2P IPTV, it is not possible to have this kind of interactive commands since the data flows are broadcasted in live.\\

The signaling and video traffics have not the same characteristics as packet size or delay constraints and they would have a different impact on the network.
We have therefore to separate video and signaling traffic and to analyze them separately.
\subsection{Signaling traffic filtering heuristic}
\label{subsec:heuristic}
\begin{table}[t]
\begin{tiny}
\renewcommand{\arraystretch}{1.7}
\caption{Signaling Traffic Ratio}
\label{tab:ovh}
\begin{center}
\begin{tabular}{|c|c|c|c|c|}
\hline
& \small{PPLive} & \small{PPStream} & \small{SOPCast} & \small{TVAnts} \\ \hline
\small{Total} & \small{4.1\%} & \small{13.6\%} & \small{19.3\%} & \small{10.2\%} \\ \hline
\small{Upload} & \small{2.2\%} & \small{10.8\%} & \small{13.6\%} & \small{7.8\%} \\ \hline
\small{Download} & \small{19.2\%} & \small{25.8\%} & \small{48.5\%} & \small{18.0\%} \\ \hline
\end{tabular}
\end{center}
\end{tiny}
\end{table}
By observing that the video packets size  should be larger than 1000 Bytes and signaling packets should be much smaller ($<$ 200 Bytes), 
we used the simple heuristic proposed by Hei.~\cite{pplive-mesure} to separate the video traffic from the overall traffic.
The heuristic works as follows: for a session (same IP addresses and ports), we counted the number of packets bigger or equal than 1000 Bytes. If a session had at least 10 large packets, then it was labeled as a video session
and we removed small packets ($<$ 1000 Bytes) for this session. 
At the end, we removed all the non video-sessions from the trace to obtain video traffic.\\
At the end, we removed all the non video-sessions from the trace and all the labeled video sessions compose the video traffic.\\

Table~\ref{tab:ovh} summarizes the signaling ratio for all the applications in the entire traces and in both upload or download directions.
As an example, for PPLive, the signaling traffic represents 4.1\% of the total traffic. If we consider both traffic directions, the signaling traffic represents 2.2\% of the upload traffic 
whereas it represents 19.2\% of the download traffic.
The signaling traffic ratios will be discussed more deeply in the results analysis section ($\S$~\ref{sec:results}).\\
The next section is dedicated to validate the heuristic used to filter the signaling traffic.
\subsection{Validation of the signaling heuristic}
The measured applications are proprietary and we do not have any implementation detail. 
We do not know exactly if a packet is for signaling traffic or video traffic. 
The heuristic used  to filter the signaling traffic relies mainly on the packet size.
The heuristic will perform well if it removes only signaling packets without removing any part of the video traffic.\\

From Table~\ref{tab:ovh}, we notice that the  download signaling traffic ratio computed by the heuristic for SOPCast is 48.5\%, which is a high ratio indicating that half of the traffic would have been signaling traffic.
This high ratio of download signaling traffic may eventually come from the inaccuracy of the heuristic.\\ 

To validate the filtering heuristic, we compute the resulting video bitrate after removing the signaling traffic.
Realistic computed video bitrate will confirm the efficiency of the heuristic.
We compute the video bitrate for download traffic  because it is the only video traffic we can deduce.
The  download traffic is provided by other peers on the Internet and 
all the video flows received by our controlled nodes compose the downloaded video.
The video upload traffic is not provided to a single consumer peer. 
We can not estimate the video bitrate received by remote peers because they mix video flows from many other providers peers to receive the entire video.
In the following, we compute for all the traces the video download bitrate received by our controlled nodes by removing signaling traffic with the presented heuristic.\\

Fig.~\ref{subfig:histosopcast} indicates that SOPCast has more than 60\% of its packets smaller than 100~Bytes. 
These packet are certainly signaling packets according to their size. 
The design of SOPCast introduces a large amount of signaling packets and the heuristic removes these packets from the video traffic.\\
Table~\ref{tab:summary} shows that the SOPCast download traffic represents 16.13\% of the collected traffic (5,475 MBytes) and the download signaling ratio represents 48.5\% of the overall download traffic.
The average bandwidth used  by SOPCast to download the video at its bitrate can be computed by dividing the amount of received video data by the measurement experiment duration (12,198s).
\begin{displaymath}
\frac{(5475*2^{20}*8*0.1613)*(1-0.485)}{12198}\approx305Kbps
\end{displaymath}

For SOPCast, the video download speed (305Kbps) computed with the heuristic is realistic for downloading a video broadcasted in the Internet.\\
 
Table~\ref{tab:compute} shows the computed download speeds for all the applications.
\begin{table}[t]
\begin{tiny}
\renewcommand{\arraystretch}{1.7}
\caption{Video Download Bitrate computed with the signaling filtering heuristic}
\label{tab:compute}
\begin{center}
\begin{tabular}{|c|c|c|c|c|}
\hline
& \small{PPLive} & \small{PPStream} & \small{SOPCast} & \small{TVAnts} \\ \hline
\small{Video Bitrate} & \multirow{2}{*}{\small{445}} & \multirow{2}{*}{\small{415}} & \multirow{2}{*}{\small{305/360}} & \multirow{2}{*}{\small{500}} \\  
\small{(Kbps)} & & & & \\ \hline
\end{tabular}
\end{center}
\end{tiny}
\end{table}
All the  computed download speeds are realistic and
it confirms that the heuristic used to filter signaling traffic performs well and is efficient to distinguish signaling  traffic from the overall traffic.
Our node running SOPCast receives really a lot of signaling packets from other peers in the Internet and this high ratio is not distorted by the heuristic.\\

To have a finer understanding of the high signaling traffic ratio of SOPCast, 
we show in our previous work~\cite{silverston-nossdav07} that SOPCast received no video traffic during 30 minutes. The video source suffered troubles during this period of time.
In this period, our node was only receiving signaling traffic from other peers requesting video data chunks, but no video chunks were received.
Since our node has high bandwidth capabilities, lots of peers were requesting video generating a large amount of signaling traffic. 
This phenomenon increased artificially the SOPCast signaling traffic and explains why the signaling traffic ratio SOPCast experiments in the download direction was so high.
If we take into account this lack of video data  during 30mn in our calculation, we obtain 360Kbps video download bitrate, which is a more realistic bitrate closer to all the other computed video bitrate.\\ 

In this section, we validate the choice of the heuristic used to filter signaling traffic from the entire traces. 
In the following, for each trace, we will consider the overall traffic generated by the applications and the video traffic
without any signaling packet deduced by using the signaling filtering heuristic.\\

To characterize the correlation structure of the generated traffic at different time scales, we analyzed the traffic using a wavelet based transform method. 
To this end, we used a tool that is presented in the next section. 
\subsection{Multiscale traffic analysis}
\label{ldestimate}
We analyze the measured P2P IPTV traffic at different time scales to characterize this traffic and its properties.
To this end, we compute the energy spectrum of the traffic at different time scales using a wavelet based transform method. 
The smaller time scales analyzed is the 20 milliseconds intervals as we observed that inter arrivals are rarely below this value (our bin duration is 20ms).
In each interval, we counted the number of packet arrivals in both directions (i.e. upload and download).
We only counted arrivals of packets with data payload and do not take into account empty TCP Acknowledgment packets.\\

Logscale Diagram Estimate~\cite{LDestimate}~(LDestimate) is based on discrete wavelet transform and allows to analyze the scaling behavior of the packet traffic.
LDestimate produces a logarithmic plot of the data energy spectrum.\\ 
For all the produced logscale diagrams, the X-axes are the octaves of the traffic, which are the time scales of the packet arrivals. The right-most part of the graph is relative to large time scales
and the left part is relative to small time scales. 
The Y-axes are the data energy spectra.
A logscale diagram can be understood as follows: an octave $j$ (X-axes) is a time scale of the packet traffic energy spectrum. Since our bin is 20 ms, the octave $j=8$ means the time scale $t=2^8*20ms=5.12s$.\\
LDEstimate is a tool that allow to visually observed the properties of the measured traffic.
In a produced diagram, a bump in the energy spectrum indicates a possible periodic behavior of the traffic, 
a constant energy spectrum a possible memoryless process and a linear increase indicates a possible long-range dependency of the traffic (LRD).
\section{Results Analysis}
\label{sec:results}
\begin{figure*}[!t]
\begin{center}
\subfigure[Upload: overall packet traffic (signaling and video traffic)]{
\includegraphics[scale=0.30]{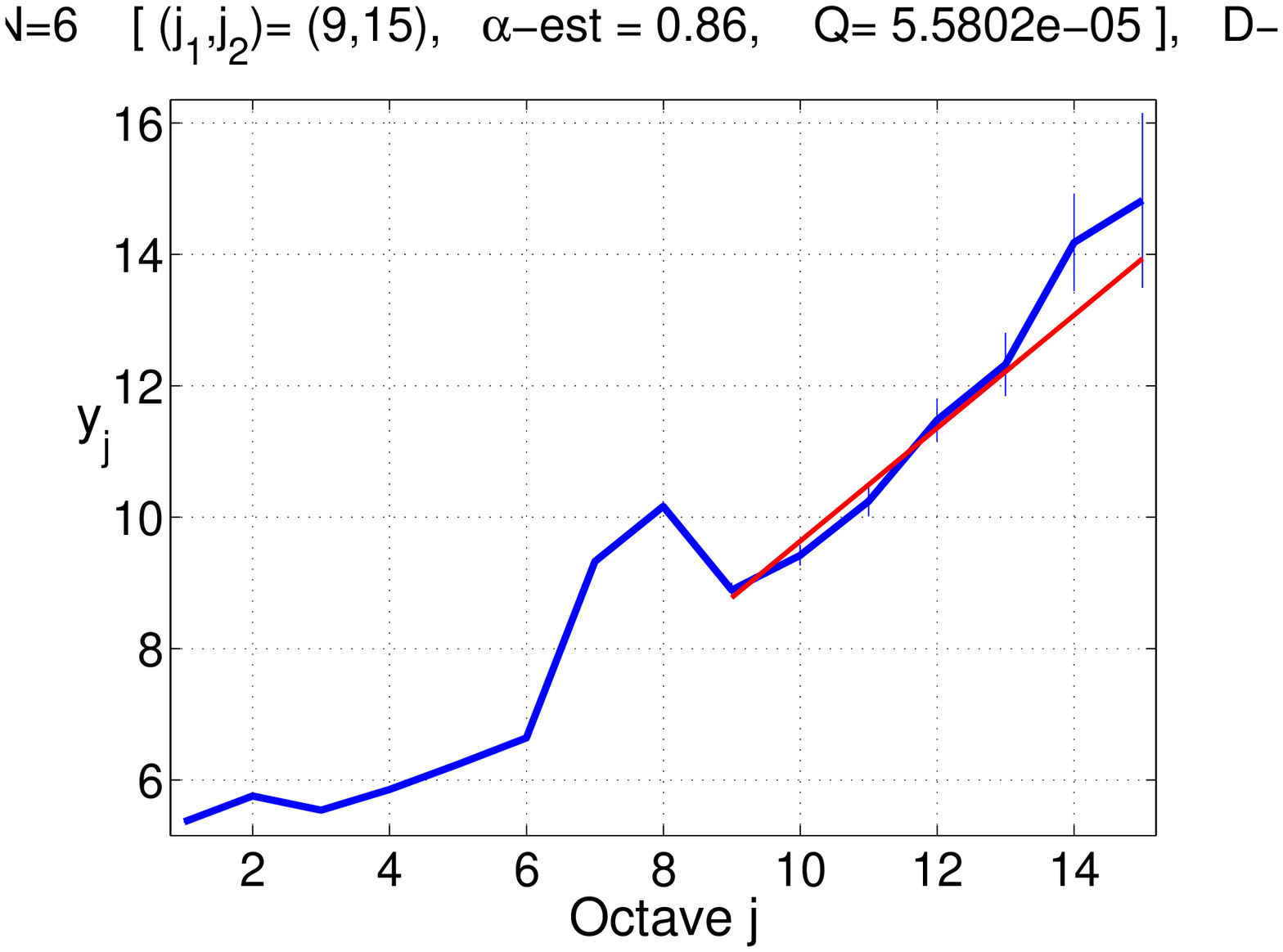}
\label{subfig:pplive_pakup}
}
\subfigure[Upload: video packet traffic]{
\includegraphics[scale=0.30]{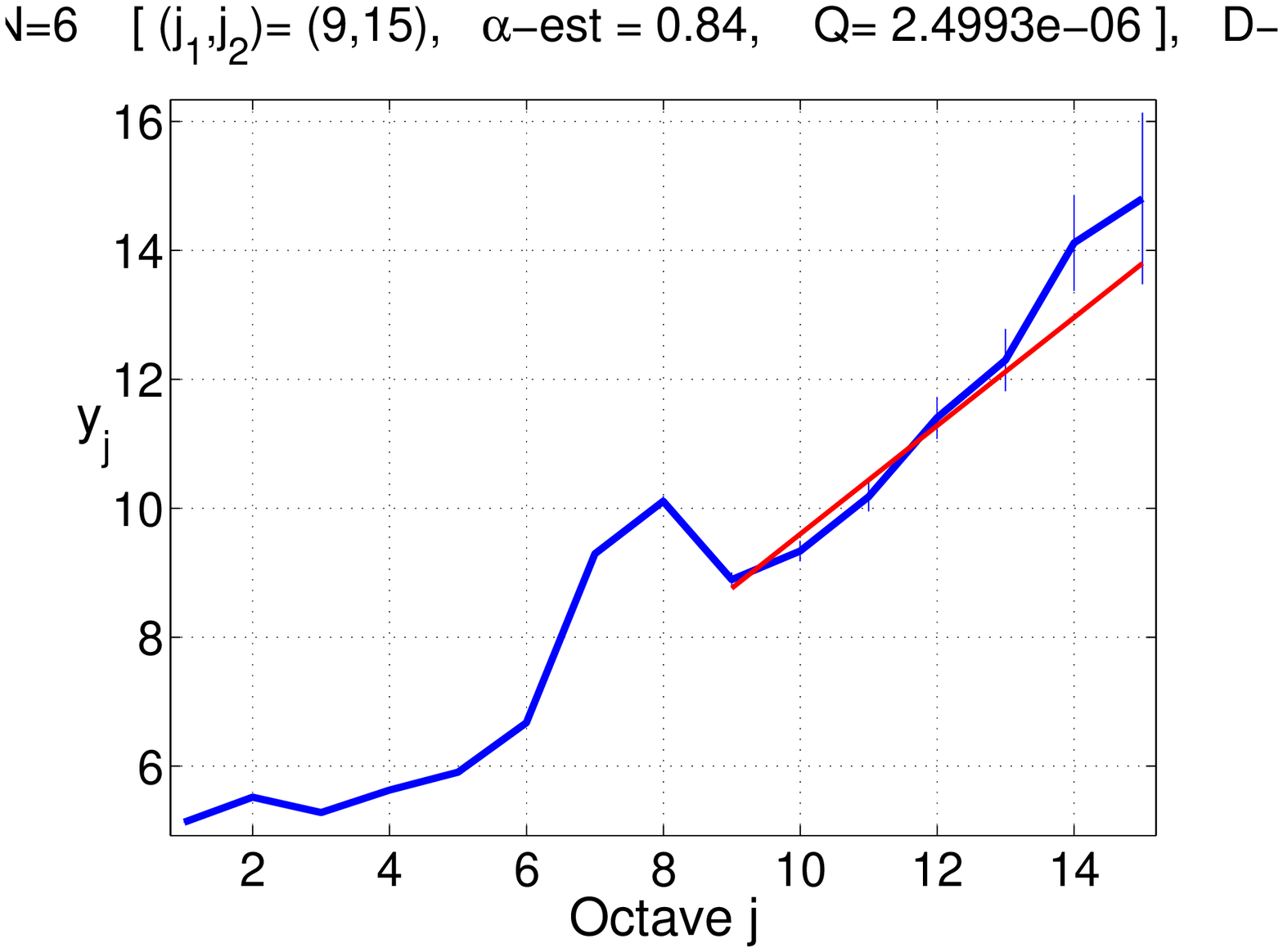}
\label{subfig:pplive_pakup_video}
}
\subfigure[Download: overall packet traffic (signaling and video traffic)]{
\includegraphics[scale=0.30]{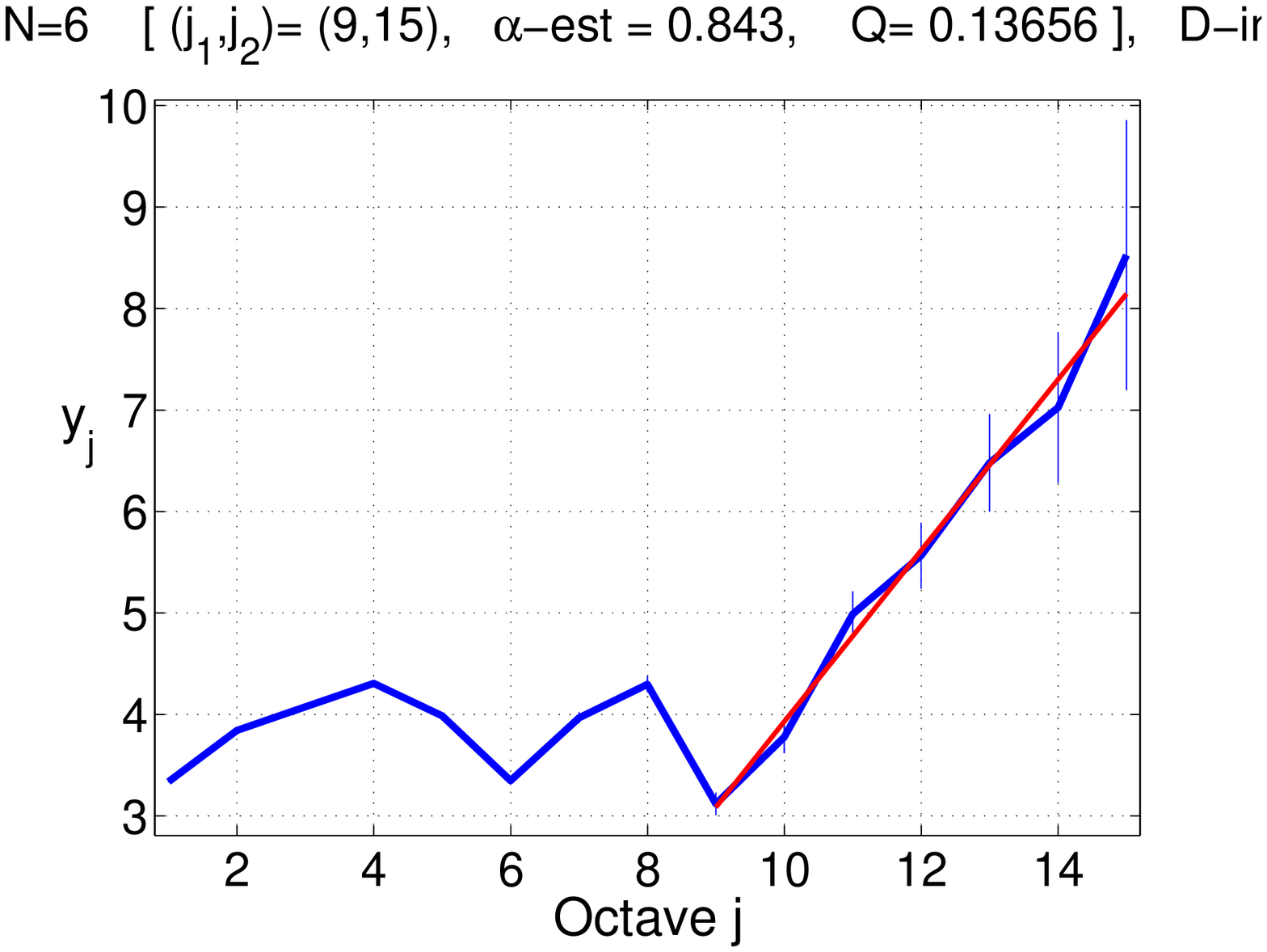}
\label{subfig:pplive_pakdown}
}
\subfigure[Download: video packet traffic]{
\includegraphics[scale=0.30]{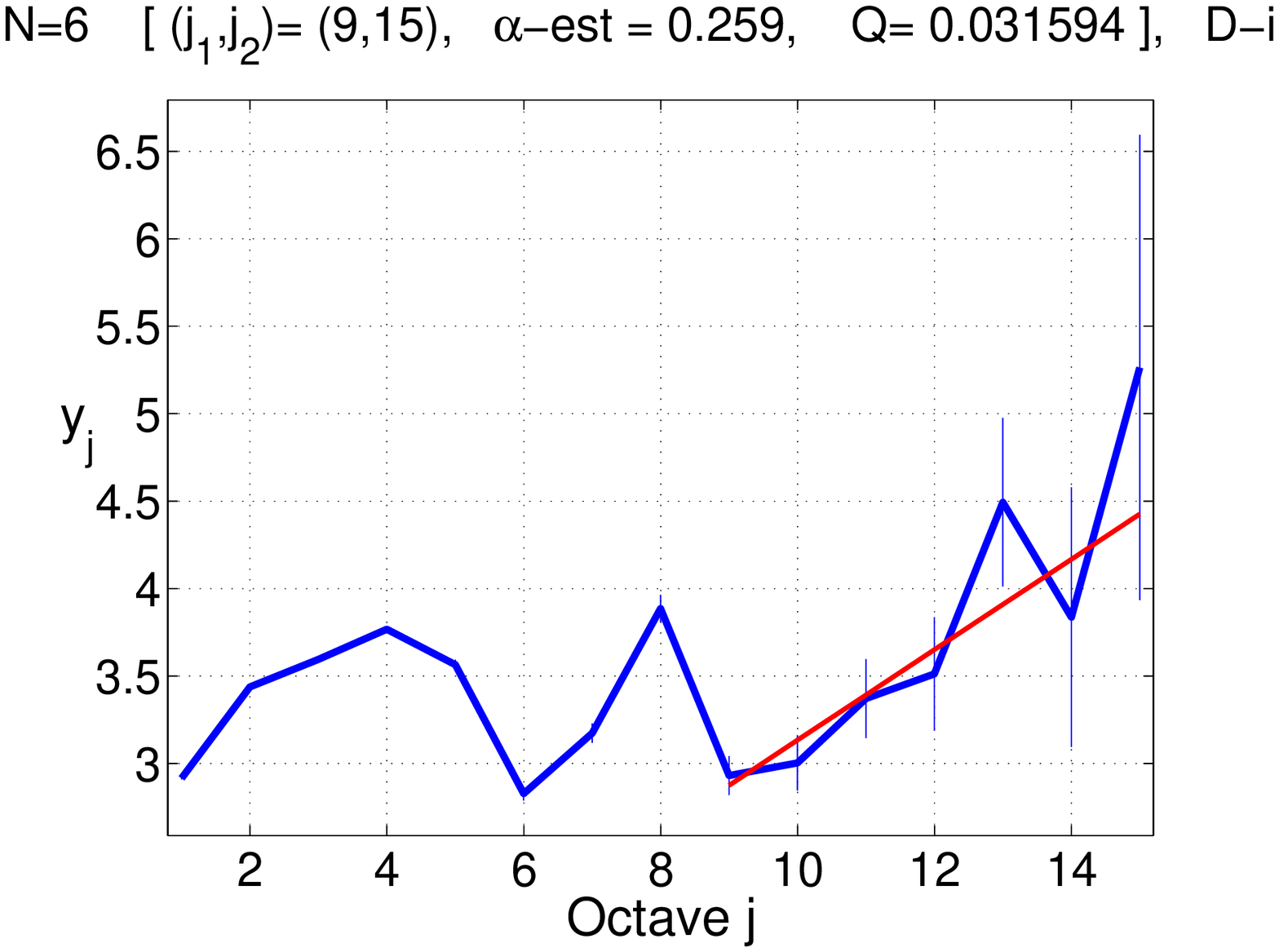}
\label{subfig:pplive_pakdown_video}
}
\caption{PPlive packet traffics energy spectra. Bin duration is 20ms. (Ex: Octave $j=8$ is for scale process at $t=2^8*20ms=5.12s$).}
\label{fig:pplive}
\end{center}
\end{figure*}
\begin{figure*}[!t]
\begin{center}
\subfigure[Upload: overall packet traffic (signaling and video traffic)]{
\includegraphics[scale=0.30]{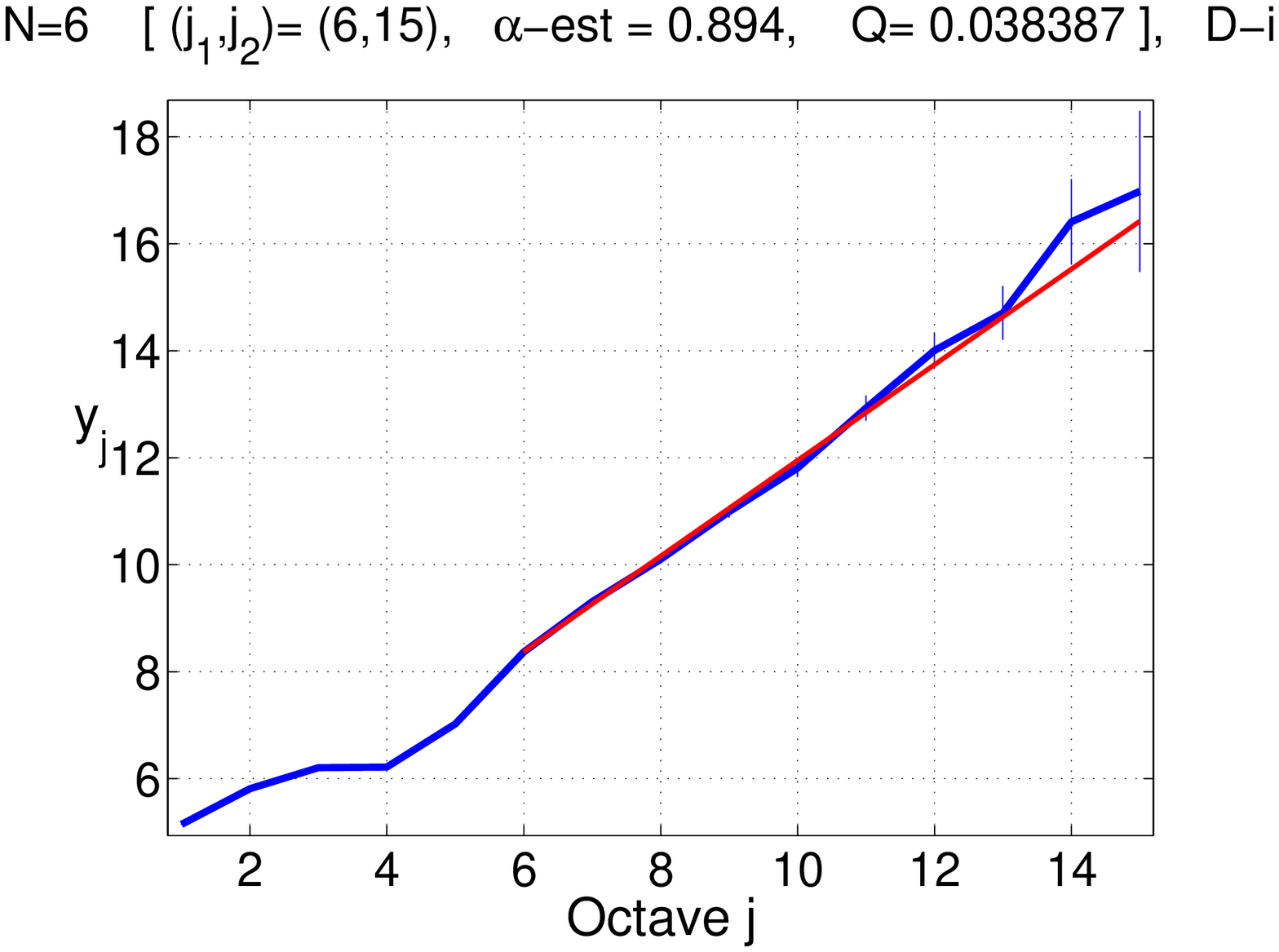}
\label{subfig:sopcast_pakup}
}
\subfigure[Upload: video packet traffic]{
\includegraphics[scale=0.30]{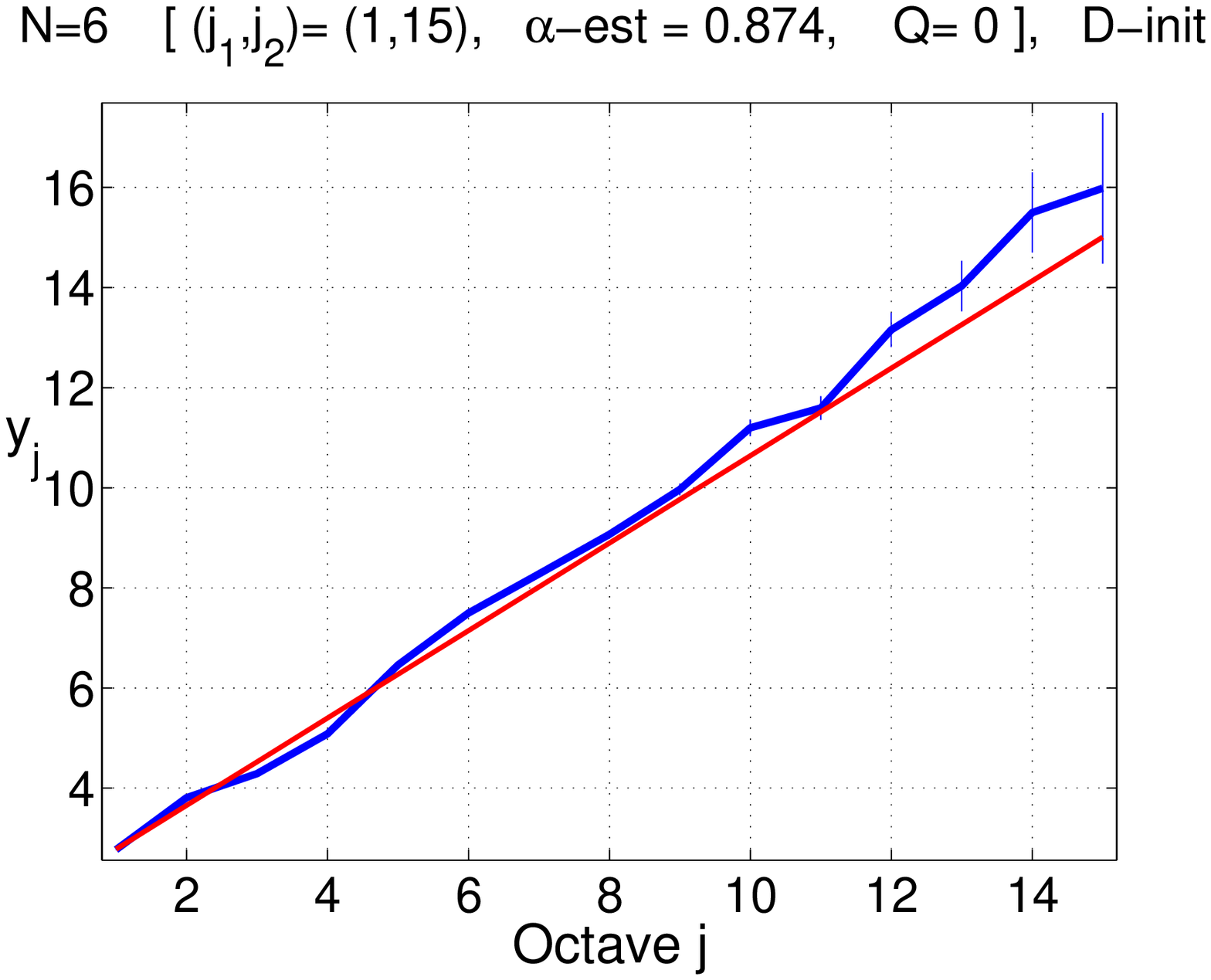}
\label{subfig:sopcast_pakup_video}
}
\subfigure[Download: overall packet traffic (signaling and video traffic)]{
\includegraphics[scale=0.30]{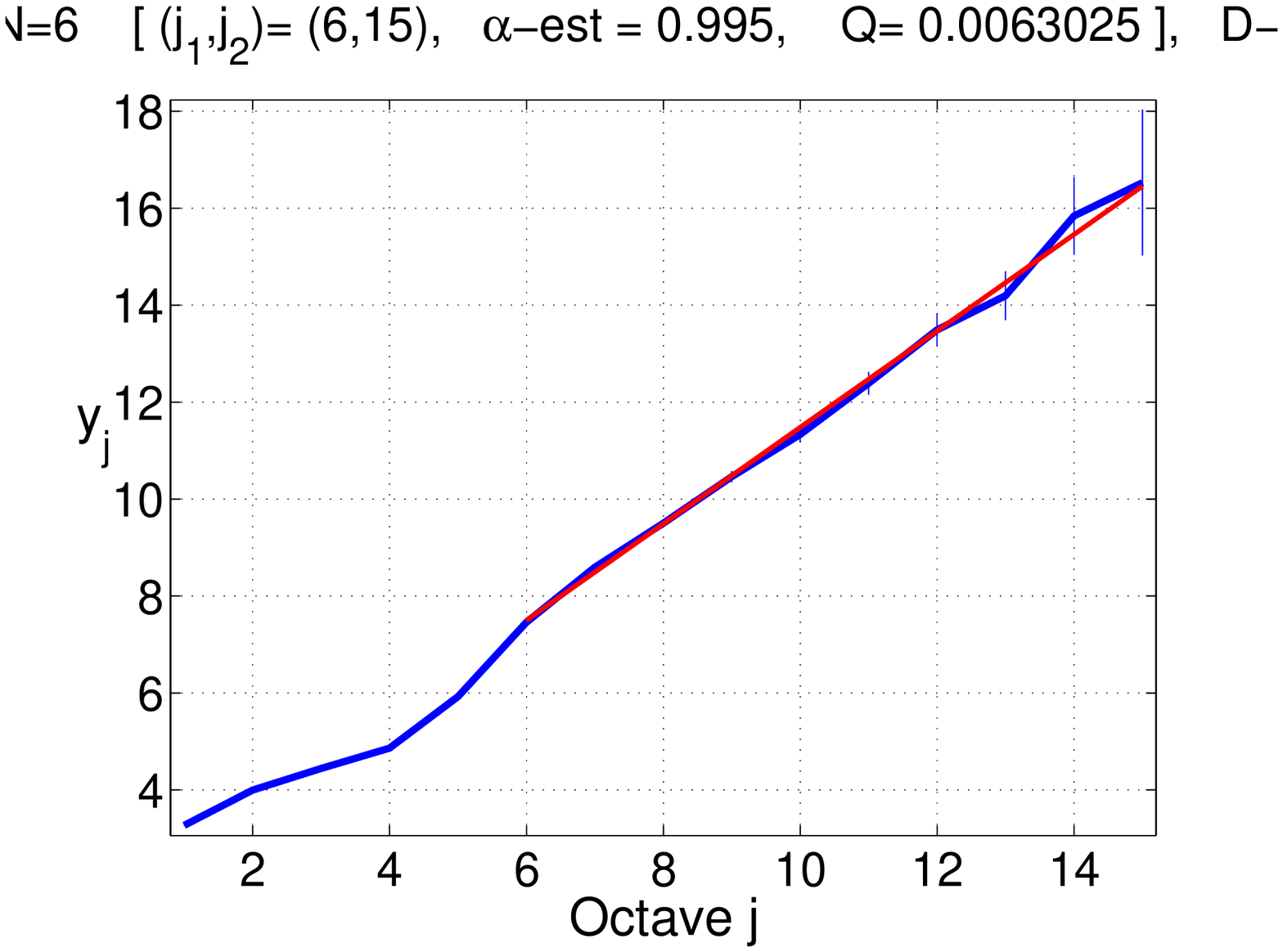}
\label{subfig:sopcast_pakdown}
}
\subfigure[Download: video packet traffic]{
\includegraphics[scale=0.30]{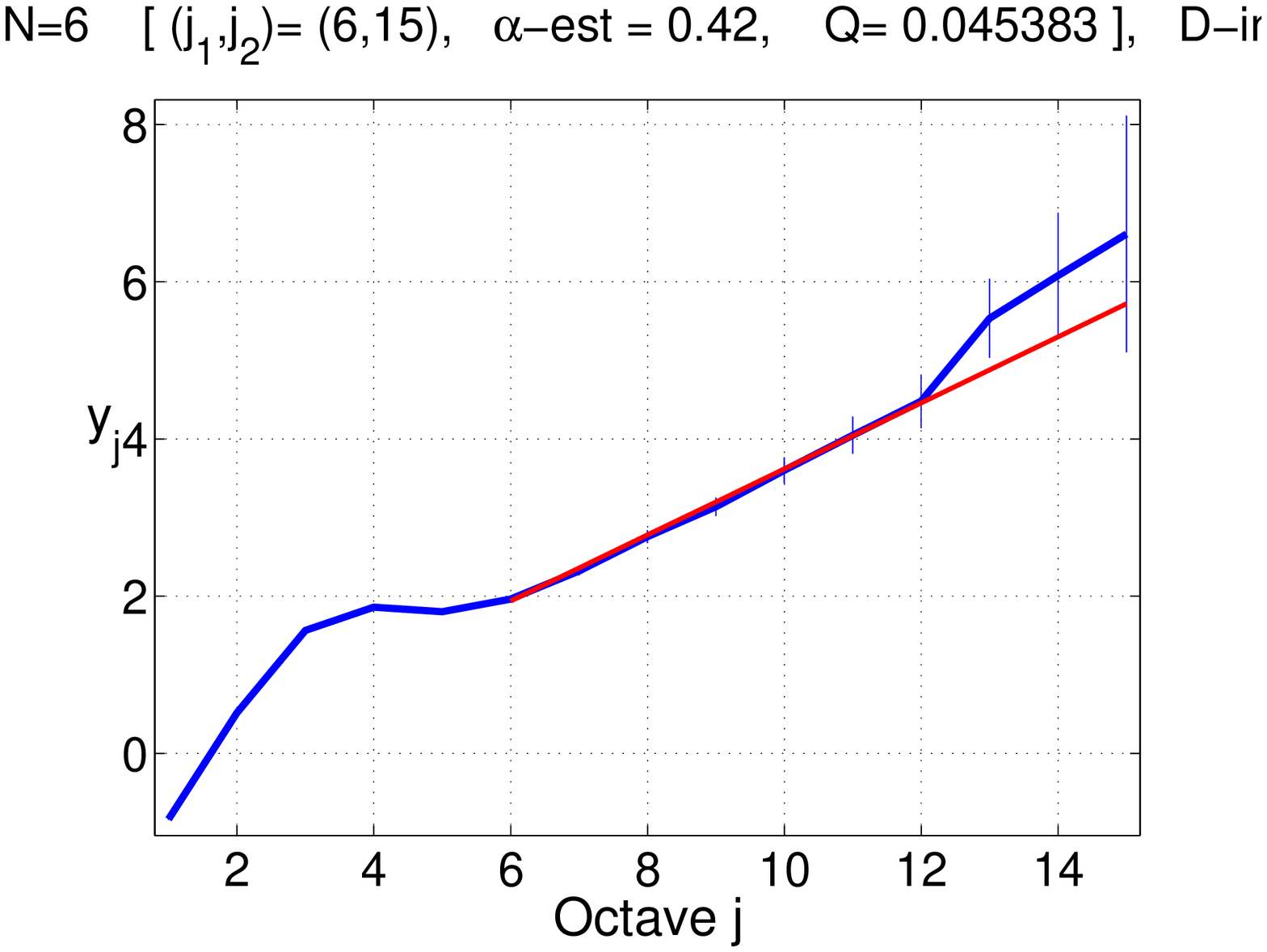}
\label{subfig:sopcast_pakdown_video}
}
\caption{SOPCast packet traffics energy spectra. Bin duration is 20ms. (Ex: Octave $j=8$ is for scale process at $t=2^8*20ms=5.12s$).}
\label{fig:sopcast}
\end{center}
\end{figure*}
\begin{figure*}[!t]
\begin{center}
\subfigure[Upload: overall packet traffic (signaling and video traffic)]{
\includegraphics[scale=0.30]{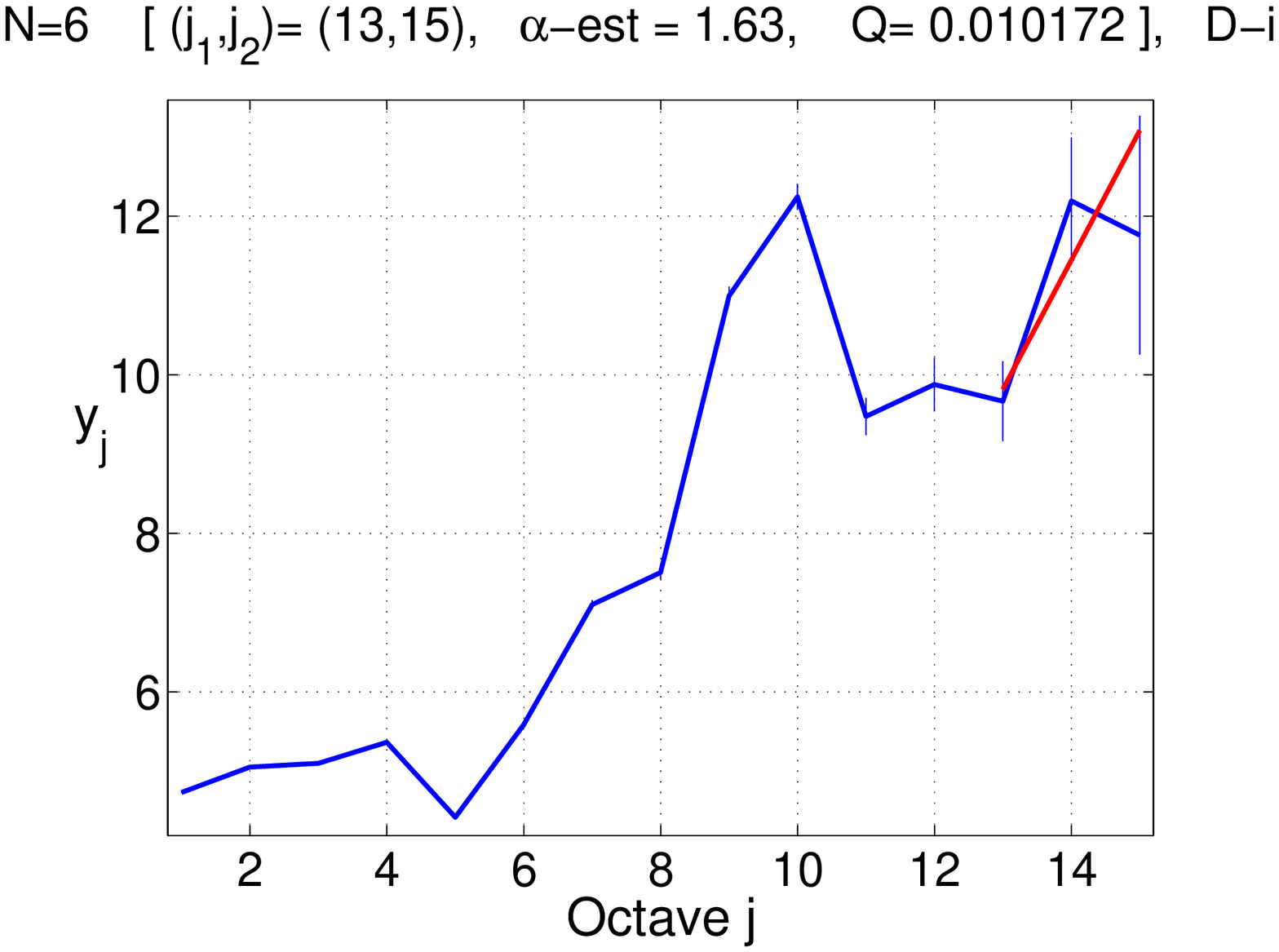}
\label{subfig:ppstream_pakup}
}
\subfigure[Upload: video packet traffic]{
\includegraphics[scale=0.30]{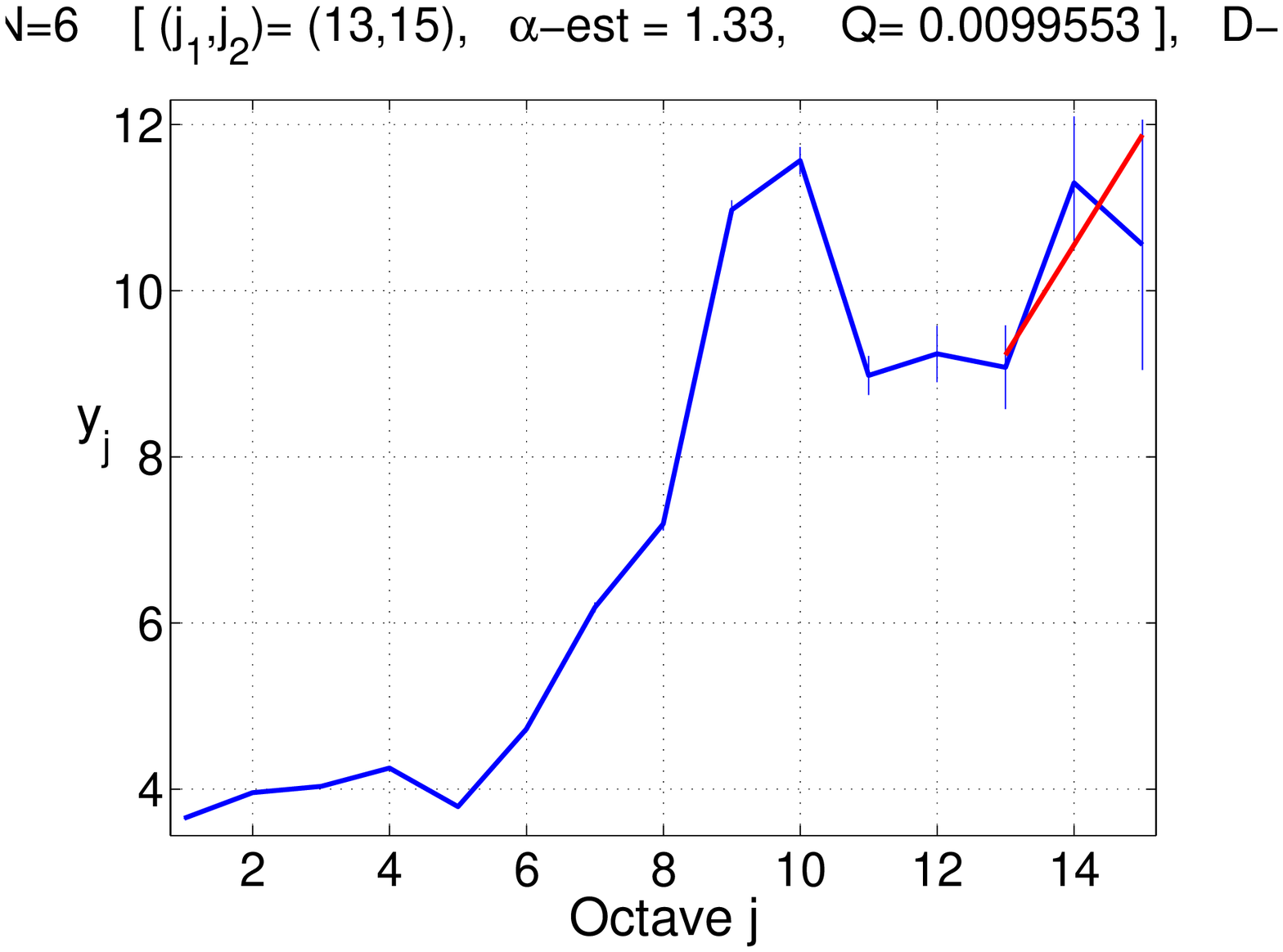}
\label{subfig:ppstream_pakup_video}
}
\subfigure[Download: overall packet traffic (signaling and video traffic)]{
\includegraphics[scale=0.30]{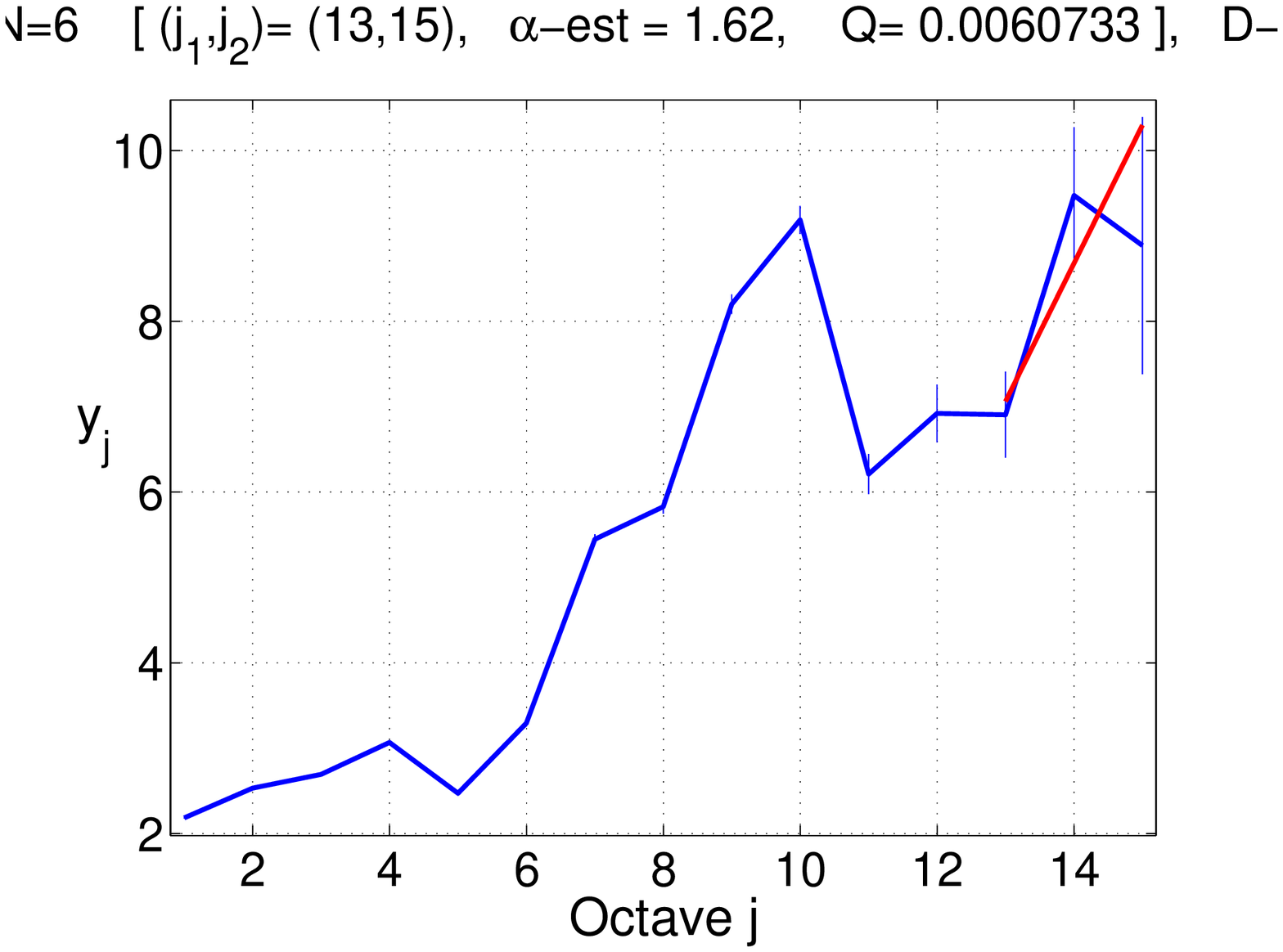}
\label{subfig:ppstream_pakdown}
}
\subfigure[Download: video packet traffic]{
\includegraphics[scale=0.30]{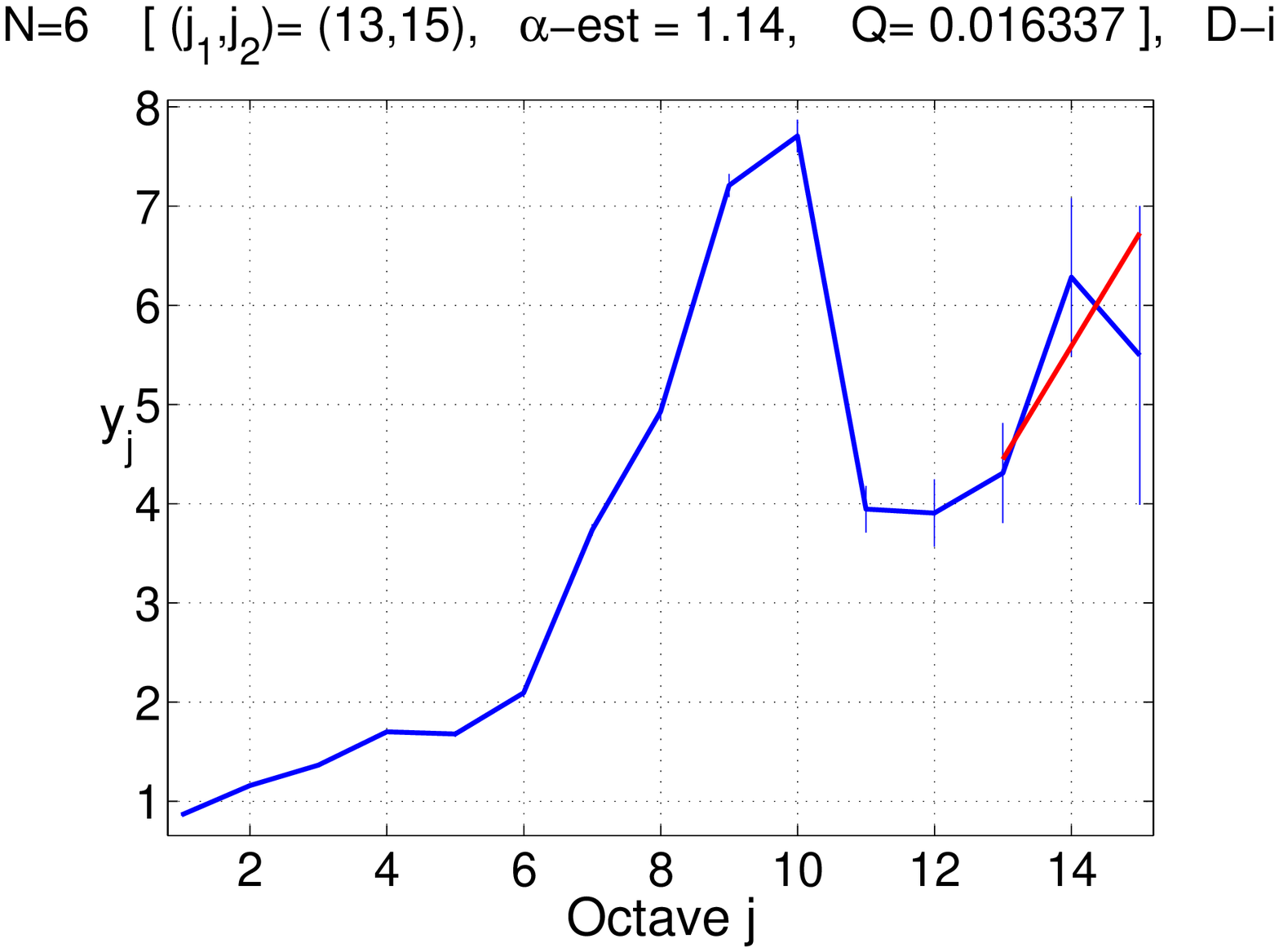}
\label{subfig:ppstream_pakdown_video}
}
\caption{PPStream packet traffics energy spectra. Bin duration is 20ms. (Ex: Octave $j=8$ is for scale process at $t=2^8*20ms=5.12s$).}
\label{fig:ppstream}
\end{center}
\end{figure*}
\begin{figure*}[!t]
\begin{center}
\subfigure[Upload: overall packet traffic (signaling and video traffic)]{
\includegraphics[scale=0.30]{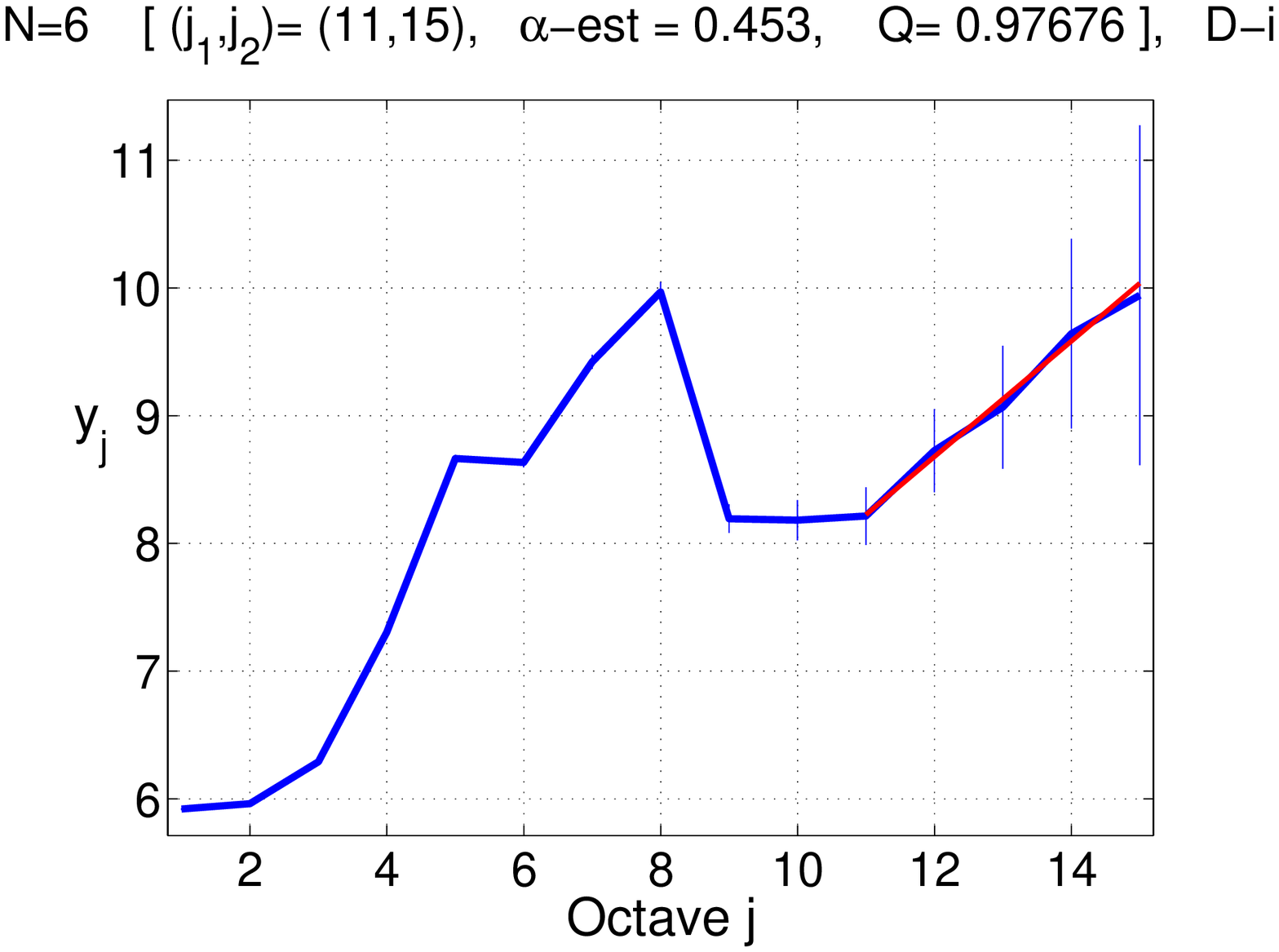}
\label{subfig:tvants_pakup}
}
\subfigure[Upload: video packet traffic]{
\includegraphics[scale=0.30]{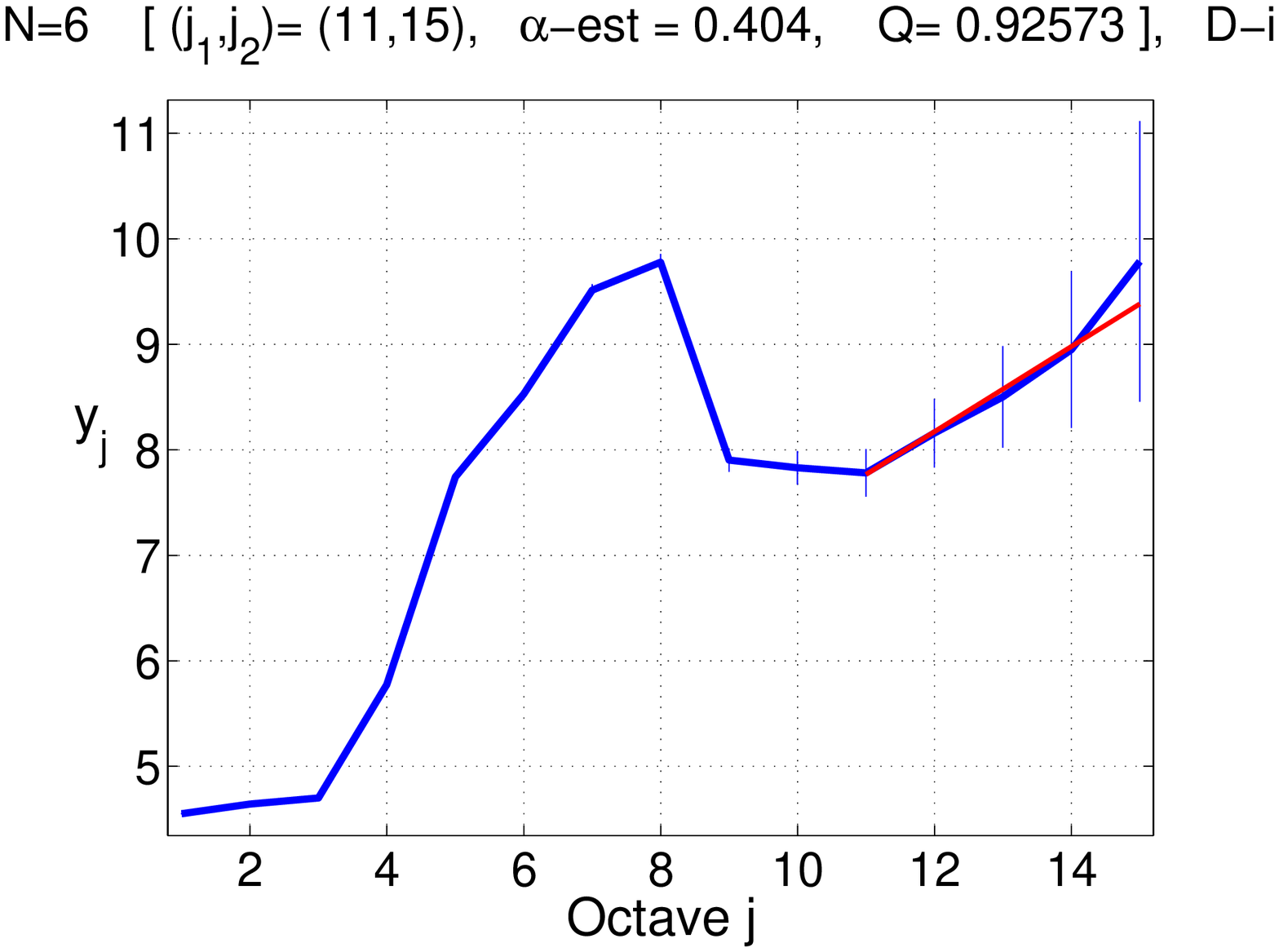}
\label{subfig:tvants_pakup_video}
}
\subfigure[Download: overall packet traffic (signaling and video traffic)]{
\includegraphics[scale=0.30]{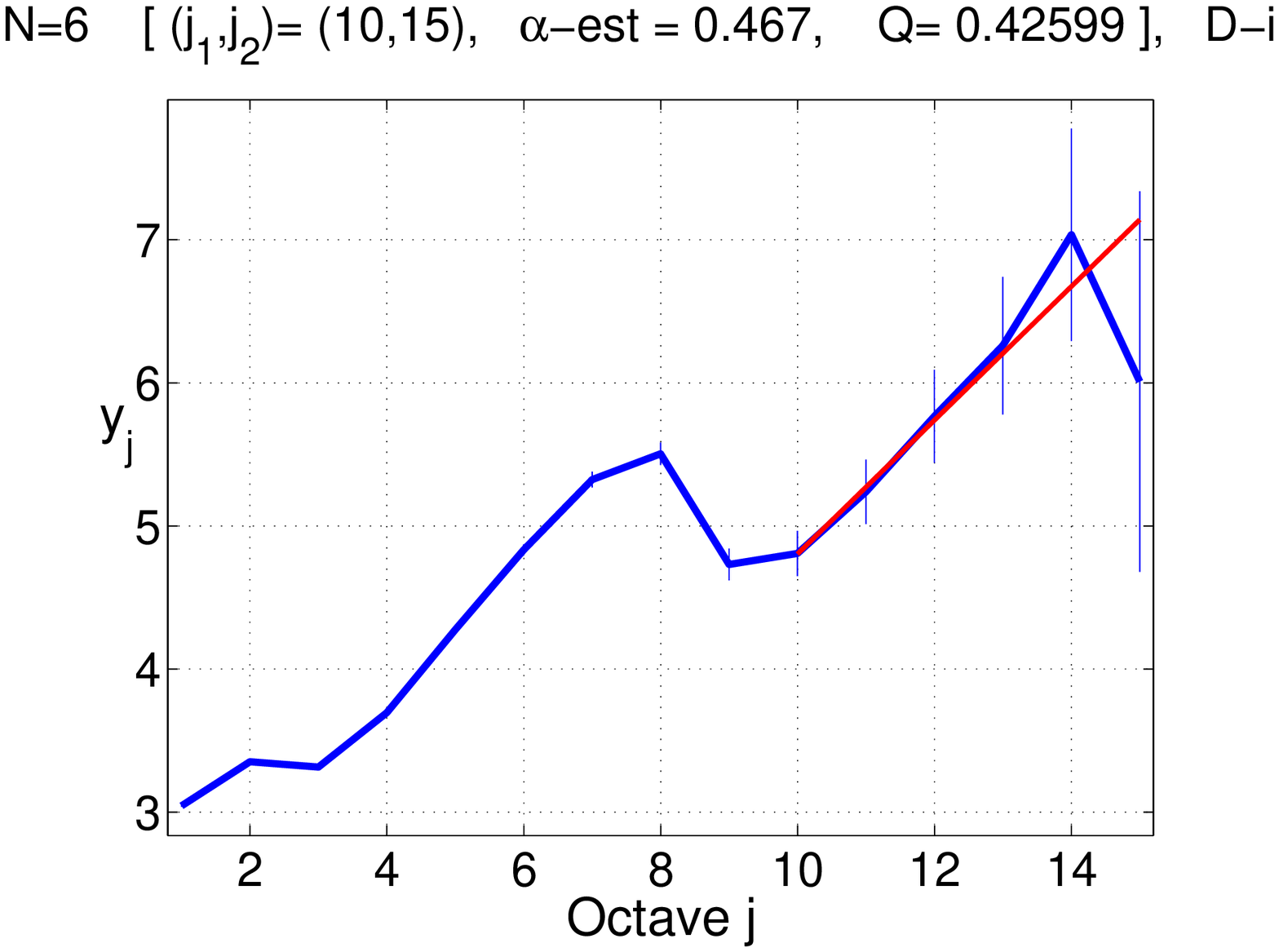}
\label{subfig:tvants_pakdown}
}
\subfigure[Download: video packet traffic]{
\includegraphics[scale=0.30]{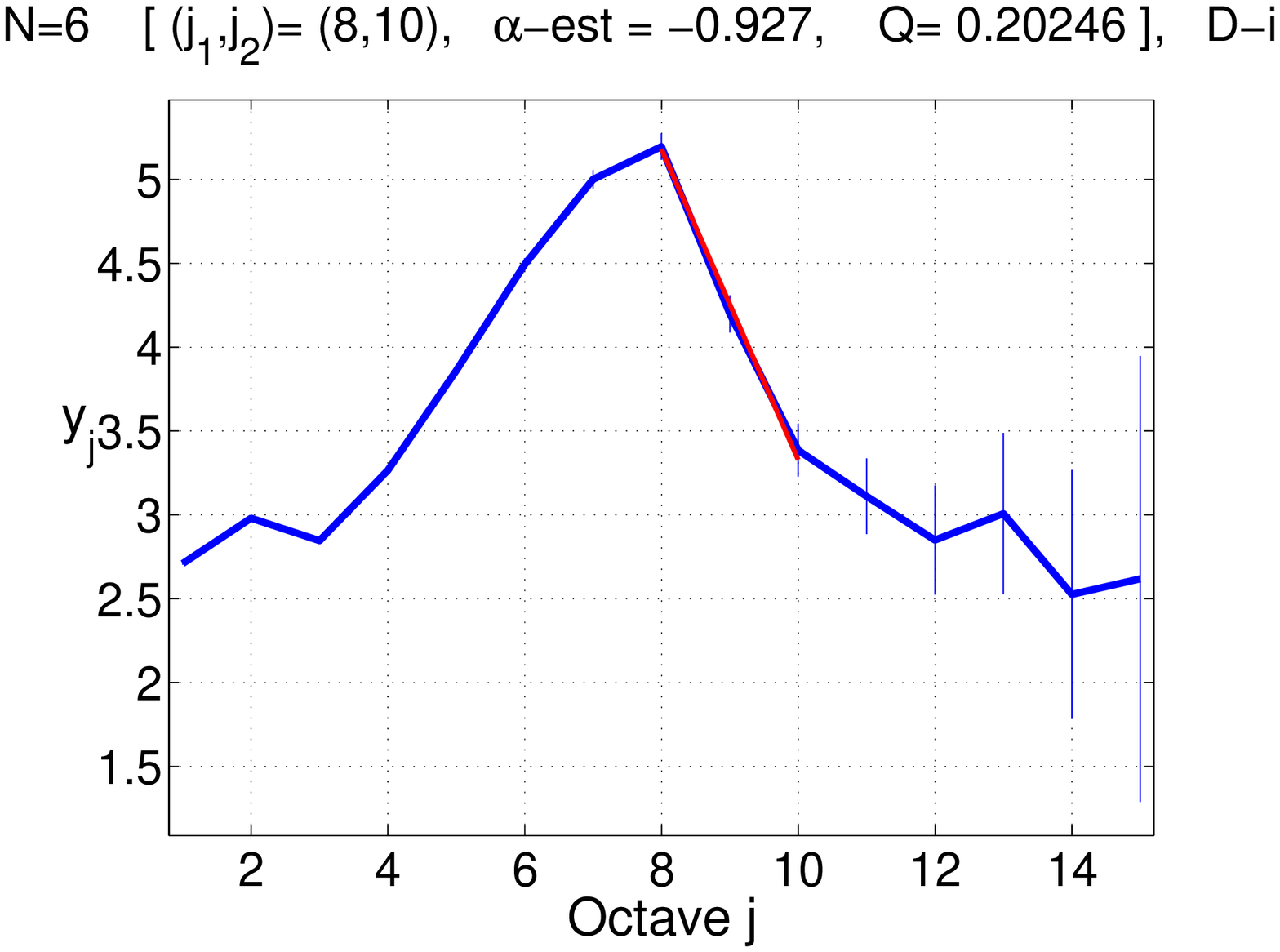}
\label{subfig:tvants_pakdown_video}
}
\caption{TVAnts packet traffics energy spectra. Bin duration is 20ms. (Ex: Octave $j=8$ is for scale process at $t=2^8*20ms=5.12s$).}
\label{fig:tvants}
\end{center}
\end{figure*}
\subsection{Presentation}
In the rest of the paper, we will refer to the traffic trace of an application by the name of the application. For example, we will refer to SOPCast packet trace by SOPCast.\\
For each application, we study the traffic by separating the upload traffic and the download traffic. 
In both traffic directions, we separated the video traffic from the overall traffic by using the filtering heuristic presented in section~\ref{subsec:heuristic}.
Then, each application is characterized by four distinct logscale diagrams: overall upload traffic, video upload traffic, overall download traffic and video download traffic.
Fig.~\ref{fig:pplive} presents the logscale diagrams of the energy spectra for PPLive,  Fig.~\ref{fig:sopcast} the energy spectra for SOPCast, Fig.~\ref{fig:ppstream} for PPStream and Fig.~\ref{fig:tvants} for TVAnts.\\

Table~\ref{tab:summary} indicated previously that three of the measured applications use massively TCP (PPLive, PPStream and TVAnts) whereas only SOPCast uses mainly UDP.
We will refer to an application using massively TCP as \textit{TCP application} and \textit{UDP application} for an application using UDP.\\

In the following, we will present traffic differences between TCP and UDP applications in section~\ref{sec:versus}. In section~\ref{subsec:signaling} we will highlight the impact of the signaling traffic for P2P IPTV applications.
Then we discuss the stationnarity of the traffic in section~\ref{subsec:stationarity}.
We will extend our findings in section~\ref{sec:topdl}. The results are summarized and discussed in section~\ref{sec:discussion}.
\subsection{TCP applications vs. UDP applications}
\label{sec:versus}
For the TCP applications (PPLive, PPStream and TVAnts), the two upload energy spectra look similar for all the time scales 
(Fig.~\ref{subfig:pplive_pakup} and \ref{subfig:pplive_pakup_video},  
Fig.~\ref{subfig:ppstream_pakup} and \ref{subfig:ppstream_pakup_video},  
Fig.~\ref{subfig:tvants_pakup} and \ref{subfig:tvants_pakup_video}) 
The two download energy spectra look similar until $j=9$ and after they are different 
(Fig.~\ref{subfig:pplive_pakdown} and \ref{subfig:pplive_pakdown_video},  
Fig.~\ref{subfig:ppstream_pakdown} and \ref{subfig:ppstream_pakdown_video},  
Fig.~\ref{subfig:tvants_pakdown} and \ref{subfig:tvants_pakdown_video}). 
The upload energy spectra of TCP applications 
(\ref{subfig:pplive_pakup} and \ref{subfig:pplive_pakup_video},
\ref{subfig:ppstream_pakup} and \ref{subfig:ppstream_pakup_video}, 
\ref{subfig:tvants_pakup} and \ref{subfig:tvants_pakup_video})
 do not look like their download energy spectra 
(\ref{subfig:pplive_pakdown} and \ref{subfig:pplive_pakdown_video},
\ref{subfig:ppstream_pakdown} and \ref{subfig:ppstream_pakdown_video},
(\ref{subfig:tvants_pakdown} and \ref{subfig:tvants_pakdown_video}).\\
All the TCP applications have similar energy spectra whatever the kind of traffic  and direction (e.g. overall or video).
The observations for TCP applications may be generalized for all of those we measured (PPLive, PPStream and TVAnts).\\
For UDP applications (SOPCast Fig.~\ref{fig:sopcast}),  the four energy spectra look similar whatever the traffic direction or the traffic nature (e.g. overall or video traffics).\\
The energy spectra of TCP applications (Fig.~\ref{fig:pplive},~\ref{fig:ppstream}~and~\ref{fig:tvants}) are different from the energy spectra of UDP applications (Fig.~\ref{fig:sopcast}).\\
 
Regarding TCP applications energy spectra more precisely, we can observe an energy bump in all the logscale diagrams about time scale $j=8$ (5.12s). 
The energy bump is more clearly defined in upload traffic 
(Fig.~\ref{subfig:pplive_pakup} and \ref{subfig:pplive_pakup_video}, 
Fig.~\ref{subfig:ppstream_pakup} and \ref{subfig:ppstream_pakup_video},  
Fig.~\ref{subfig:tvants_pakup} and \ref{subfig:tvants_pakup_video}) 
than download traffic 
(Fig.~\ref{subfig:pplive_pakdown} and \ref{subfig:pplive_pakdown_video},
Fig.~\ref{subfig:ppstream_pakdown} and \ref{subfig:ppstream_pakdown_video},
Fig.~\ref{subfig:tvants_pakdown} and \ref{subfig:tvants_pakdown_video}).\\
The energy bump exists for all the applications using mostly TCP. 
The energy bump indicates a possible periodic behavior at these time scales whatever the traffic direction or its nature.
The energy bump phenomenon has to be confirmed by studying the stationnarity of the traffic.
We study traffic stationnarity in section~\ref{subsec:stationarity}. 
However, the stationnarity analysis shows that the energy bumps observed in the spectra are essential phenomena and not simply artifact coming from non-stationnarity.\\
The energy bumps are observed for all the TCP applications but not for the UDP applications.
The well known TCP mechanisms used to transport data and TCP retransmission mechanisms could lead to such periodic traffic behaviors.
However, the periodic behaviors are observed for time scale $j=8$ (5.12s). A 5 seconds duration is a very long duration for TCP mechanisms.
It is not so obvious that this periodic behavior is provided by TCP mechanisms.\\
The periodic behaviors could also come from the video broadcasted in the network. 
However, SOPCast does not show any energy bump in its energy spectra and SOPCast broadcasts also video in the network.\\ 
Currently, we can not surely establish the source of these periodic behaviors. It does not seem to come from TCP mechanisms nor broadcasted video.
We are still investigating the periodic behavior we observed in the TCP applications energy spectra.\\

The energy bumps are characteristics shared by all the measured applications using massively TCP.
They illustrate how the application design may impact the properties of the generated network traffic. 
\subsection{Impact of the signaling traffic}
\label{subsec:signaling}
For all the  applications, whatever the transport protocol they use, their video upload energy spectra look like their overall upload energy spectra. 
This is illustrated on Fig.~\ref{subfig:pplive_pakup} and \ref{subfig:pplive_pakup_video}, \ref{subfig:sopcast_pakup} and \ref{subfig:sopcast_pakup_video}, \ref{subfig:ppstream_pakup} and \ref{subfig:ppstream_pakup_video},
 \ref{subfig:tvants_pakup} and \ref{subfig:tvants_pakup_video}.
Removing the signaling traffic has no impact on the upload traffic.\\
Regarding the download traffic, the video download energy spectra 
(Fig.~\ref{subfig:pplive_pakdown_video} \ref{subfig:sopcast_pakdown_video} \ref{subfig:ppstream_pakdown_video} and \ref{subfig:tvants_pakdown_video}) 
are different from their corresponding overall download energy spectra 
(Fig.~\ref{subfig:pplive_pakdown} \ref{subfig:sopcast_pakdown} \ref{subfig:ppstream_pakdown} and \ref{subfig:tvants_pakdown}).
Removing signaling traffic from the download traffic has clearly an impact on the download traffic because it modifies the download energy spectra.\\

Table~\ref{tab:ovh} shows the signaling traffic ratio for all the applications in upload and download. 
For all the applications, the signaling traffic represents a larger part in the download traffic than the upload traffic.\\
The upload signaling traffic is only provided by our controlled node to other peers in the Internet.
Since our node has high bandwidth capabilities, it serves video to many other peers. Table~\ref{tab:summary} indicates that the amount of upload traffic is 3 to 6 times larger than the amount of download traffic.
The signaling traffic sent by our node to other peers in the Internet counts only for a small part of the overall upload traffic.\\
The download signaling traffic is provided by other peers in the Internet to our controlled node. Our node just needs to download the video at the video bitrate. 
The download signaling traffic coming from many other peers counts for a large part of the overall download traffic. 
This explains the impact of signaling traffic on the download traffic.\\

To summarize our observations, the signaling traffic has no impact on the upload traffic but it has an impact on the download traffic.
In the following, we will discuss the stationnarity of the traffic that will help to better characterize the properties of the P2P IPTV signaling traffic.
\subsection{Traffic stationnarity}
\label{subsec:stationarity}
\begin{figure*}[!t]
\begin{center}
\subfigure[Upload: overall packet traffic (signaling and video traffic)]{
\includegraphics[scale=0.30]{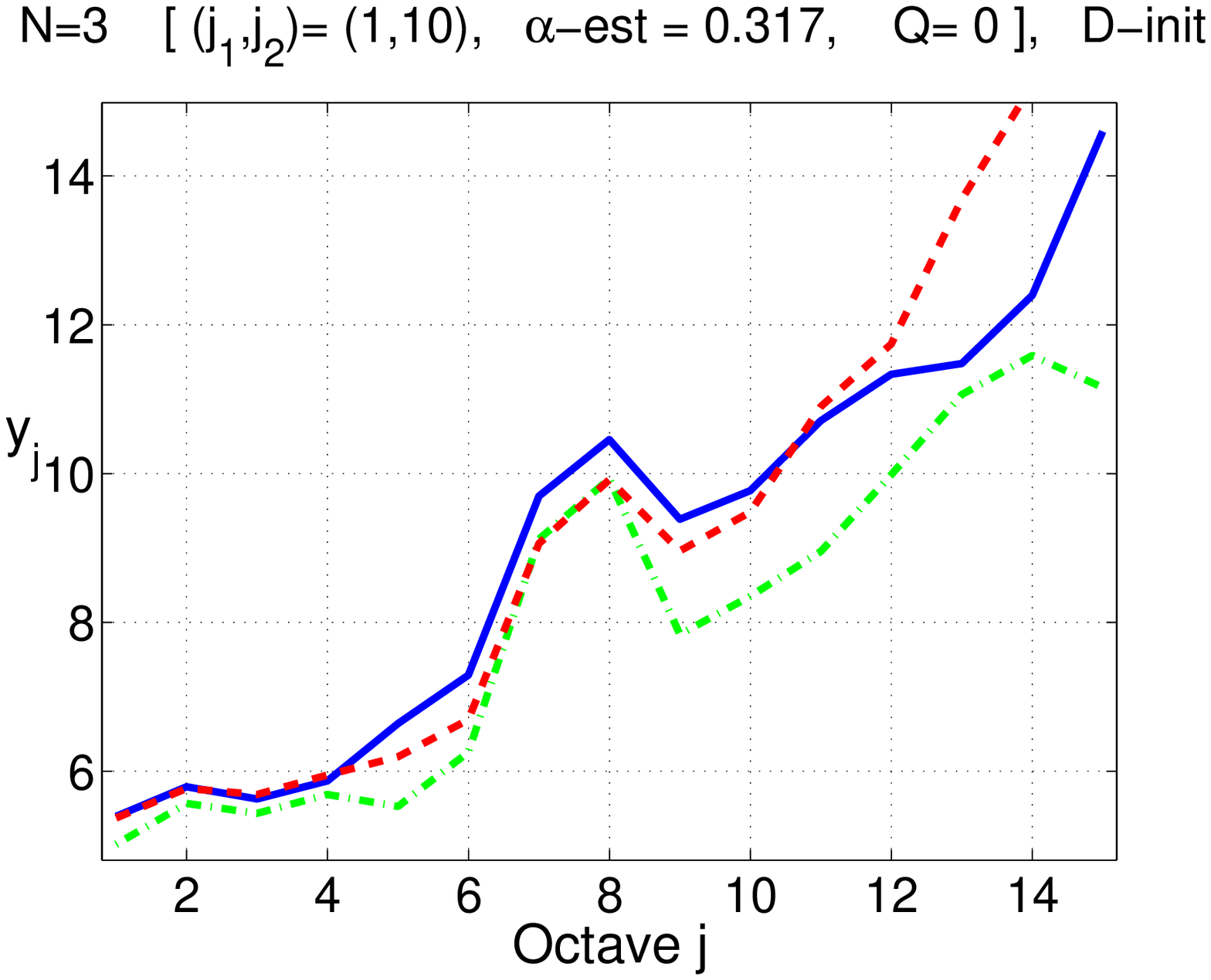}
\label{subfig:pplive_stat_pakup}
}
\subfigure[Upload: video packet traffic]{
\includegraphics[scale=0.30]{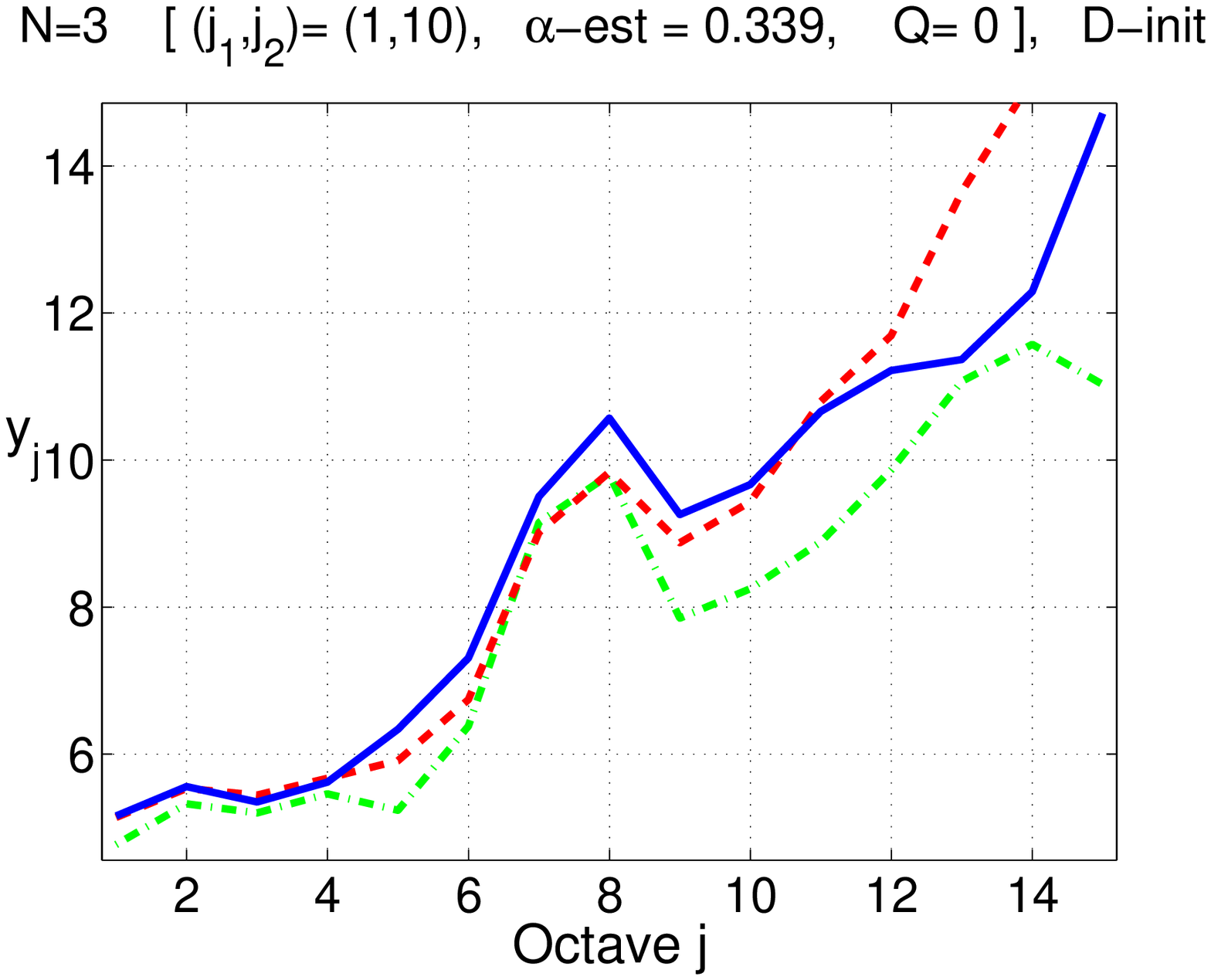}
\label{subfig:pplive_stat_pakup_video}
}
\subfigure[Download: overall packet traffic (signaling and video traffic)]{
\includegraphics[scale=0.30]{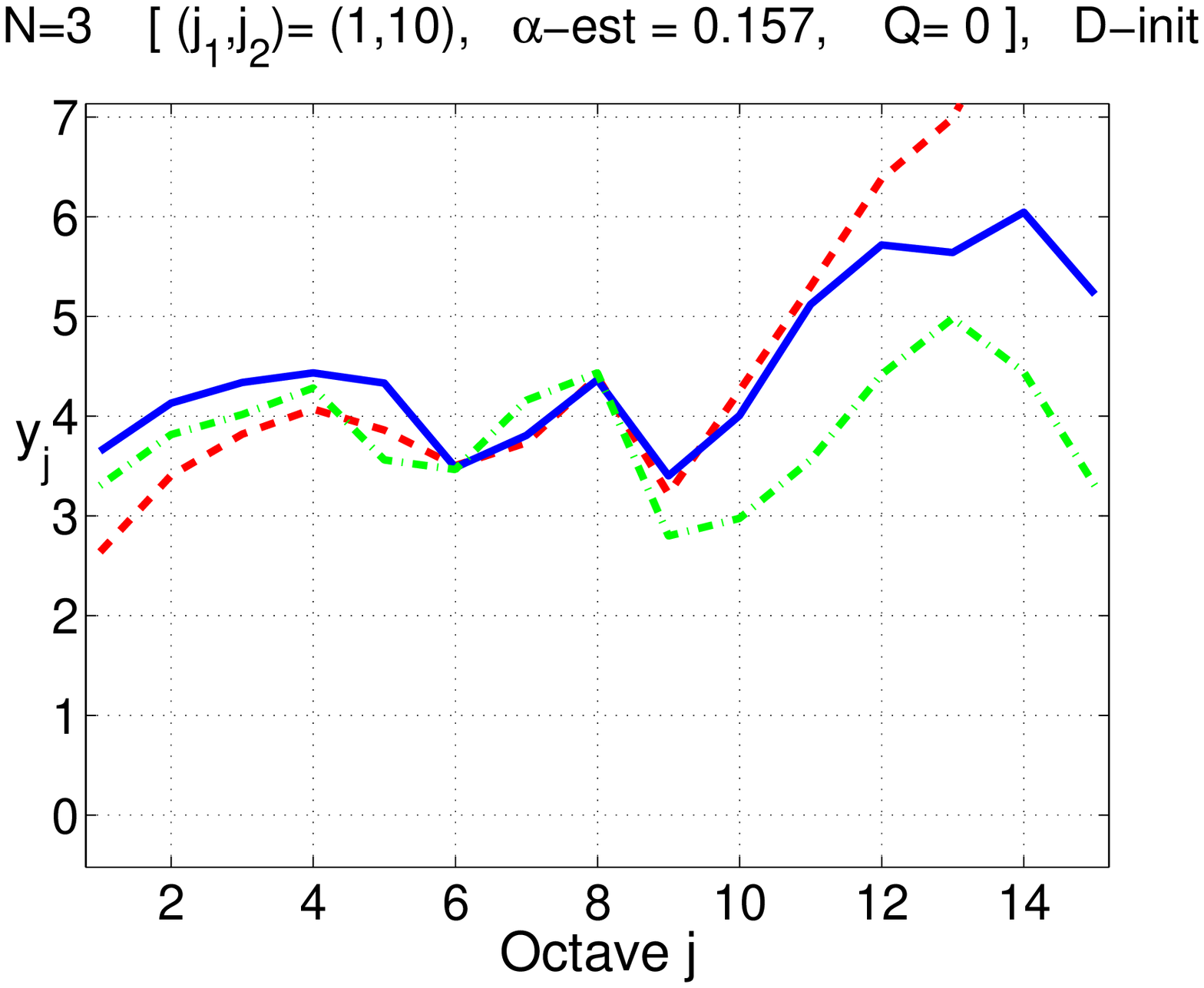}
\label{subfig:pplive_stat_pakdown}
}
\subfigure[Download: video packet traffic]{
\includegraphics[scale=0.30]{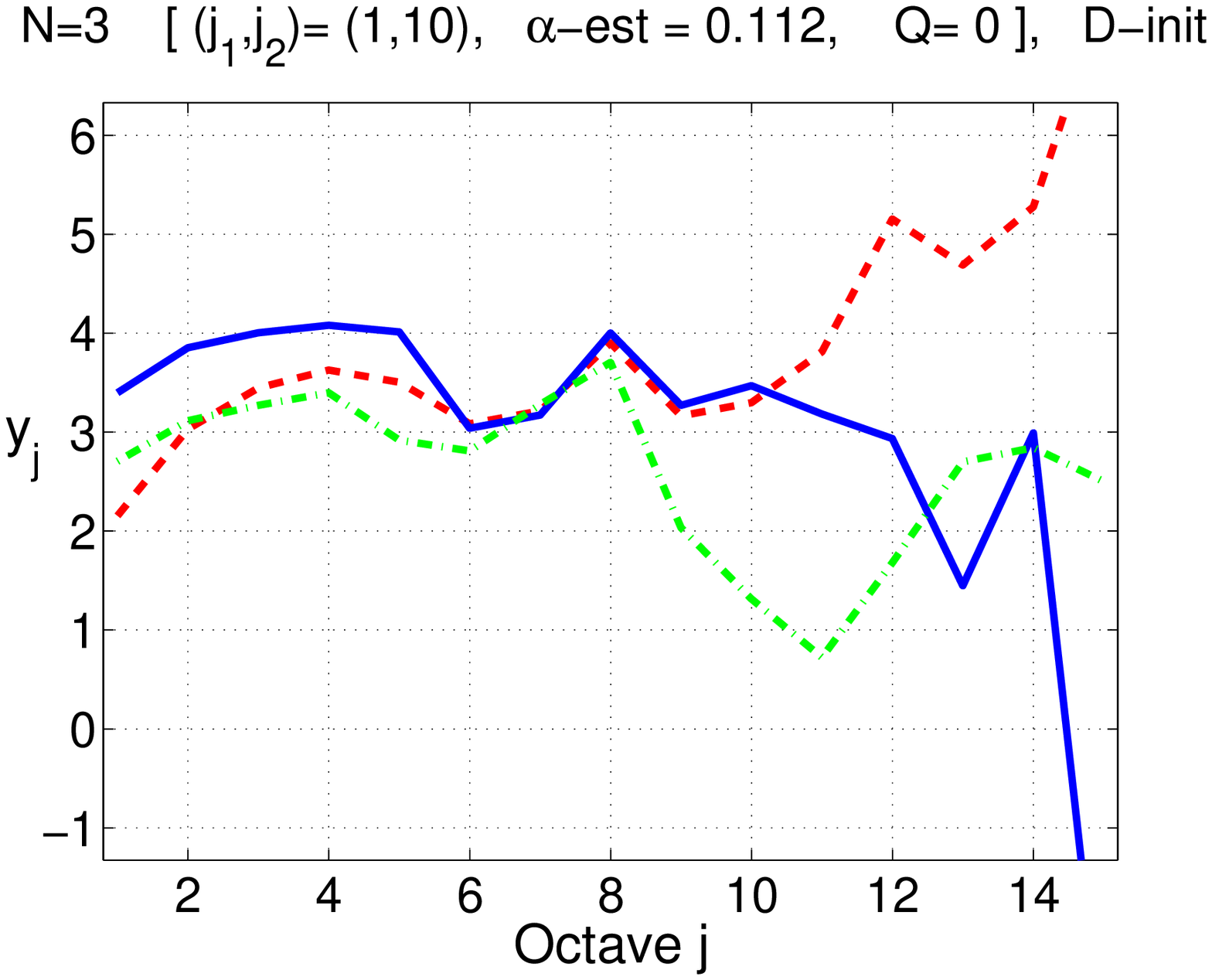}
\label{subfig:pplive_stat_pakdown_video}
}
\caption{TCP applications: PPLive. Traffic stationarity for the three equal parts of the traffic. The blue solid line is the first part, the dashed red line is the second part and the dash-dotted green line is the third part of the traffic.}
\label{fig:pplive_stat}
\end{center}
\end{figure*}
\begin{figure*}[!t]
\begin{center}
\subfigure[Upload: overall packet traffic (signaling and video traffic)]{
\includegraphics[scale=0.30]{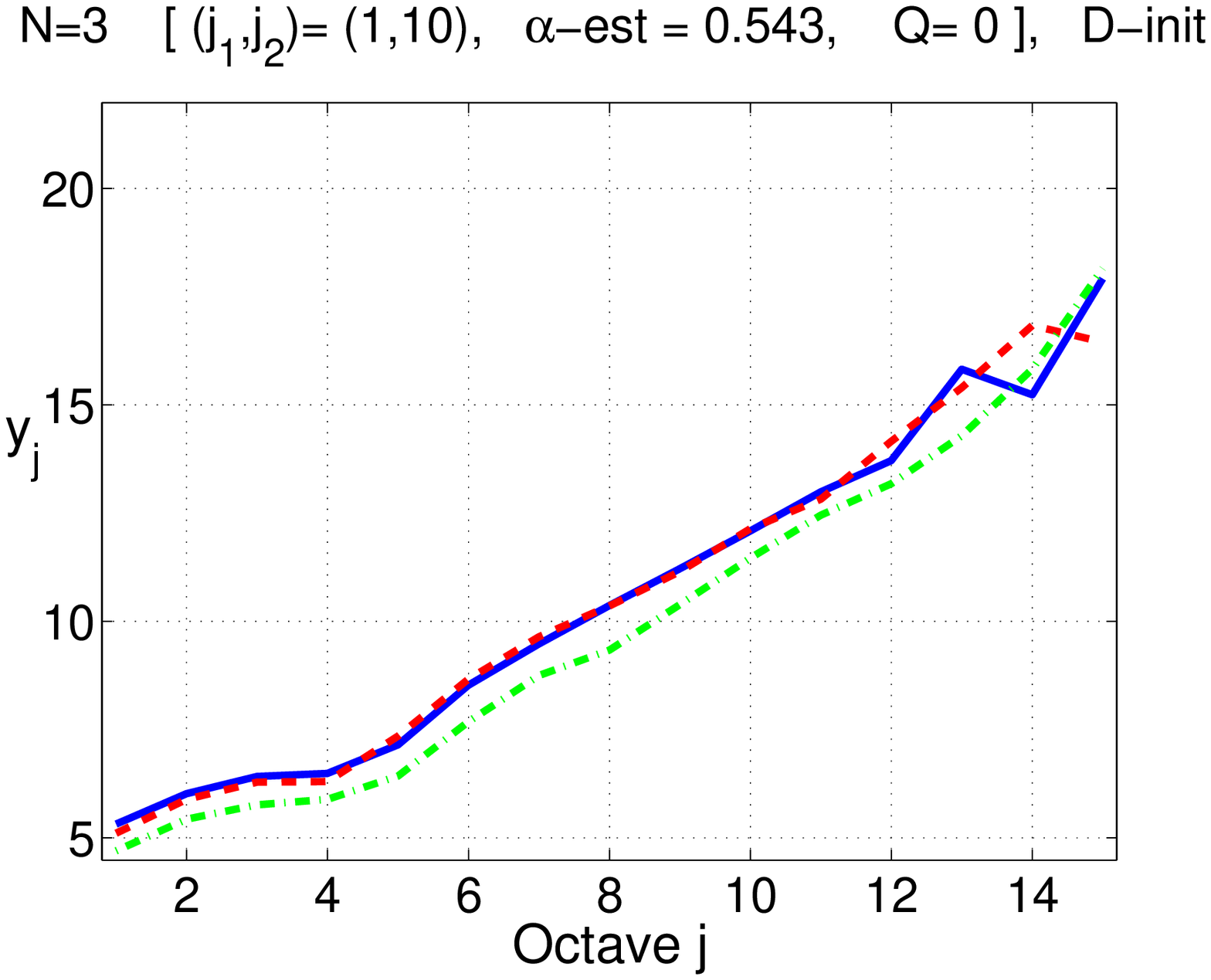}
\label{subfig:sopcast_stat_pakup}
}
\subfigure[Upload: video packet traffic]{
\includegraphics[scale=0.30]{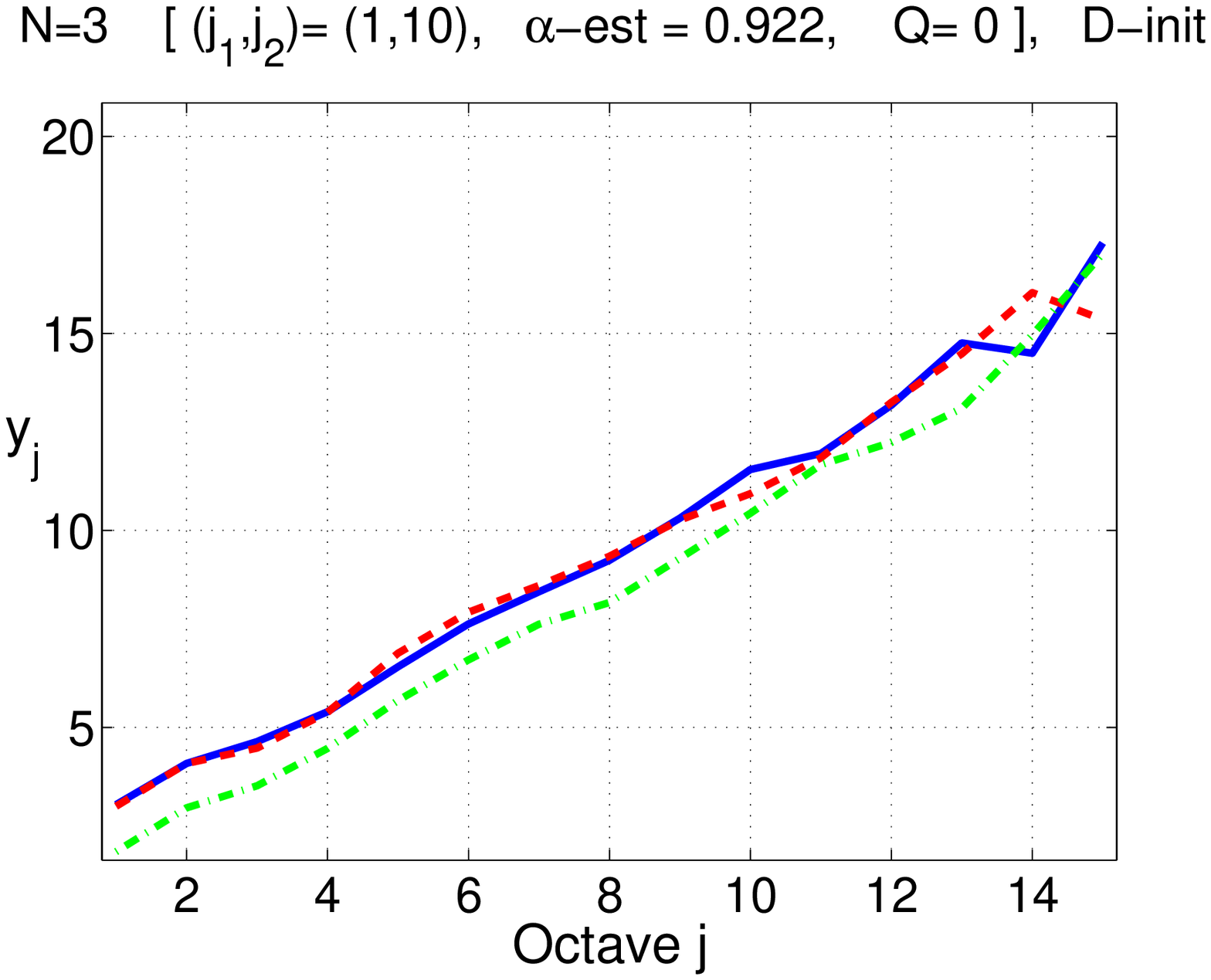}
\label{subfig:sopcast_stat_pakup_video}
}
\subfigure[Download: overall packet traffic (signaling and video traffic)]{
\includegraphics[scale=0.30]{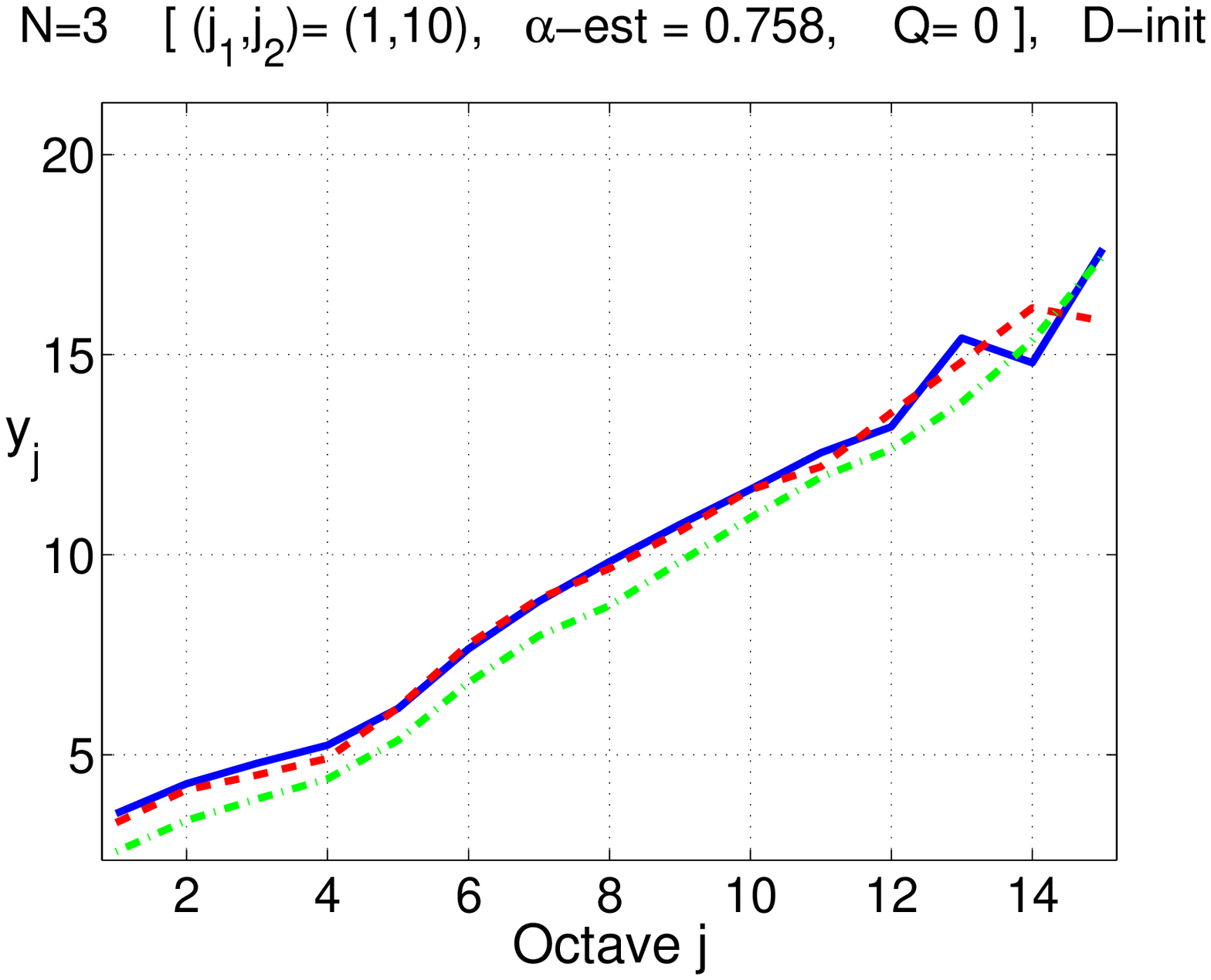}
\label{subfig:sopcast_stat_pakdown}
}
\subfigure[Download: video packet traffic]{
\includegraphics[scale=0.30]{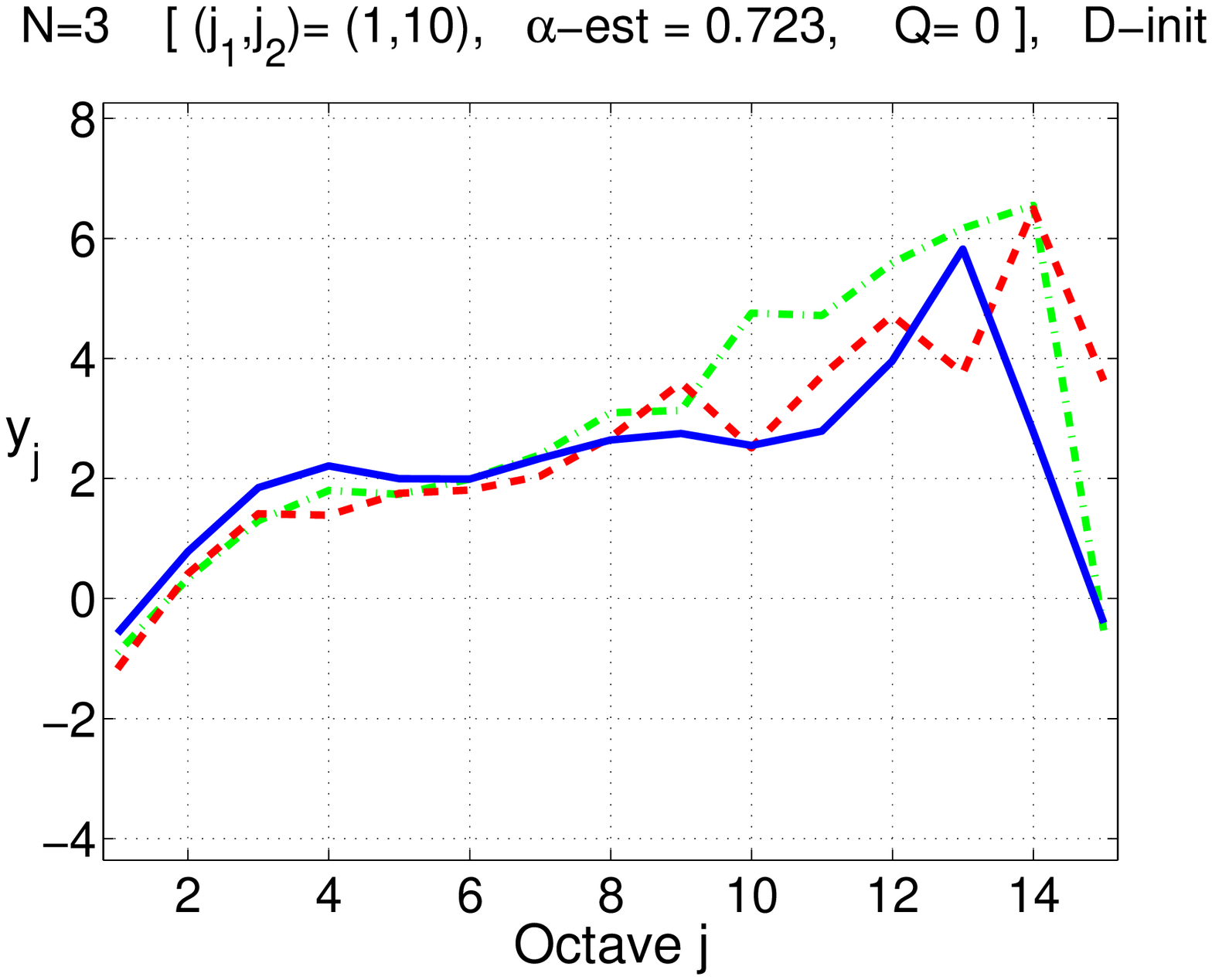}
\label{subfig:sopcast_stat_pakdown_video}
}
\caption{UDP applications: SOPCast. Traffic stationarity for the three equal parts of the traffic. The blue solid line is the first part, the dashed red line is the second part and the dash-dotted green line is the third part of the traffic.}
\label{fig:sopcast_stat}
\end{center}
\end{figure*}
From all the upload energy spectra 
(Fig~\ref{subfig:pplive_pakup}~\ref{subfig:pplive_pakup_video}, Fig~\ref{subfig:sopcast_pakup}~\ref{subfig:sopcast_pakup_video}, 
Fig~\ref{subfig:ppstream_pakup}~\ref{subfig:ppstream_pakup_video}, and Fig~\ref{subfig:tvants_pakup}~\ref{subfig:tvants_pakup_video}), 
we observe a linear increase, starting from $j=6$ for SOPCast, from $j=9$ for PPLive, from $j=12$ for PPStream or $j=10$ for TVAnts.
We also show that the linear increase exists in the download traffic 
(Fig.~\ref{subfig:pplive_pakdown} \ref{subfig:sopcast_pakdown} \ref{subfig:ppstream_pakdown} and \ref{subfig:tvants_pakdown}) 
but it is modified (Fig.~\ref{subfig:sopcast_pakdown_video}) or wasted (Fig.~\ref{subfig:pplive_pakdown_video} \ref{subfig:ppstream_pakdown_video} \ref{subfig:tvants_pakdown_video}) when removing signaling traffic.\\
We already show that signaling traffic has an impact on the download traffic because it count for a larger amount of data.
The signaling traffic may also lead to the linear increase in the download energy spectra.\\

A linear increase indicates a possible long-range dependency of the traffic (LRD).
It means that the traffic fluctuates largely and is not predictable.
With such traffic fluctuations, it becomes impossible to forecast the traffic behavior or to make network provisioning.\\

The signaling traffic is used by peers to get the video chunks they need.
If the signaling traffic is responsible for LRD,  the signaling traffic generates itself the troubles to download the video.
In other words, the design of the applications would be not efficient if the signaling traffic generates the long-range dependency of the traffic.\\
For a P2P network, a LRD traffic indicates, there is no stability in the network traffic. 
Then, it becomes a hard task to provide QoS parameters (delay, bandwidth, jiiter) to users because networks conditions are always changing.\\ 
Regarding P2P IPTV systems, which are delay sensitive, the traffic LRD has to be avoided because it will directly affect and decrease the quality of video reception.
For example, under a high churn of peers, each peer has always to discover new peers and to establish new partnerships with other peers to receive the video data chunks.
In this case, there would be no stability in the overlay network because of the disruptions in the peers connections and communications. 
This would lead to non-predictable traffic and long-range dependency of the traffic.\\ 

In this section, we want to know if we really observe a long-range dependency of the traffic. 
To this end, we study the stationnarity of the traffic.
We split each trace in three equal parts, and we used the previous wavelet transform method to analyze all the parts of the traffic.
We visually control the traffic stability with LDEstimate.\\
Due to space limitations, we only present one example for TCP applications (PPLive, Fig.~\ref{fig:pplive_stat}) and one example for UDP application (SOPCast, Fig.~\ref{fig:sopcast_stat}).
The other TCP applications energy spectra are similar to the TCP example (Fig.~\ref{fig:pplive_stat}) and can be found in the appendix.\\
In each logscale diagram, we plot the three parts of the trace. The blue solid line is the first part of the trace, the dashed red line is the second part and the green dash-dotted line is the third part of the trace.
The traffic is stationary if each part of the traffic looks like the other parts and has the same energy level.
\\

For TCP applications (Fig.~\ref{fig:pplive_stat}), the three parts of the energy spectra are similar at small time scales but become different at large time scales (about $j=10$).
There is stationnarity in the traffic until $j=10$. The traffic is not stationary beyond $j=10$.\\ 
As shown in section~\ref{sec:versus}, the TCP applications experiment a bump in their energy spectra at time scale $j=8$, whatever the traffic direction or its nature. 
At this time scale, the traffic is stationary and it demonstrates that the observed energy bumps are essential phenomenon and not simply
artifact from non-stationnarity. 
On the contrary, the linear increase observed from $j=9$ is not a long-range dependency of the traffic.\\

For UDP applications~(Fig.~\ref{fig:sopcast_stat}), the three parts of the traffic are similar and increase linearly. The traffic of UDP applications is stationary. 
The linear increase characterizes a long-range dependency of the UDP applications traffic represented by SOPCast.
Removing signaling traffic modifies the linear increase in the SOPCast download traffic.
For UDP applications, signaling traffic may lead the to the long-range dependency in the traffic.\\ 

In this section, we show an important design difference between TCP and UDP applications. 
The traffic of UDP applications is stationary and present long-range dependency
whereas the traffic of TCP applications is not stationary and does not experiments long-range dependency.
\subsection{Analysis of the top download flows}
\label{sec:topdl}
\begin{figure*}[!t]
\begin{center}
\subfigure[PPLive Download: overall packet traffic (signaling and video traffic)]{
\includegraphics[scale=0.30]{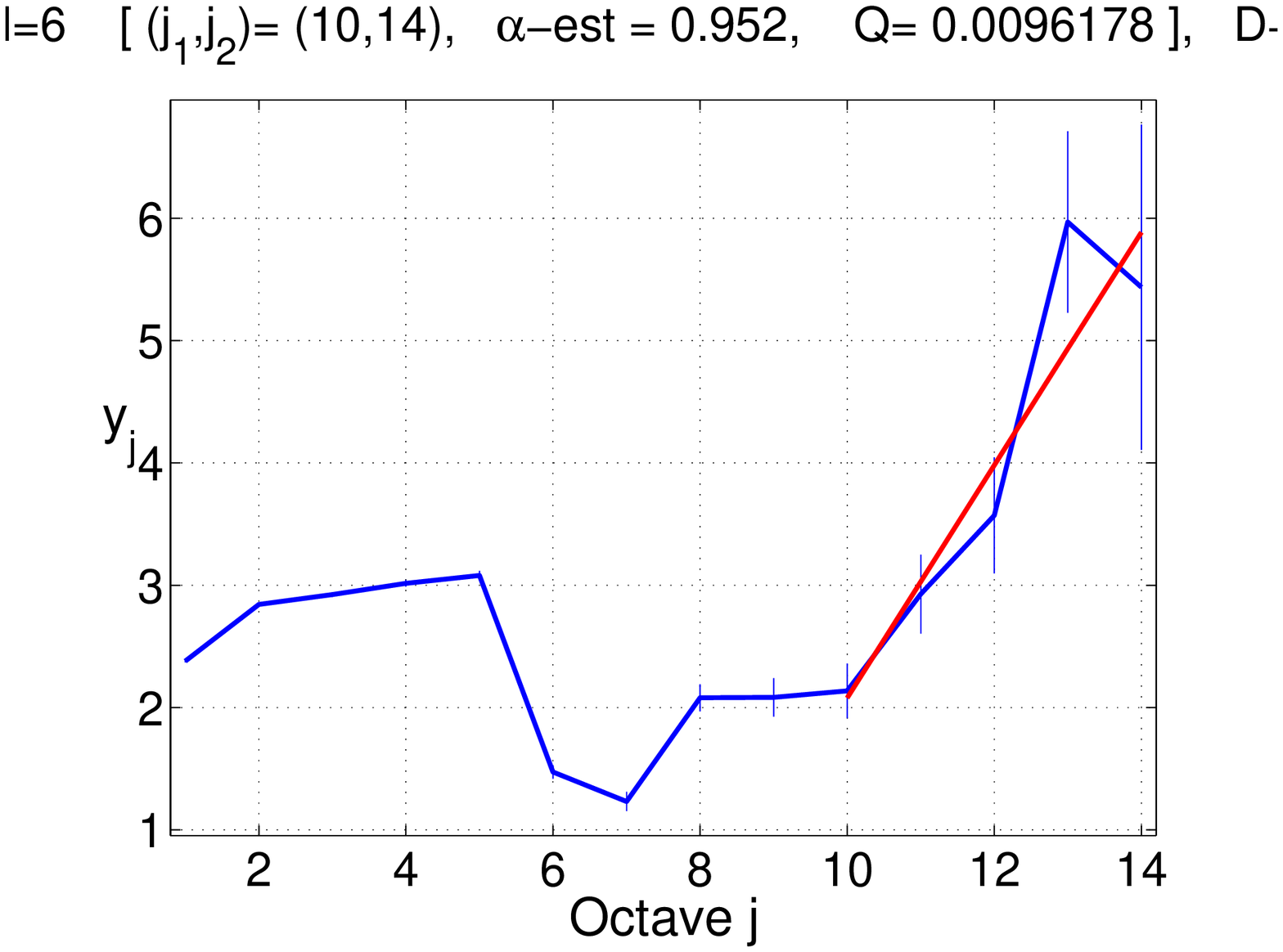}
\label{subfig:pplive_flow}
}
\subfigure[PPLive Download: video packet traffic]{
\includegraphics[scale=0.30]{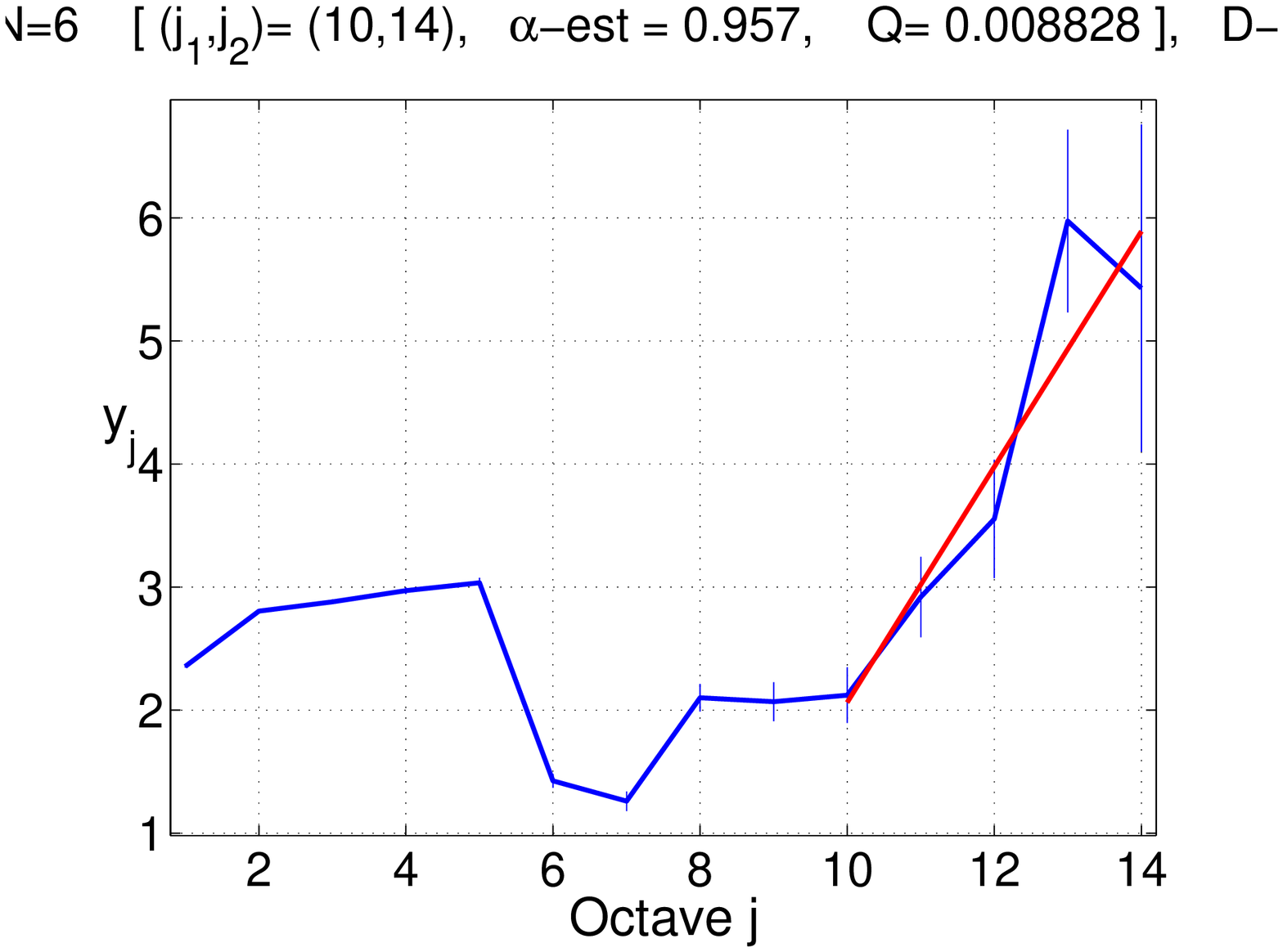}
\label{subfig:pplive_flow_video}
}
\subfigure[SOPCast Download: overall packet traffic (signaling and video traffic)]{
\includegraphics[scale=0.30]{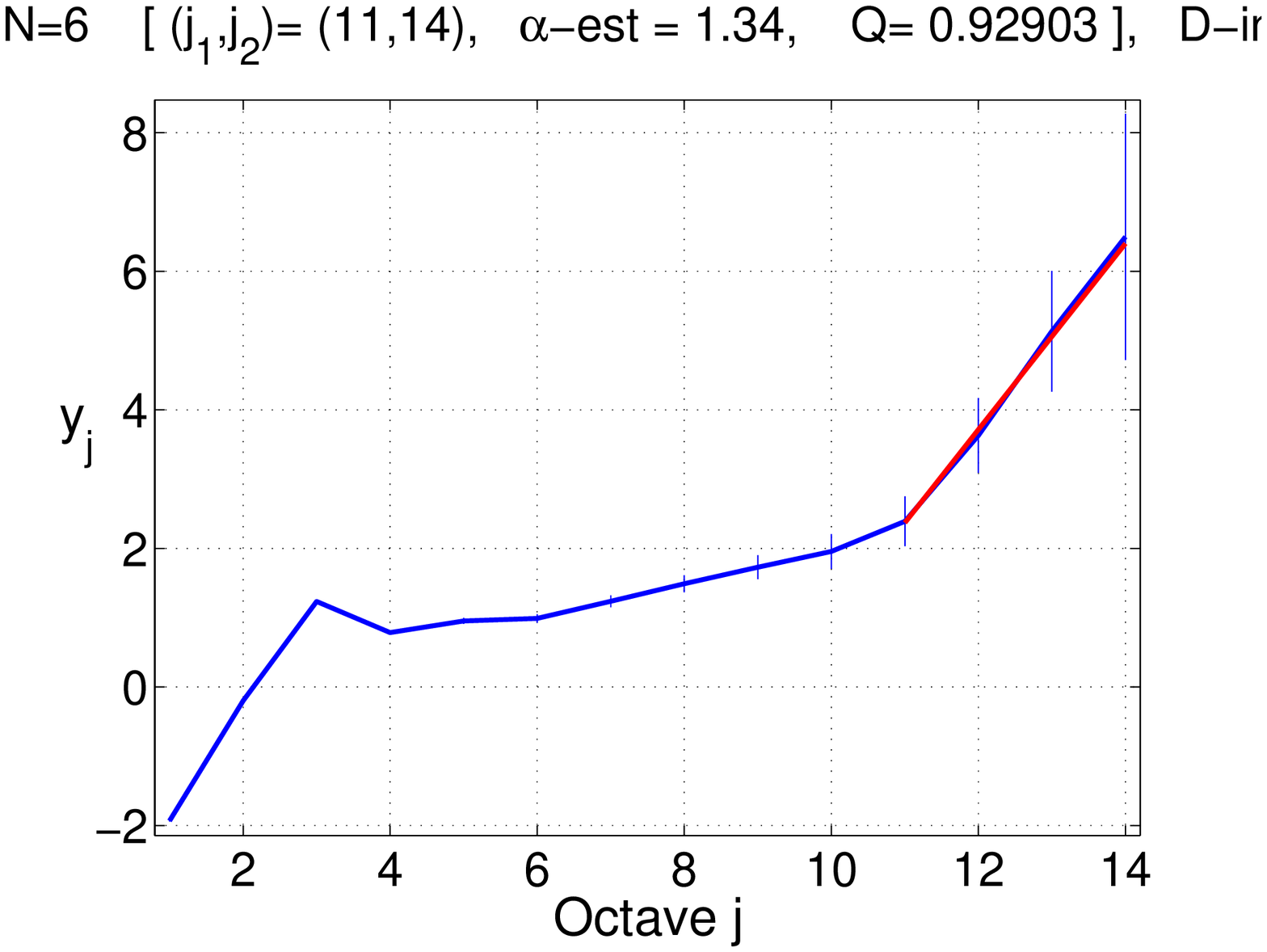}
\label{subfig:sopcast_flow}
}
\subfigure[SOPCast Download: video packet traffic]{
\includegraphics[scale=0.30]{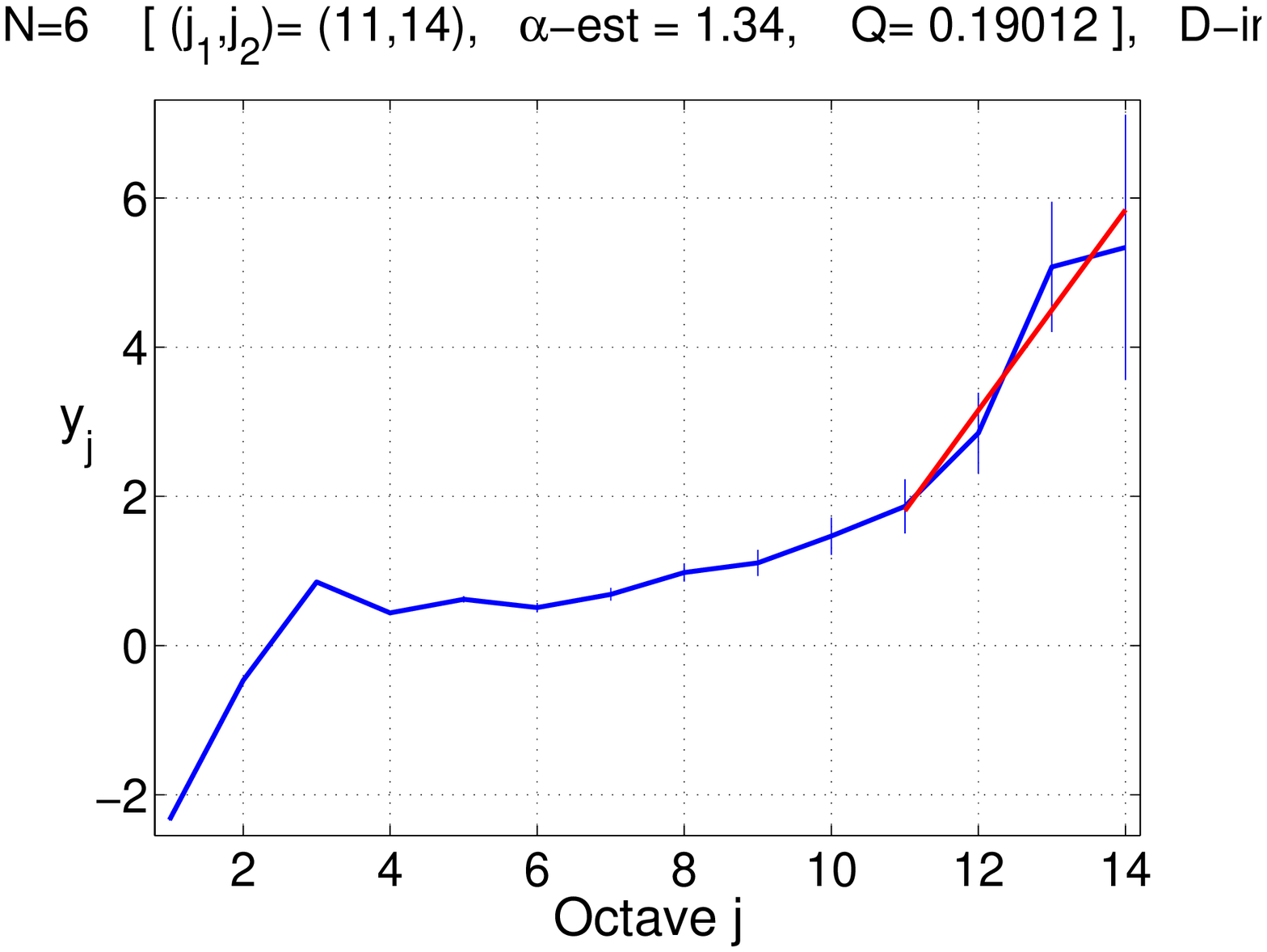}
\label{subfig:sopcast_flow_video}
}
\caption{Top download flows}
\label{fig:topflows}
\end{center}
\end{figure*}
In the previous experiments, we make our observations based on the aggregate traffic received and sent by our controlled peers situated in our campus network.
Our nodes has high bandwidth capacities and they play an important role in duplicating data to a large amount of peers.
The high-speed access of our nodes help to get the video efficiently compare to a residential access to the Internet, 
in which received packets are limited by a smaller download rate.\\

To extend our previous findings without being strongly related to our network environment, 
we want to analyze the properties of a  single flow
instead of observing the aggregate traffic in our contoled peers. 
Whatever the network environment is, a peer will try to download the video at its bitrate.
On the contrary, the number of duplicate flows a peer could upload  is directly related to the capacities of the network environment.
Thus, we limit the scope of this experiment to the download traffic flows because 
this is a general observation of the traffic that does not directly depend on the network environment.\\

For all the traces, we isolated the top peer that sent the biggest amount of data to our nodes.
We refer to these resulting flows as top download flows since the data transported by these flows are downloaded by our controlled nodes.\\
Table~\ref{tab:flows} summarizes the amount of data carried by each top flow for all the applications.
\begin{table}[t]
\begin{tiny}
\renewcommand{\arraystretch}{1.7}
\caption{Top download flows}
\label{tab:flows}
\begin{center}
\begin{tabular}{|c|c|c|c|c|}
\hline
& \multicolumn{2}{|c|}{\scriptsize{Overall traffic}} & \multicolumn{2}{|c|}{\scriptsize{Video traffic}} \\ \hline
& \tiny{Volume (MB)} & \tiny{\#Packets} & \tiny{Volume (MB)} & \tiny{\#Packets} \\ \hline
\scriptsize{PPLive} & \tiny{106.30} & \tiny{82,428} &  \tiny{105.97} & \tiny{76,797} \\ \hline
\scriptsize{PPStream} & \tiny{13.21} & \tiny{23,951} & \tiny{11.79} & \tiny{9,809} \\ \hline 
\scriptsize{SOPCast} & \tiny{61.91} & \tiny{63,054} & \tiny{59.20} & \tiny{47,187} \\ \hline
\scriptsize{TVAnts} & \tiny{53.86} & \tiny{41,740} & \tiny{52.13} & \tiny{37,460} \\ \hline
\end{tabular}
\end{center}
\end{tiny}
\end{table}
We already show in~\cite{silverston-nossdav07} that all the applications do not implement the same mechanisms to download the video. 
According to the applications, the video can be received from only a few provider peers at the same time or from many peers and the video peers session durations are various.
This explains why the amount of data transported by the top flows are different for all the applications.\\ 

The top download flows sent by the top peer to our nodes will be analyzed by using wavelet based transform method with LDEstimate, similarly to the previous experiments.
Due to space limitation, we only present on Fig.~\ref{fig:topflows}, the energy spectra for a single TCP application (PPLive Fig~\ref{subfig:pplive_flow} and \ref{subfig:pplive_flow_video}) 
and for UDP application (SOPCast Fig.~\ref{subfig:sopcast_flow} and \ref{subfig:sopcast_flow_video}).
The other TCP applications plots are similar to the presented TCP application and can be found in the appendix.\\

We notice for the two applications that their video energy spectra look similar to their overall energy spectra.
This was expected because these flows are sent by the top contributor peers and transport almost entirely video packets and not signaling packets.
Removing signaling traffic on these flows can only have a limited impact, depending on the signaling packets in the flows.\\ 
For example, the top download flow of SOPCast transport 15,867 packets of signaling traffic ($63054-47187$) counting for 2.71~MBytes whereas PPLive top download flow transport only 5,731 signaling packets ($82428-76797$)
counting for 0.33~MBytes.\\

Regarding TCP applications Fig.~\ref{subfig:pplive_flow} and \ref{subfig:pplive_flow_video}, 
until time scale $j=10$, the energy spectra of the top flow look similar to the aggregate traffic. Beyond this time scale, the energy spectra of the top download flow are 
different from the aggregate traffic because the energy spectra of the top flow are increasing.\\
With UDP applications, until time scale $j=11$, the energy spectra of the top flow are different from the aggregate traffic. 
Fig.~\ref{subfig:sopcast_flow} and \ref{subfig:sopcast_flow_video} show an energy bump at time scale $j=3$, then the energy spectra
increase slightly from $j=4$ to $j=11$. Beyond $j=11$, we observe the linear increase usually observed for the UDP energy spectra.

In this experiment, we observe that the top flows in the download traffic do not have the same scaling properties as the aggregate download traffic.
We did the same experiments for the 10th top download flows (i.e. the 10th flow according to data volume transported).
The 10th top flows present the same scaling properties as the top flows. The plots for the 10th top flows can be shown in the appendix.\\ 
The aggregate traffic is not only the mix of every single flow.
The granularity of the P2P IPTV traffic has to be taken into account when designing P2P IPTV traffic models.
\section{Results Discussion}
\label{sec:discussion}
In this work, we analyzed the P2P IPTV traffic by using a wavelet based transform method.
This allows us to characterize this traffic and to understand its properties and impact on the network.
Thanks to our original P2P IPTV traffic analysis, we have many new findings and observations that have to be summarized and discussed.\\

First of all, we observed that the energy spectra of TCP applications are different from the energy spectra of UDP applications (section~\ref{sec:versus}).
One of the most relevant difference is the energy bump observed in the spectra of TCP applications at time scale $j=8$ (5.12s), which  indicates a possible periodic behavior in the traffic.
Intuitively, we could believe these differences come from the two different transport protocols used.
However, a 5 seconds periodic behaviors is a very long duration for TCP mechanisms and TCP should not be the responsible of this periodic behavior.
With a simple application design difference, the scaling properties of the generated traffic do not have the same impact on the network.\\

Secondly, for all the applications, the signaling traffic represents a larger part in the download traffic than the upload traffic (section~\ref{subsec:signaling}). 
The signaling traffic has clearly an impact on the scaling properties of the download traffic and has no impact on the upload traffic.
This observation is important since signaling traffic is necessary to coordinate the data exchanges in such P2P systems. 
For scalability reasons, the amount of signaling traffic has to be kept as low as possible.
The download signaling traffic comes from other peers on the Internet that request the video data. 
Efforts have to be made to reduce the number of packets sent by the signaling protocol to get the video data and to preserve the scalability of these systems in the network.\\

Then,  the previous observation highlights an important point when modeling P2P IPTV traffic: the download traffic has not the same properties as the upload traffic. 
The differences between both sides of the traffic (i.e. upload and download) have to be taken into account carefully when designing synthetic traffic generation models.\\

The generated traffic of TCP applications is not stationary beyond time scale $j=10$. On the contrary, the traffic of UDP applications is stationary.
As shown in section~\ref{subsec:stationarity}, the stationnarity experiment proves that the signaling traffic of UDP application involves long-range dependency in the download traffic.
The UDP application experiments also long-range dependency in the upload traffic.
In presence of traffic LRD, the network conditions are always changing and it becomes a hard task to provide QoS parameters as delay for users to get
good quality video.\\ 
This finding highlights the not so trivial choice of transport protocols for P2P IPTV traffic.
It is usually admitted that the non-elastic data transfer -as video- has to rely on UDP but we show that UDP traffic may lead to trouble in the network traffic.\\ 

Finally, the aggregate download traffic of P2P IPTV systems has not exactly the same scaling properties as the top download flow (section~\ref{sec:topdl}).
The granularity of the traffic has to be taken into account when designing P2P IPTV traffic models.
A P2P IPTV traffic model based only on flows properties would fail to capture the global characteristics of the aggregate traffic.
The use of an inappropriate traffic model would lead to wrong results when simulating new architectures with such significant input parameter.
\section{Conclusion}
\label{sec:conclusion}
In this paper, we analyzed network traffic generated by P2P IPTV applications. 
We performed an extensive measurement campaign during the 2006 FIFA World Cup and we measured the most popular P2P IPTV applications on the Internet.
We used wavelet transform based method to study the P2P IPTV traffic at different time scales and to characterize its properties.\\

Our multiscale traffic analysis show how different are the scaling properties of the TCP and UDP traffics.
For all the applications, the signaling traffic has a significant impact on the download traffic but not on the upload traffic. 
It involves scalability concerns regarding the P2P IPTV signaling protocols used to download the video data.
The UDP traffic is stationary and leads to long-range dependency of the traffic. 
The choice of UDP as transport protocol for non-elastic transfers in P2P networks becomes not so trivial since the traffic LRD indicates that the traffic is not predictable in the network.
The scaling properties of the download traffic are different from the upload traffic.
The traffic granularity and both traffic directions have to be taken into account to model P2P IPTV traffic accurately.\\

Currently, we are analyzing the traffic collected during other games and under different network environments to extend our observations. 
It will allow us to have a finer analysis of our findings and could also help to answer to the open questions introduced by this work.
In a long-term work, the characterization of the P2P IPTV traffic will help us to accurately model and simulate such systems. 

\clearpage
\section*{Appendix}

\begin{figure*}[!h]
\begin{center}
\subfigure[Upload: overall packet traffic (signaling and video traffic)]{
\includegraphics[scale=0.30]{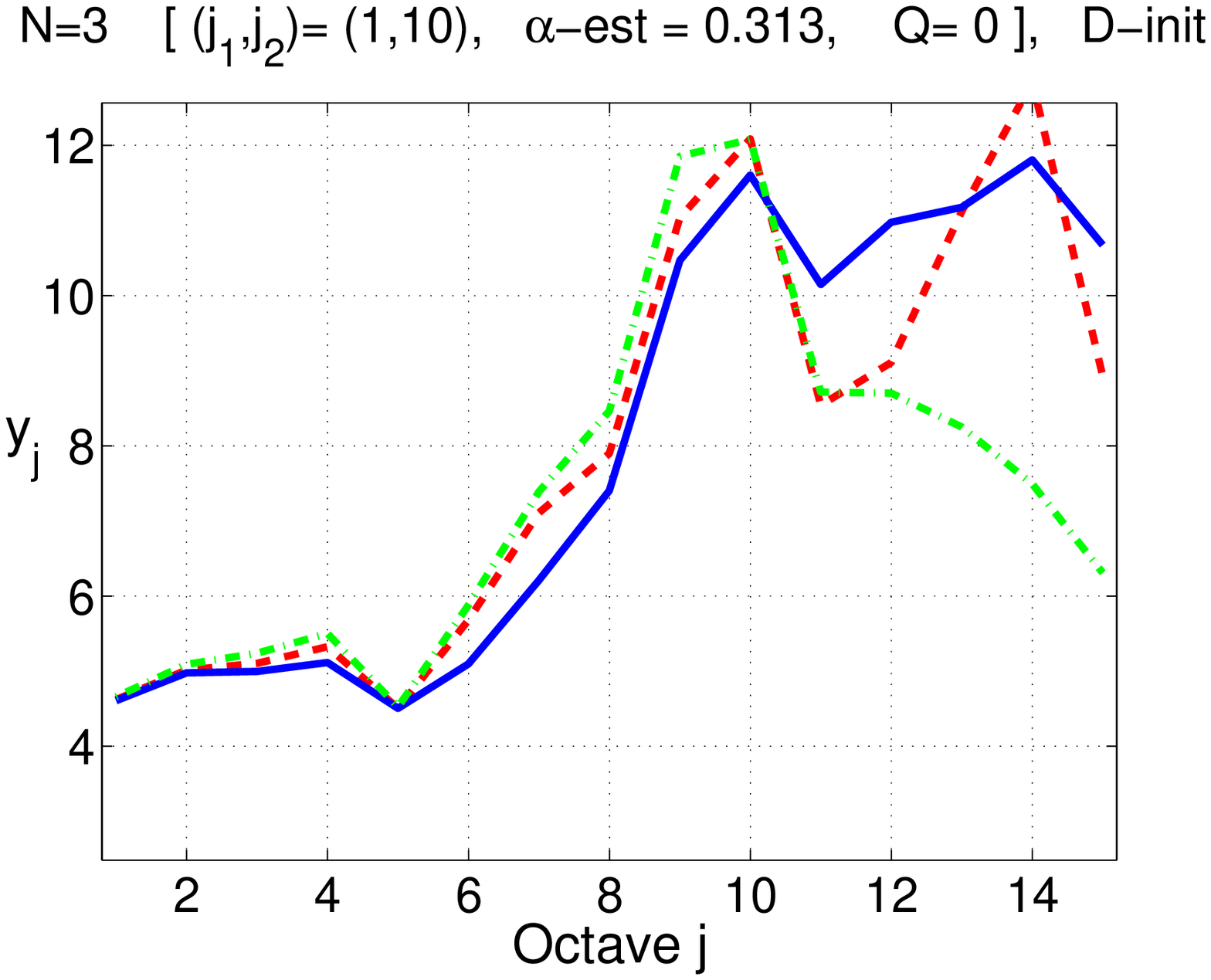}
\label{subfig:ppstream_stat_pakup}
}
\subfigure[Upload: video packet traffic]{
\includegraphics[scale=0.30]{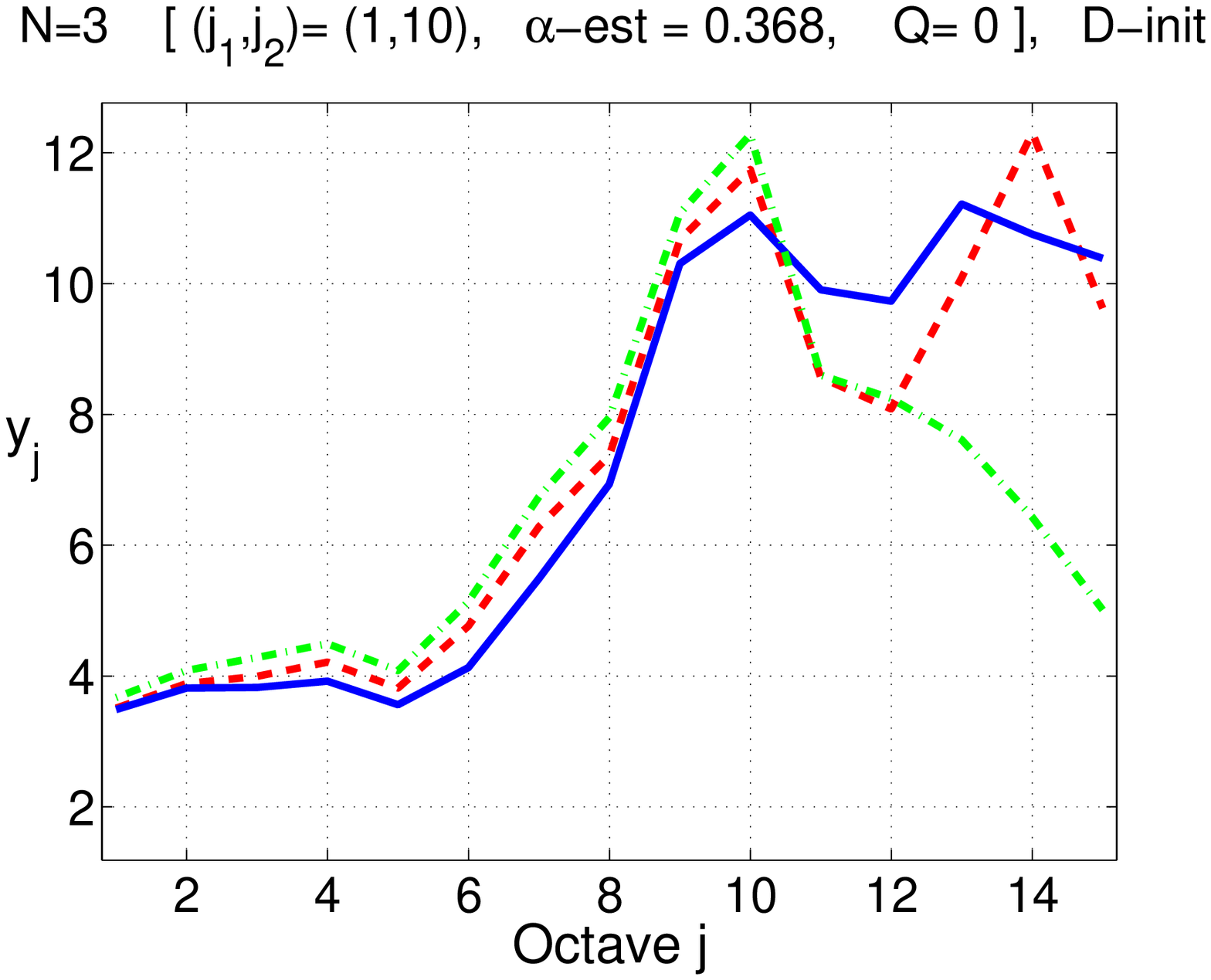}
\label{subfig:ppstream_stat_pakup_video}
}
\subfigure[Download: overall packet traffic (signaling and video traffic)]{
\includegraphics[scale=0.30]{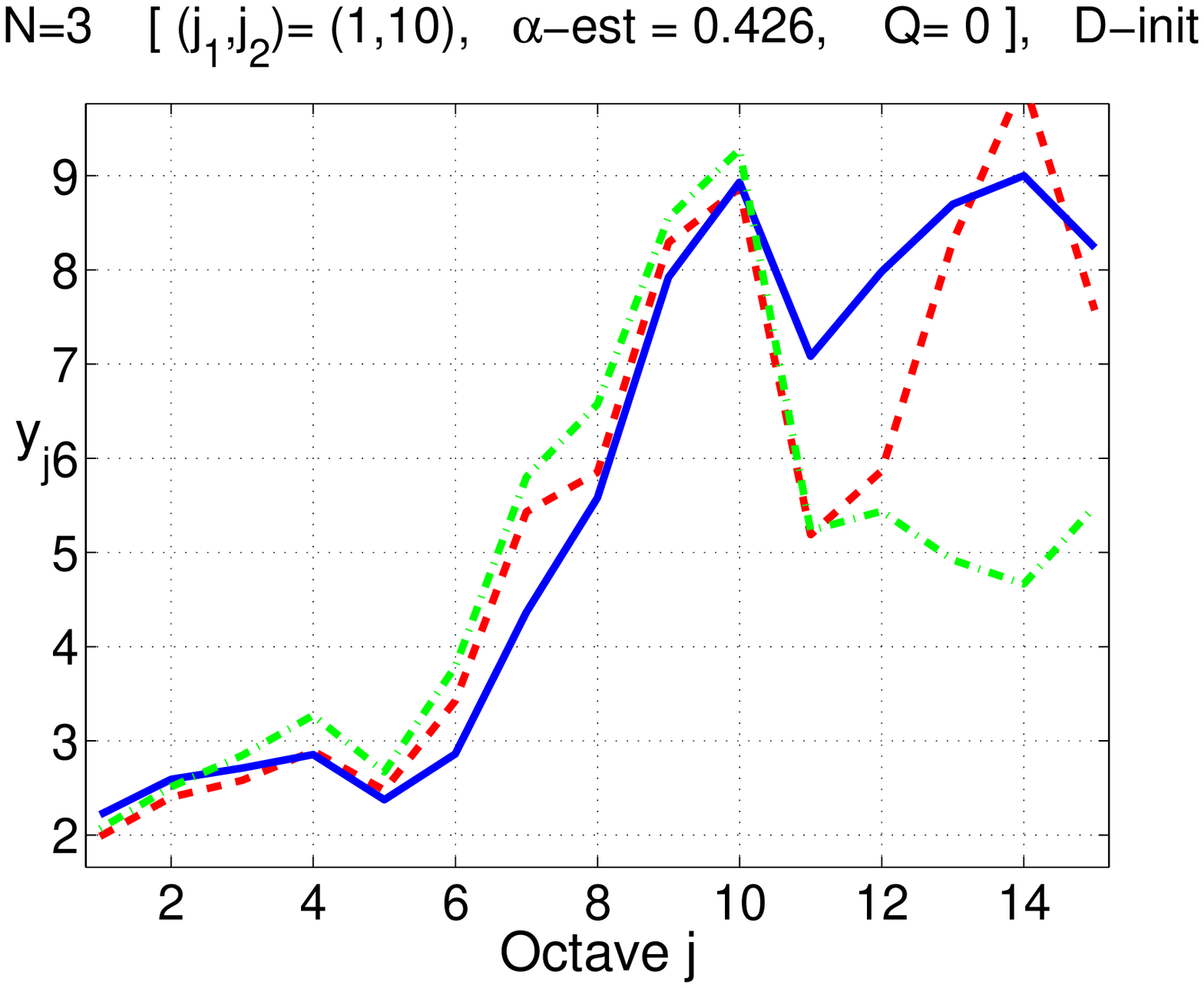}
\label{subfig:ppstream_stat_pakdown}
}
\subfigure[Download: video packet traffic]{
\includegraphics[scale=0.30]{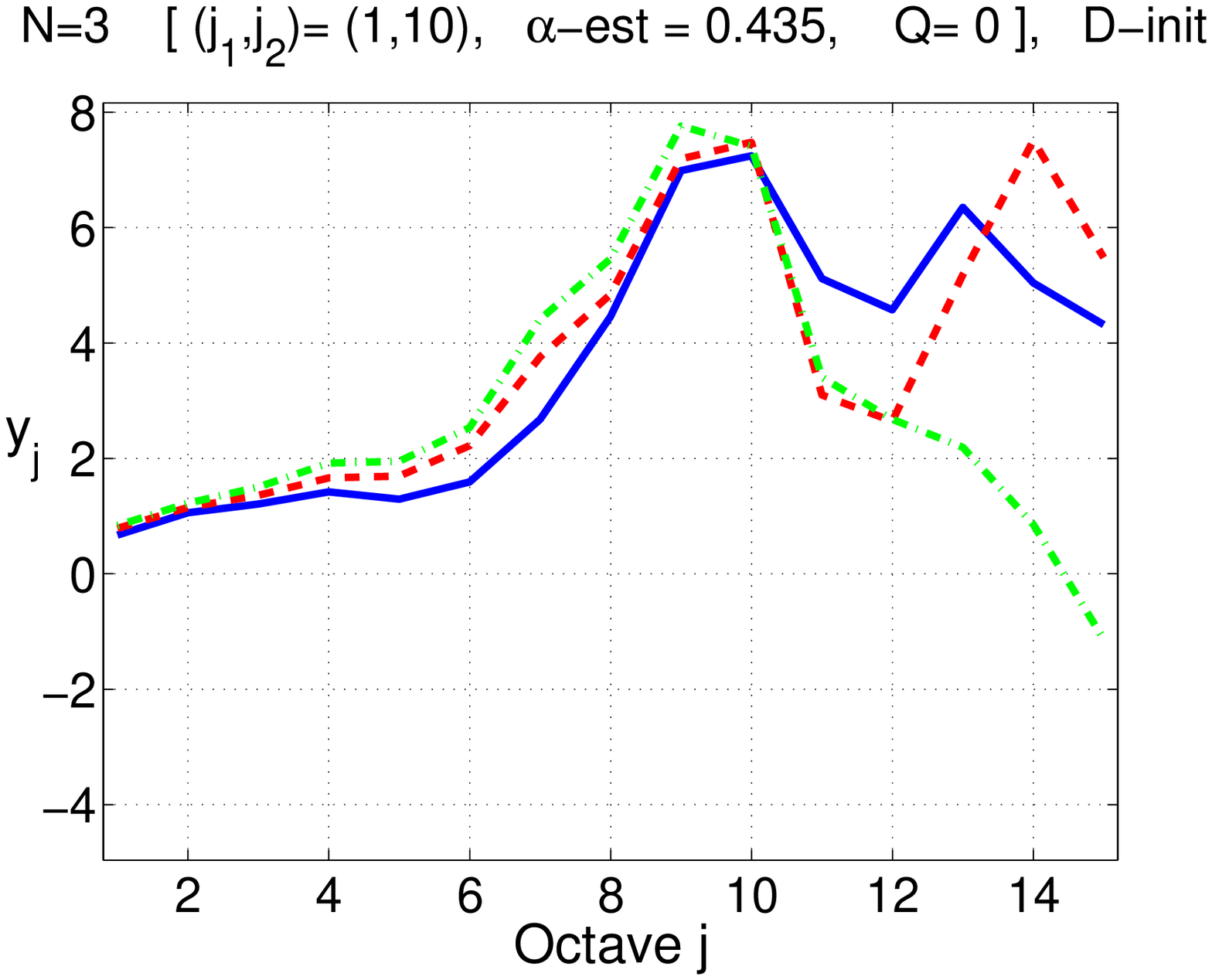}
\label{subfig:ppstream_stat_pakdown_video}
}
\caption{PPStream: Traffic stationarity for the three equal parts of the traffic. The blue solid line is the first part, the dashed red line is the second part and the dash-dotted green line is the third part of the traffic.}
\label{fig:pplive_stat}
\end{center}
\end{figure*}
\begin{figure*}[!h]
\begin{center}
\subfigure[Upload: overall packet traffic (signaling and video traffic)]{
\includegraphics[scale=0.30]{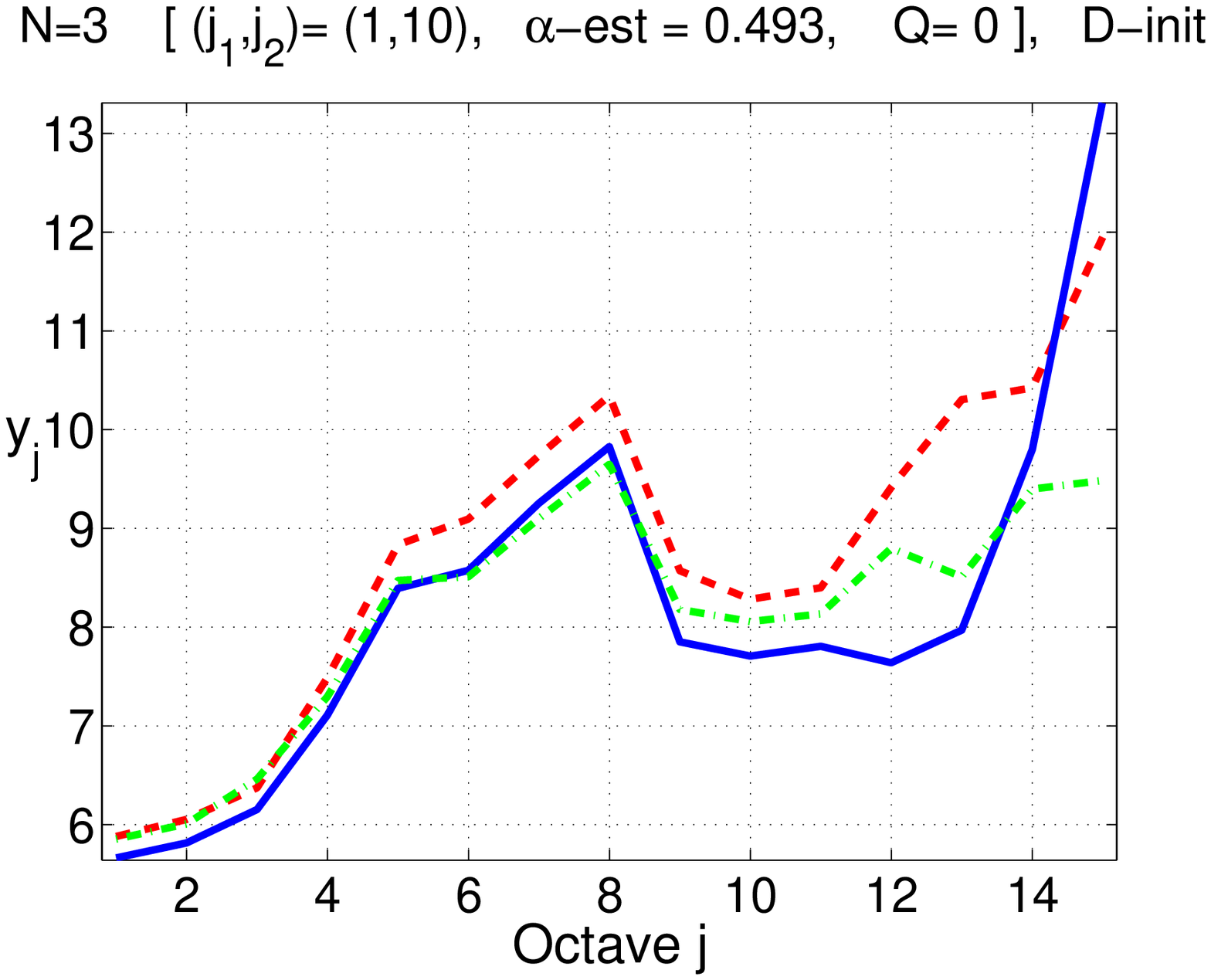}
\label{subfig:tvants_stat_pakup}
}
\subfigure[Upload: video packet traffic]{
\includegraphics[scale=0.30]{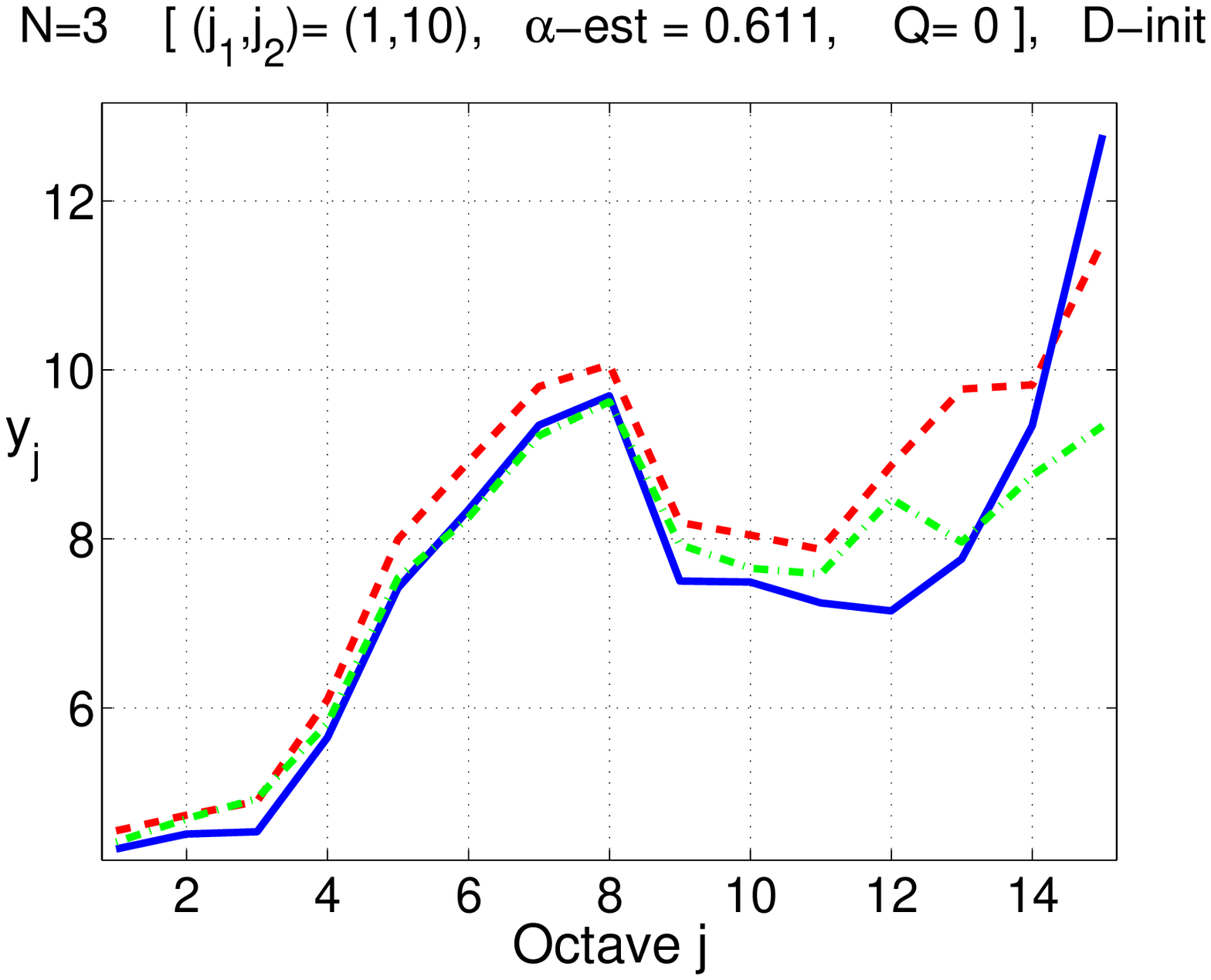}
\label{subfig:tvants_stat_pakup_video}
}
\subfigure[Download: overall packet traffic (signaling and video traffic)]{
\includegraphics[scale=0.30]{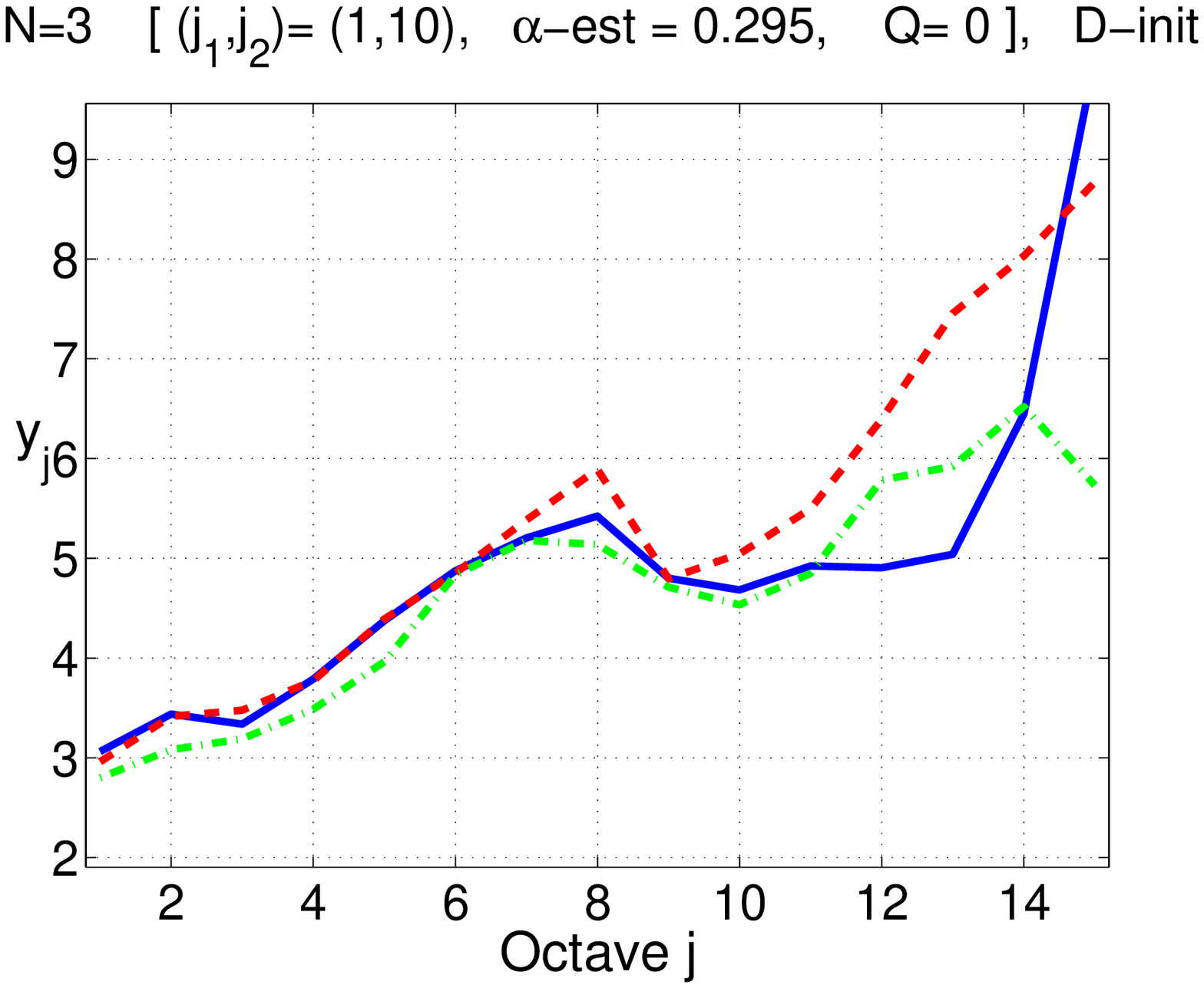}
\label{subfig:tvants_stat_pakdown}
}
\subfigure[Download: video packet traffic]{
\includegraphics[scale=0.30]{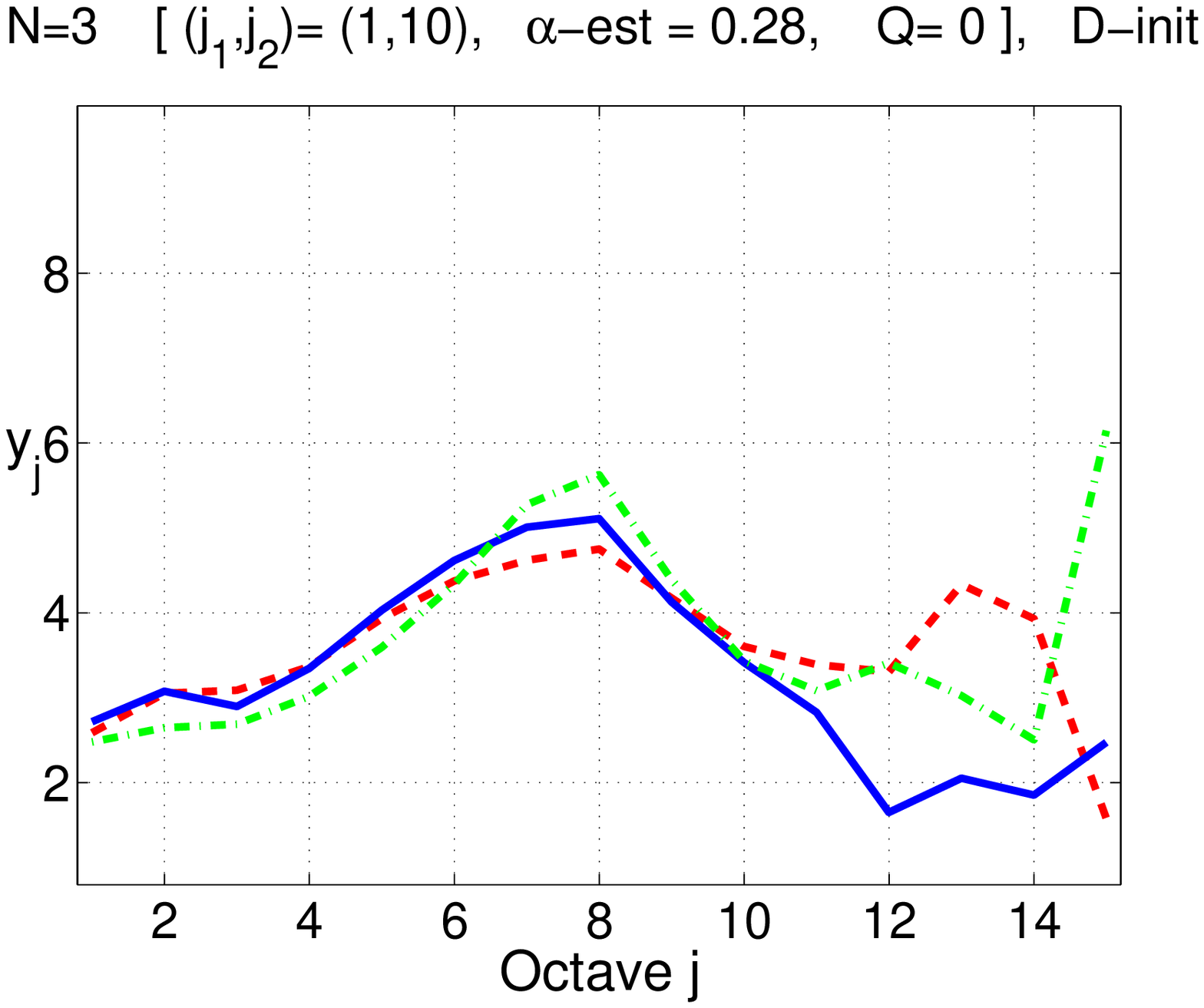}
\label{subfig:tvants_stat_pakdown_video}
}
\caption{TVAnts: Traffic stationarity for the three equal parts of the traffic. The blue solid line is the first part, the dashed red line is the second part and the dash-dotted green line is the third part of the traffic.}
\label{fig:sopcast_stat}
\end{center}
\end{figure*}
\begin{figure*}[!h]
\begin{center}
\subfigure[PPStream Download: overall packet traffic (signaling and video traffic)]{
\includegraphics[scale=0.30]{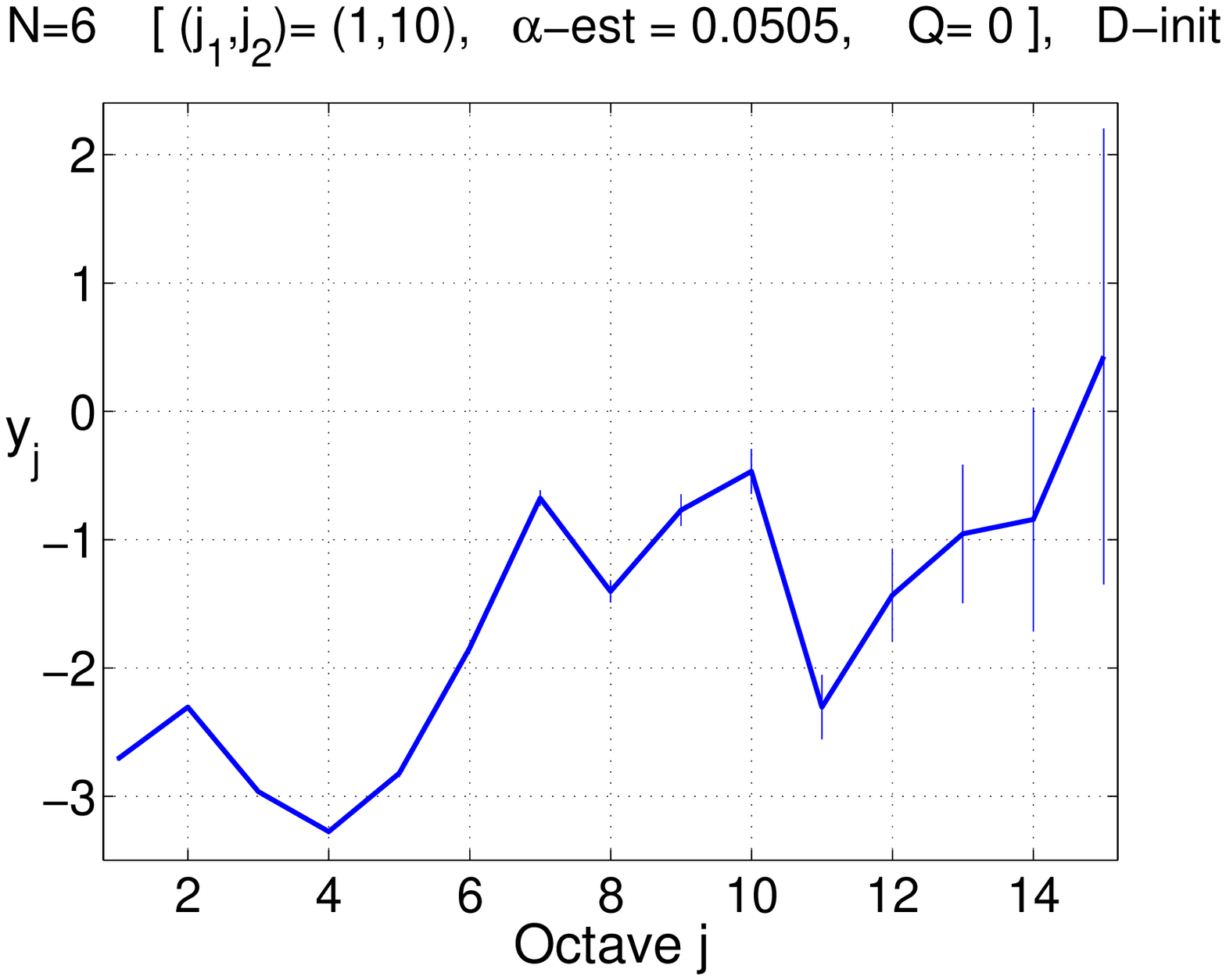}
\label{subfig:pplive_flow}
}
\subfigure[PPStream Download: video packet traffic]{
\includegraphics[scale=0.30]{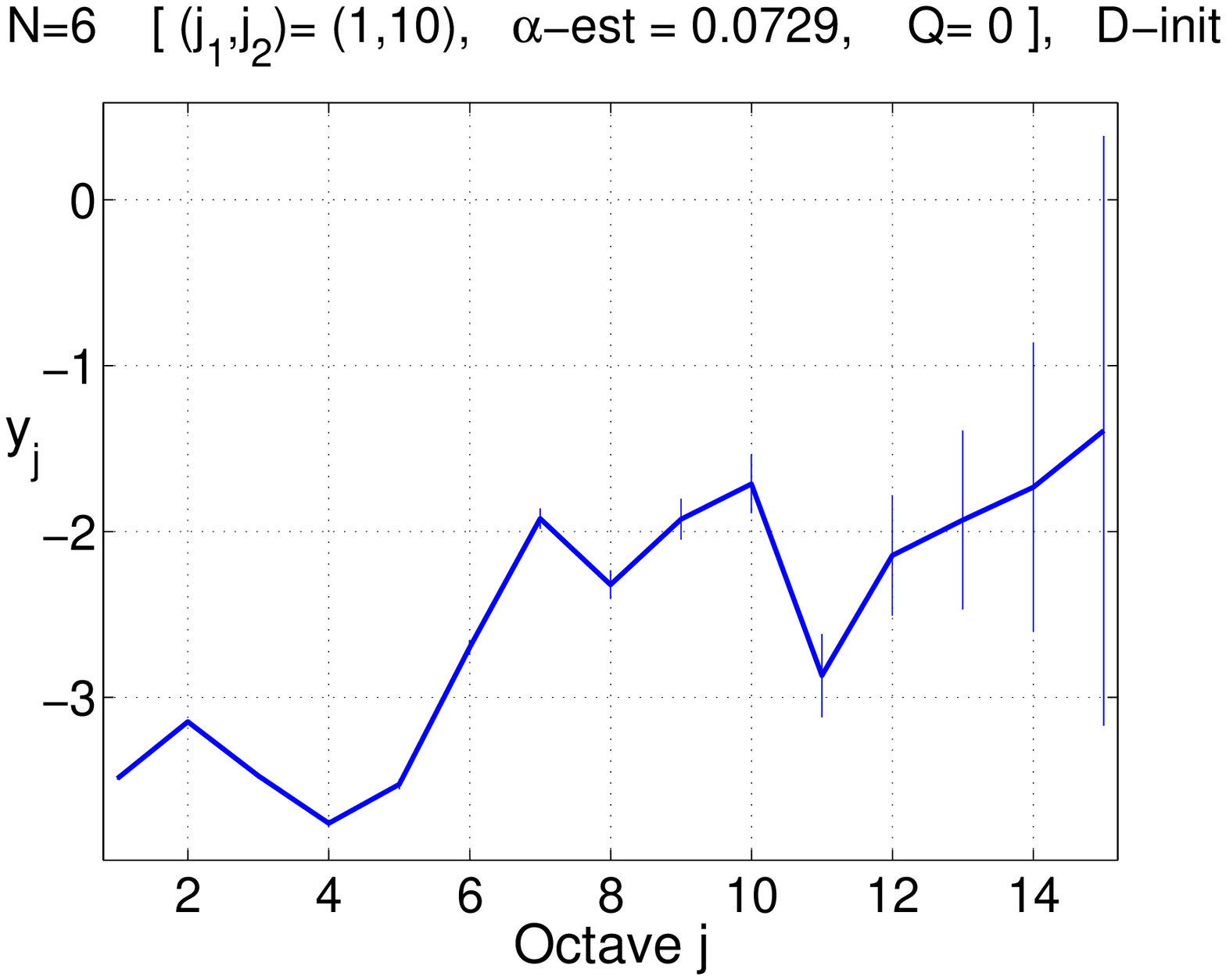}
\label{subfig:pplive_flow_video}
}
\subfigure[TVAnts Download: overall packet traffic (signaling and video traffic)]{
\includegraphics[scale=0.30]{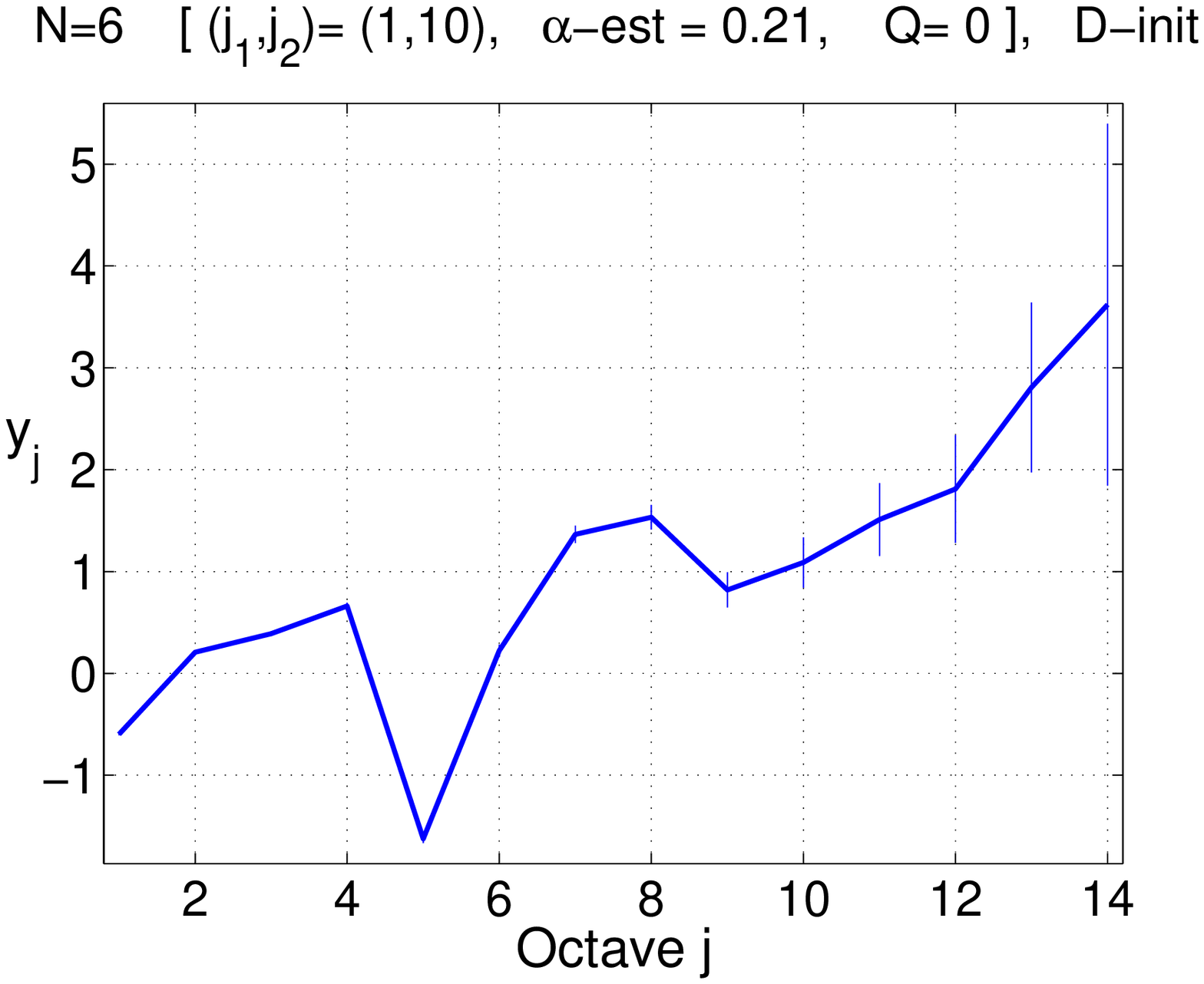}
\label{subfig:sopcast_flow}
}
\subfigure[TVAnts Download: video packet traffic]{
\includegraphics[scale=0.30]{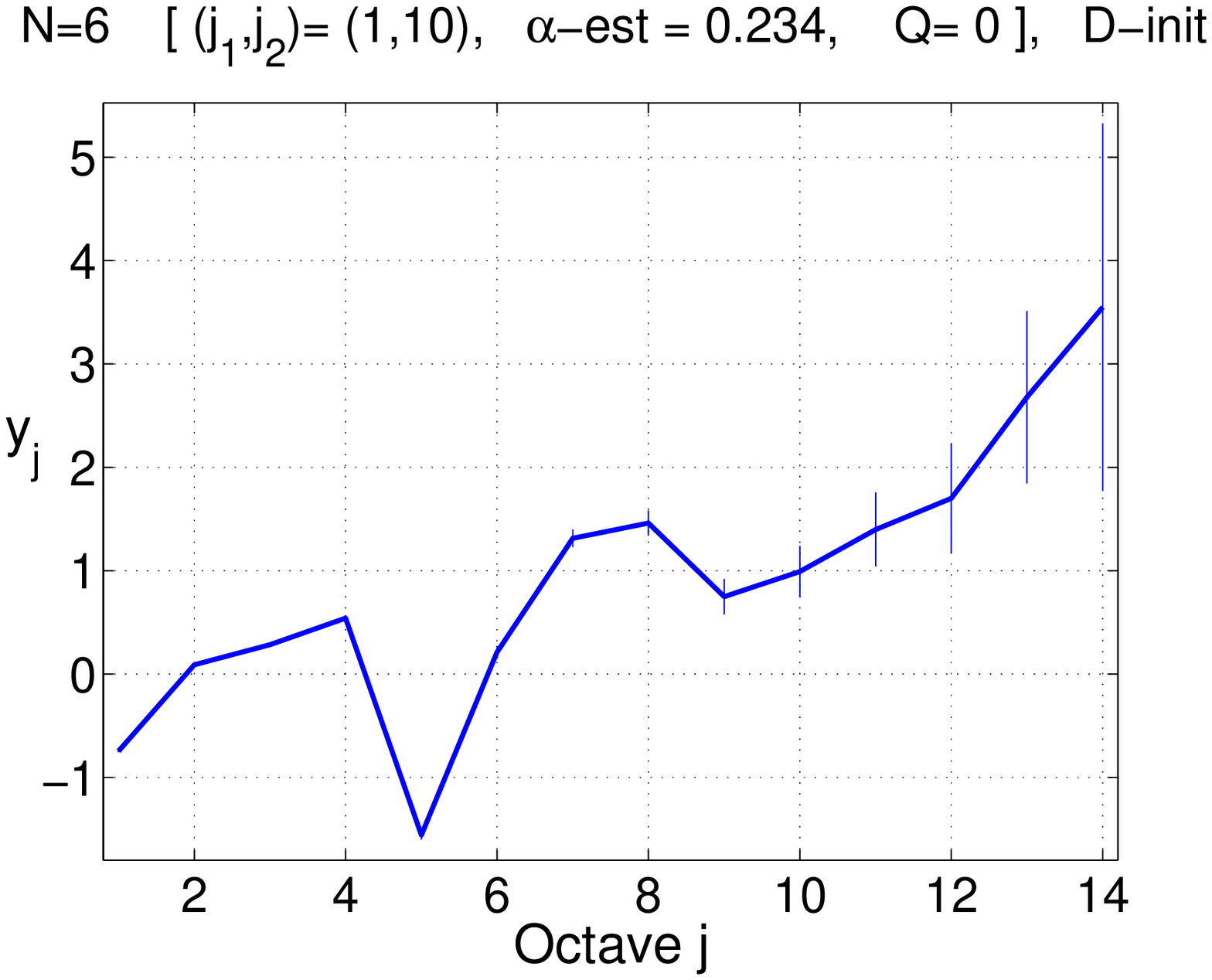}
\label{subfig:sopcast_flow_video}
}
\caption{Top download flows: PPStream: (a) and (b). TVAnts: (c) and (d)}
\label{fig:topflows}
\end{center}
\end{figure*}

\begin{figure*}[!h]
\begin{center}
\subfigure[PPLive Download: overall packet traffic (signaling and video traffic)]{
\includegraphics[scale=0.30]{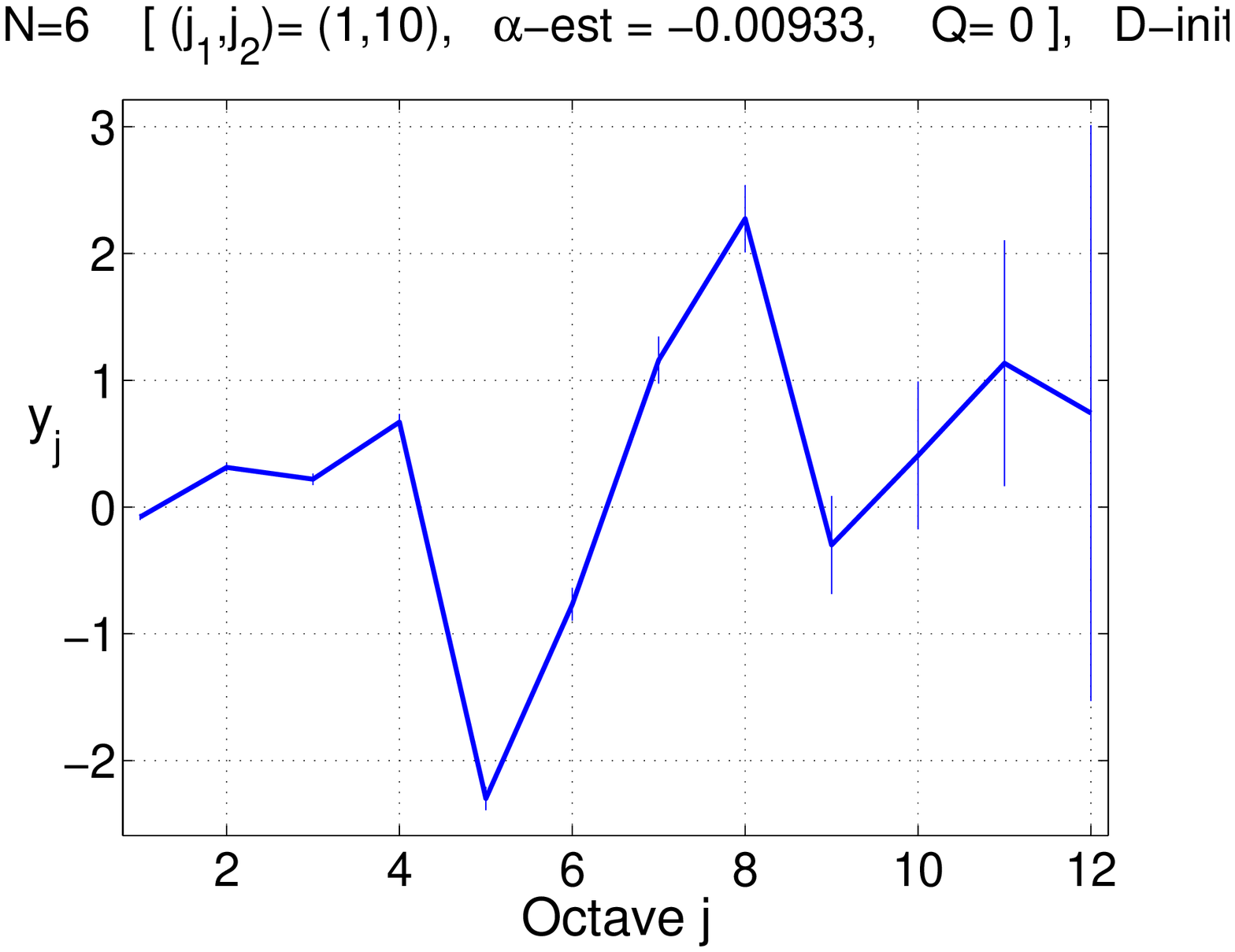}
\label{subfig:pplive_10flow}
}
\subfigure[PPLive Download: video packet traffic]{
\includegraphics[scale=0.30]{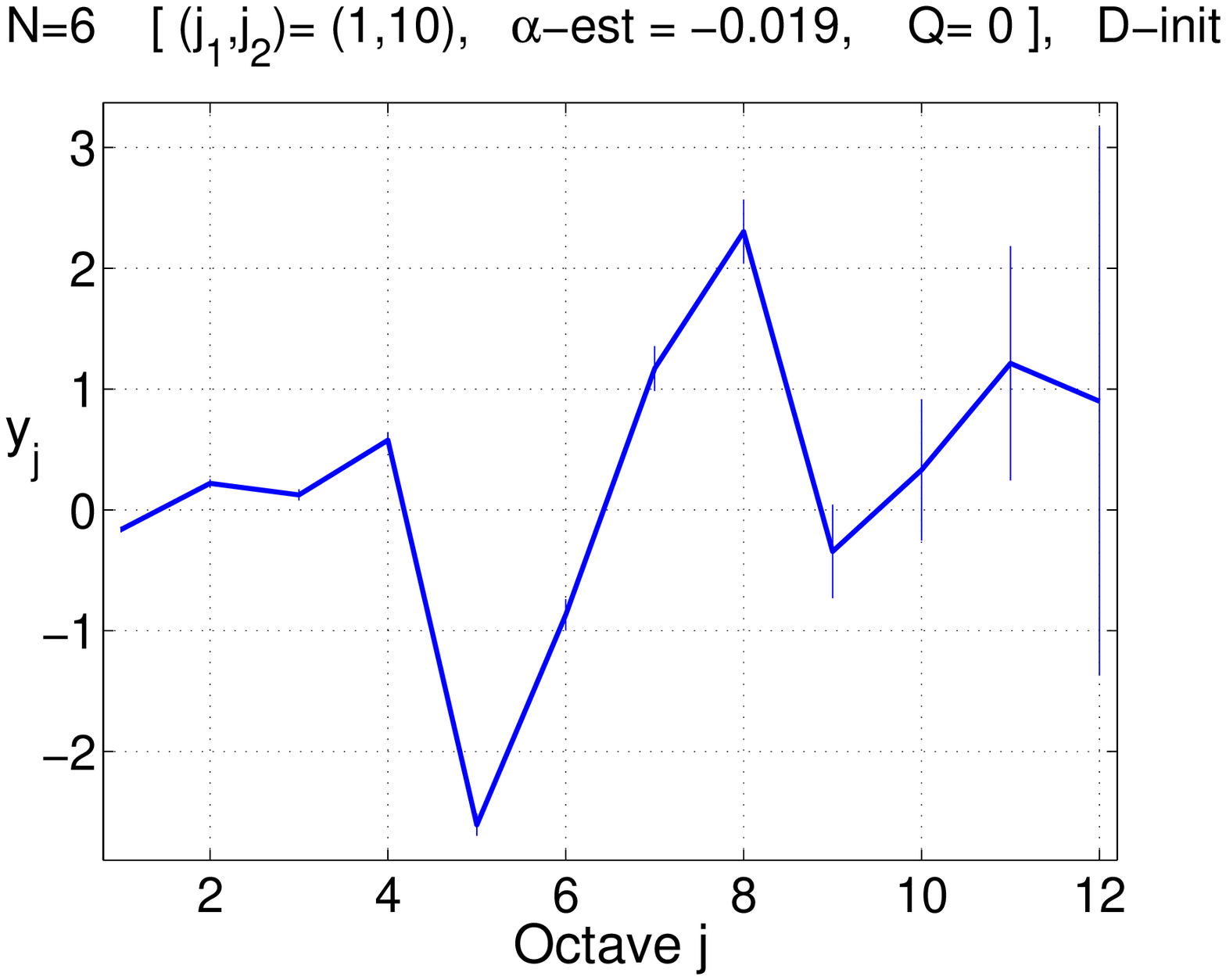}
\label{subfig:pplive_10flow_video}
}
\subfigure[SOPCast Download: overall packet traffic (signaling and video traffic)]{
\includegraphics[scale=0.30]{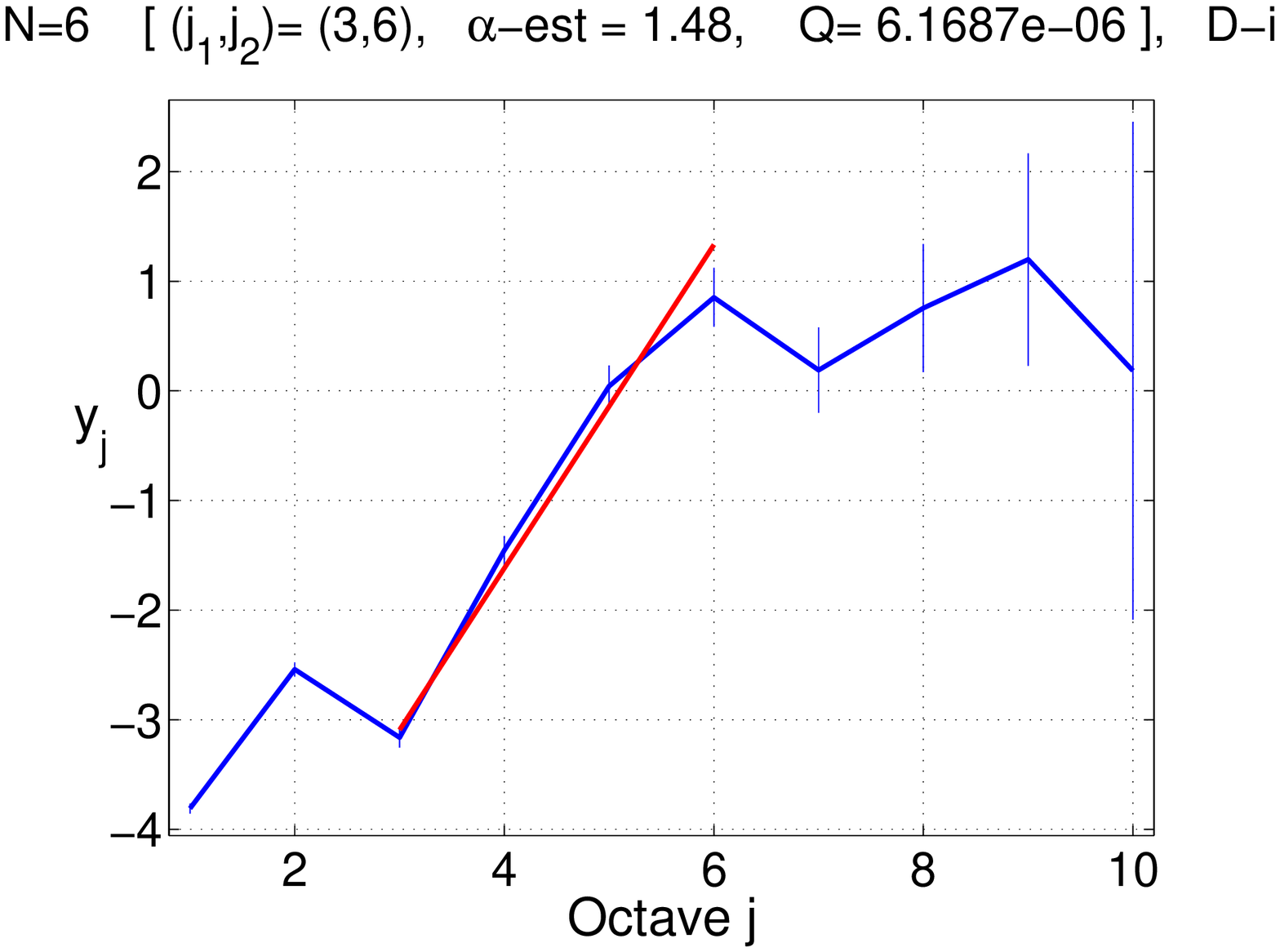}
\label{subfig:sopcast_10flow}
}
\subfigure[SOPCast Download: video packet traffic]{
\includegraphics[scale=0.30]{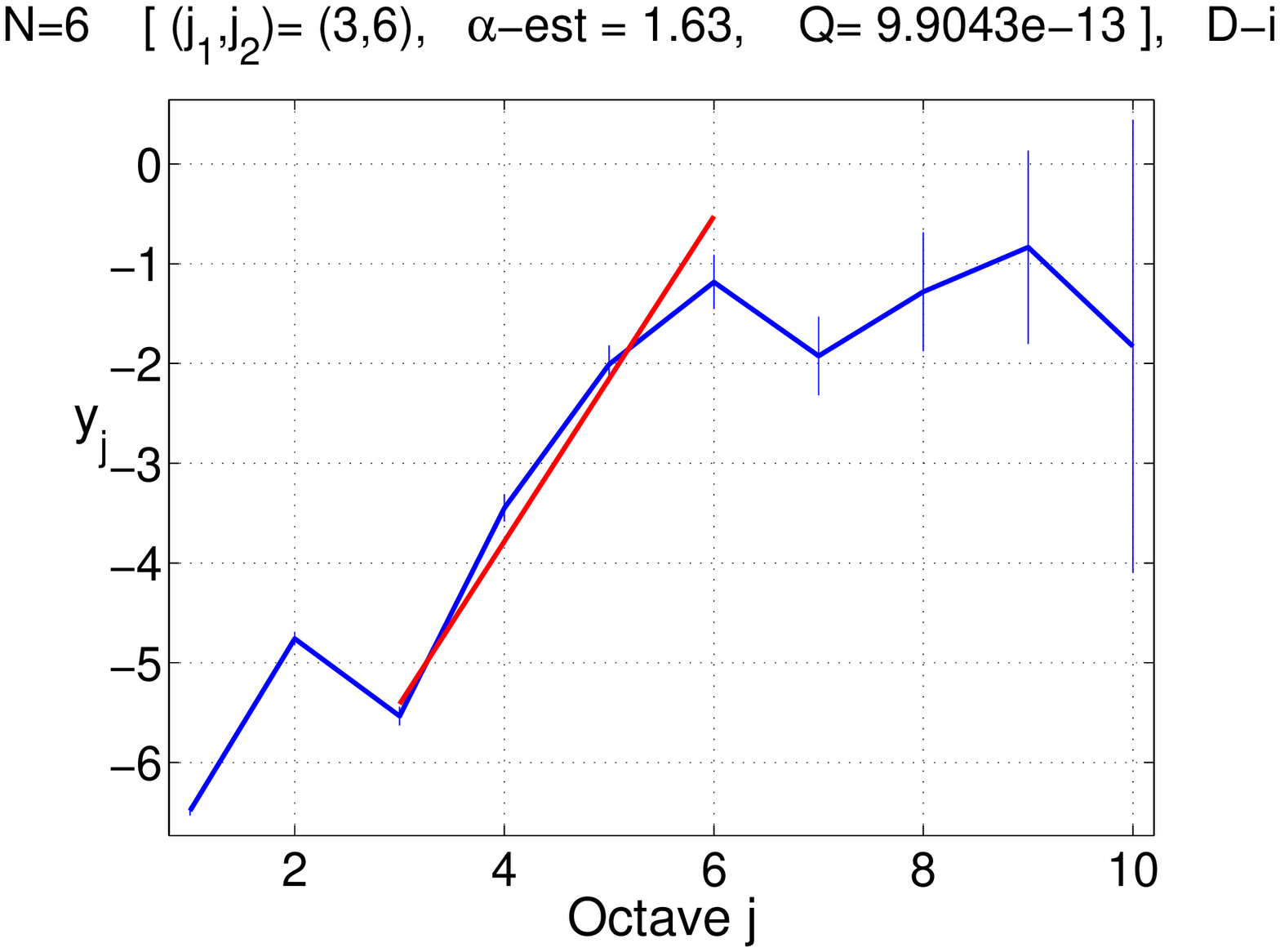}
\label{subfig:sopcast_10flow_video}
}
\subfigure[PPStream Download: overall packet traffic (signaling and video traffic)]{
\includegraphics[scale=0.30]{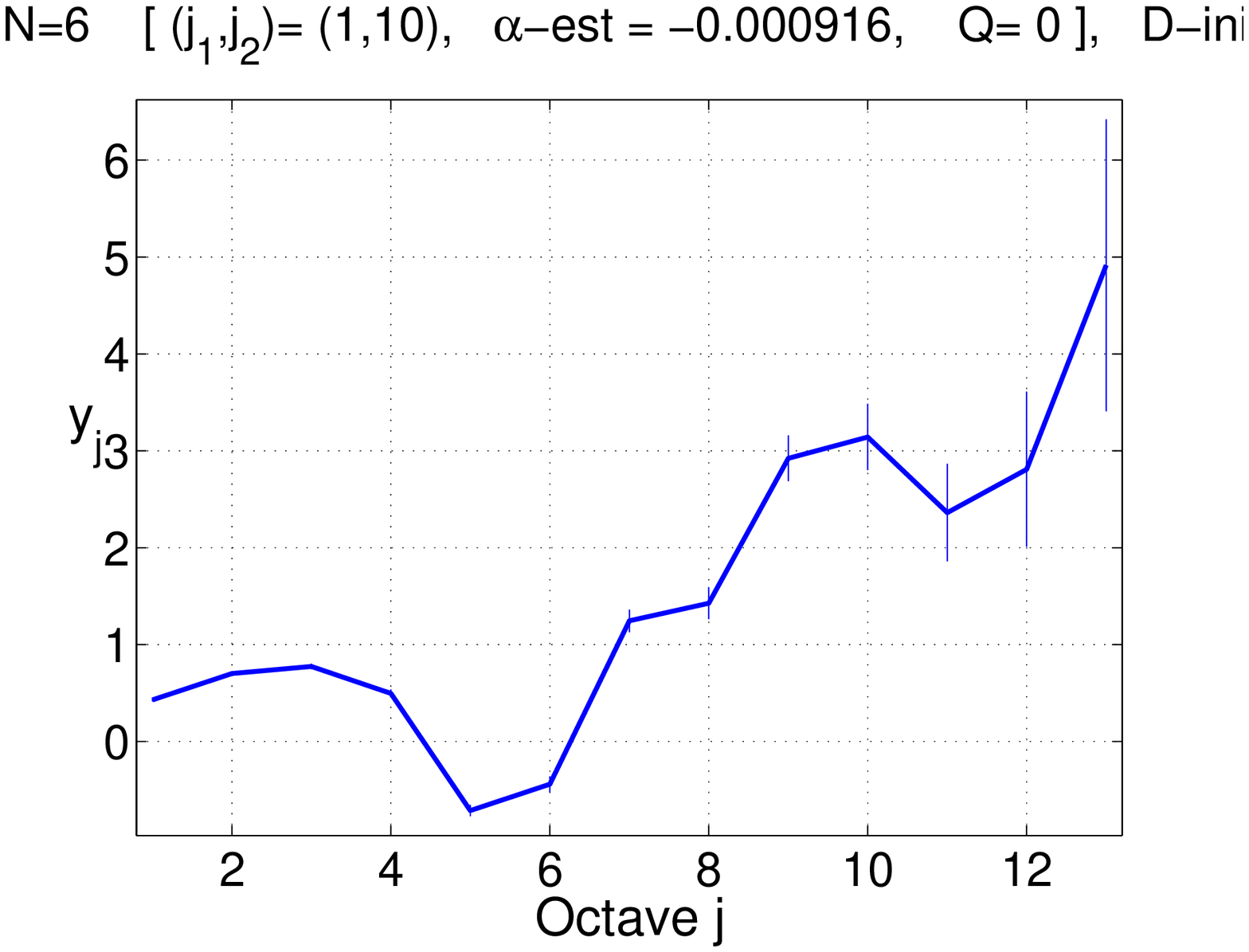}
\label{subfig:ppstream_10flow}
}
\subfigure[PPStream Download: video packet traffic]{
\includegraphics[scale=0.30]{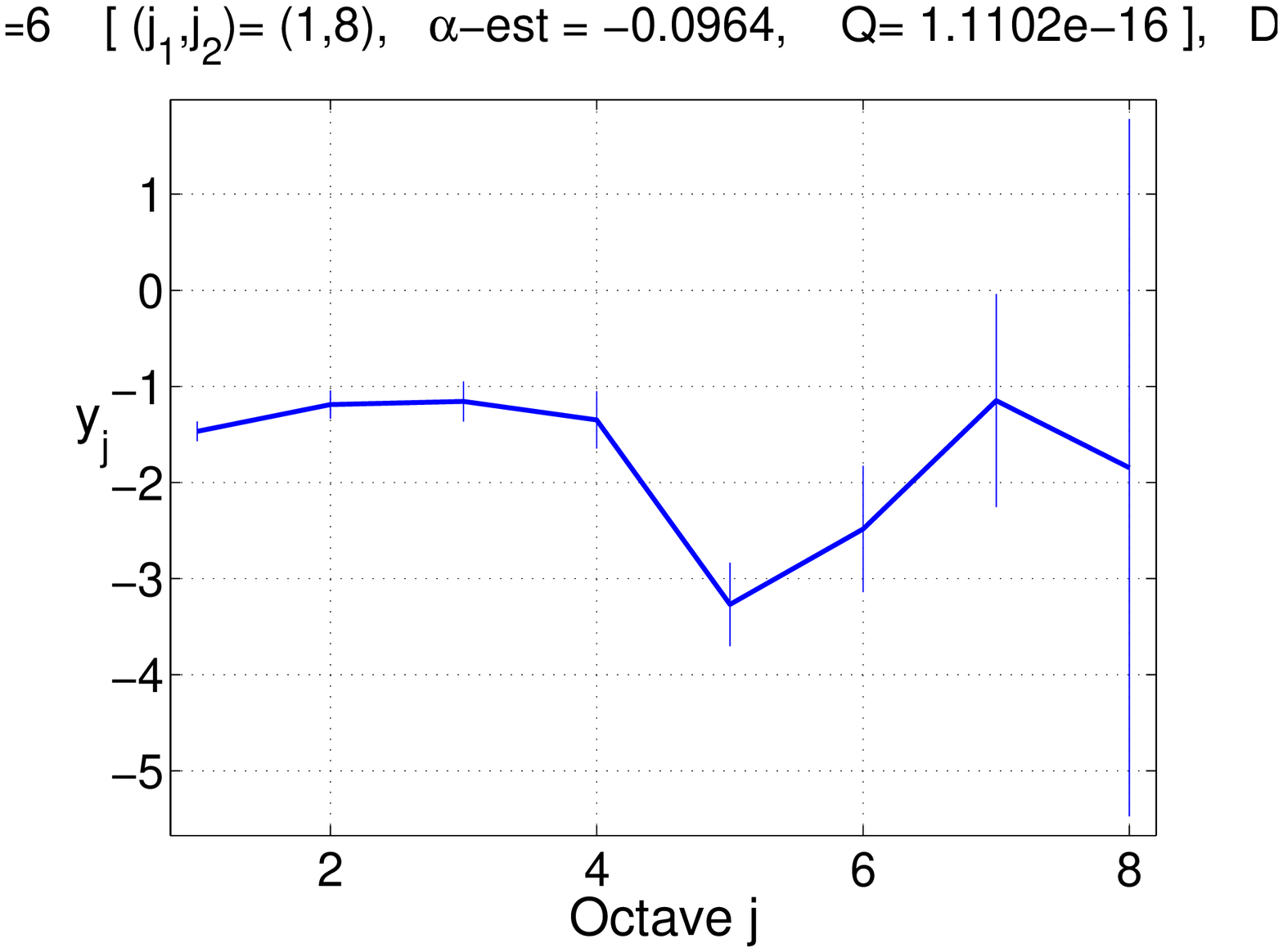}
\label{subfig:ppstream_10flow_video}
}
\subfigure[TVAnts Download: overall packet traffic (signaling and video traffic)]{
\includegraphics[scale=0.30]{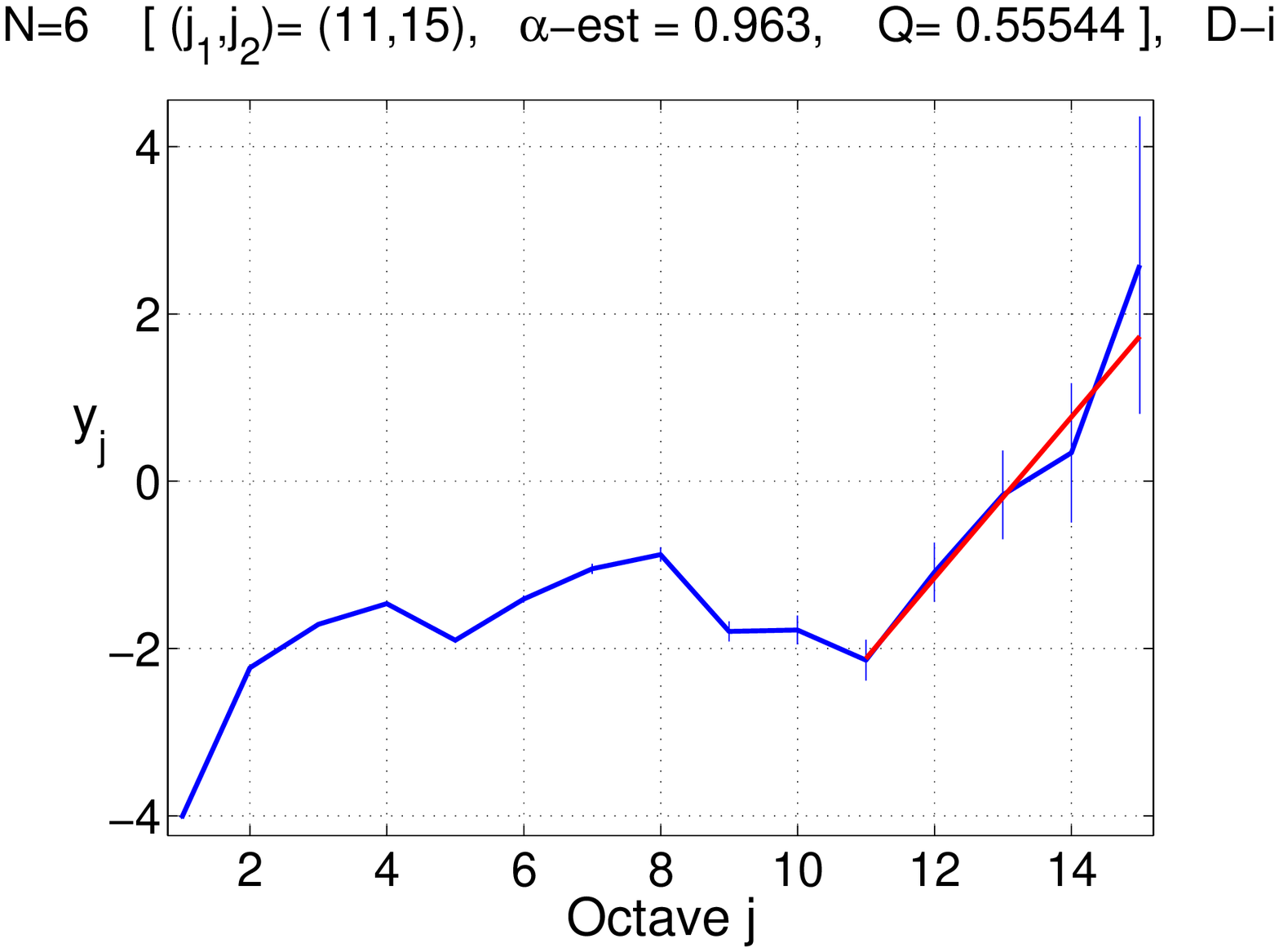}
\label{subfig:tvants_10flowo}
}
\subfigure[TVAnts Download: video packet traffic]{
\includegraphics[scale=0.30]{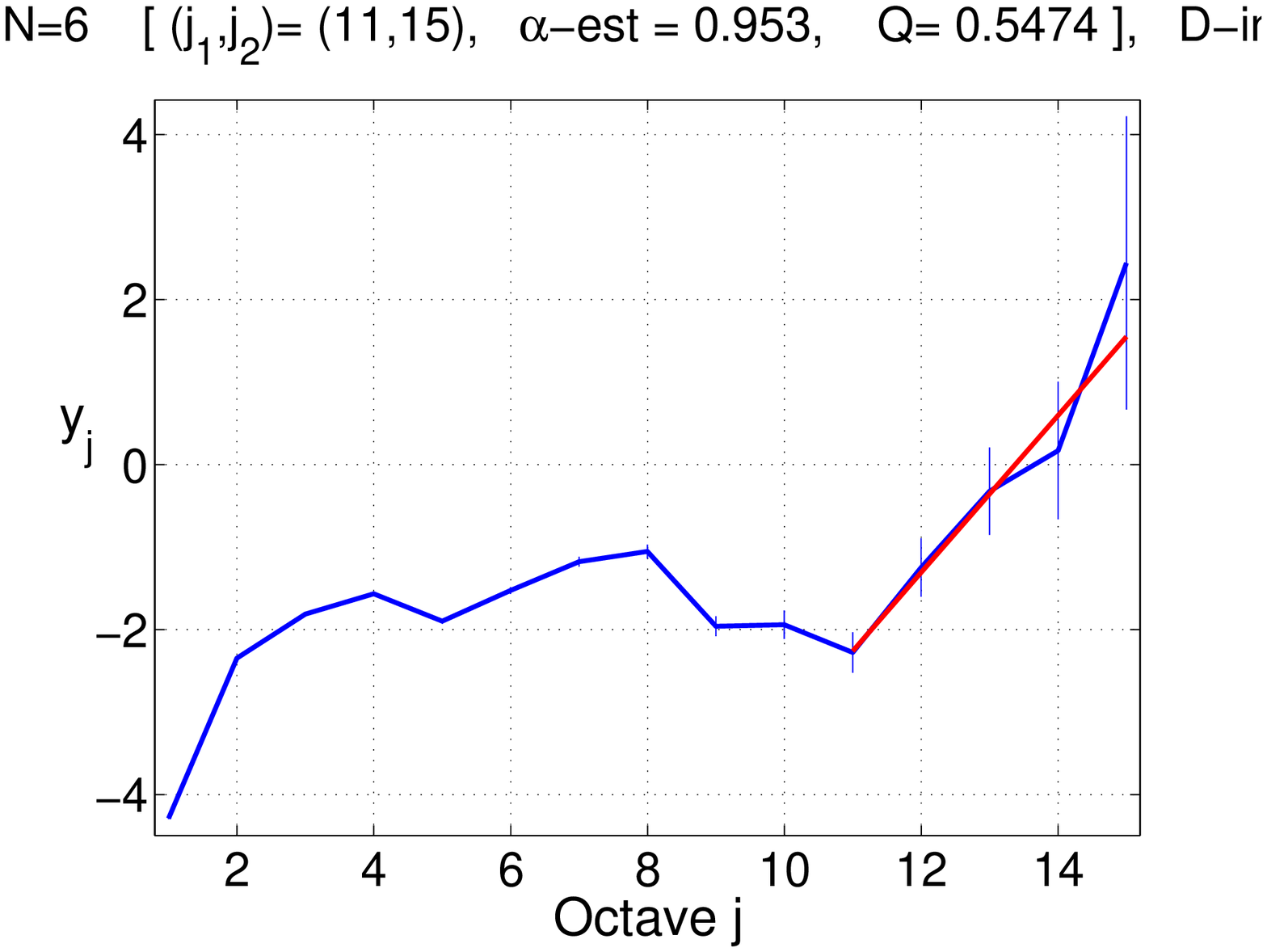}
\label{subfig:tvants_10flow_video}
}
\caption{10th top download flows: PPLive: (a) and (b). SOPcast: (c) and (d). PPStream: (e) and (f). TVAnts: (g) and (h)}
\label{fig:topflows}
\end{center}
\end{figure*}

\begin{thebibliography}{10}
\providecommand{\url}[1]{#1}
\csname url@rmstyle\endcsname
\providecommand{\newblock}{\relax}
\providecommand{\bibinfo}[2]{#2}
\providecommand\BIBentrySTDinterwordspacing{\spaceskip=0pt\relax}
\providecommand\BIBentryALTinterwordstretchfactor{4}
\providecommand\BIBentryALTinterwordspacing{\spaceskip=\fontdimen2\font plus
\BIBentryALTinterwordstretchfactor\fontdimen3\font minus
  \fontdimen4\font\relax}
\providecommand\BIBforeignlanguage[2]{{%
\expandafter\ifx\csname l@#1\endcsname\relax
\typeout{** WARNING: IEEEtran.bst: No hyphenation pattern has been}%
\typeout{** loaded for the language `#1'. Using the pattern for}%
\typeout{** the default language instead.}%
\else
\language=\csname l@#1\endcsname
\fi
#2}}

\bibitem{bittorrent}
B.~Cohen, ``Incentives build robustness in bittorrent,'' 2003.

\bibitem{edonkey}
http://www.edonkey2000.com.

\bibitem{youtube}
http://www.youtube.com.

\bibitem{pplive}
http://www.pplive.com.

\bibitem{ppstream}
http://www.ppstream.com.

\bibitem{sopcast}
http://www.sopcast.com.

\bibitem{tvants}
http://www.tvants.com.

\bibitem{invited:coolstreaming}
X.~Zhang, J.~Liu, and B.~Li, ``On large-scale peer-to-peer live video
  distribution: Coolstreaming and its preliminary experimental results,'' in
  \emph{Proc. MMSP}, 2005.

\bibitem{infocom05:donet}
X.~Zhang, J.~Liu, B.~Li, and T.~P. Yum, ``Coolstreaming/donet: A data-driven
  overlay network for peer-to-peer live media streaming,'' in \emph{Proc.
  Infocom}, 2005.

\bibitem{pplive-mesure}
X.~Hei, C.~Liang, J.~Liang, Y.~Liu, and K.~W. Ross, ``Insights into pplive: A
  measurement study of a large-scale p2p iptv system,'' in \emph{Proc. of IPTV
  Workshop, International World Wide Web Conference}, 2006.

\bibitem{pplive-tom07}
X.~Hei, C.~Liang, J.~Liang, Y.~Liu, and K.~W. Ross, ``A measurement study of a large-scale p2p iptv system,'' \textit{to appear} in \emph{IEEE
  Transactions on Multimedia}, 2007.

\bibitem{vu-qshine07}
L.~Vu, I.~Gupta, J.~Liang, and K.~Nahrstedt, ``Measurement and modeling of a
  large-scale overlay for multimedia streaming,'' in \emph{Proc. of QSHINE'07,
  International Conference on Heterogeneous Networking for Quality,
  Reliability, Security and Robustness}, 2007.

\bibitem{silverston-nossdav07}
T.~Silverston and O.~Fourmaux, ``Measuring p2p iptv systems,'' in \emph{Proc.
  of NOSSDAV'07, International Workshop on Network and Operating Systems
  Support for Digital Audio \& Video}, 2007.

\bibitem{commercial-measurement}
S.~Ali, A.~Mathur, and H.~Zhang, ``Measurement of commercial peer-to-peer live
  video streaming,'' in \emph{Proc. of Workshop in Recent Advances in
  Peer-to-Peer Streaming}, 2006.

\bibitem{pplive-methodology}
\BIBentryALTinterwordspacing
X.~Hei, Y.~Liu, and K.~W. Ross, ``Inferring network-wide quality in p2p live
  streaming systems.'' [Online]. Available:
  \url{http://cis.poly.edu/~ross/papers/buffermap.pdf}
\BIBentrySTDinterwordspacing

\bibitem{content-traces}
http://content.lip6.fr/traces/.

\bibitem{joost}
http://www.joost.com.

\bibitem{LDestimate}
\BIBentryALTinterwordspacing
D.~Veitch and P.~Abry, ``Matlab code for the wavelet based analysis of scaling
  processes.'' [Online]. Available:
  \url{http://www.cubinlab.ee.mu.oz.au/$\sim$darryl}
\BIBentrySTDinterwordspacing

\end{thebibliography}
\end{document}